\documentclass[defaultstyle,10pt,Helvetica]{istulthesis}

\usepackage{url}
\usepackage[section]{placeins}
\usepackage{comment}

\usepackage{amsthm}
\theoremstyle{definition}
\newtheorem{definition}{Definition}[section]

\newtheorem{theorem}{Theorem}
\newtheorem{lemma}{Lemma}
\newtheorem{corollary}{Corollary}

\usepackage{ucs}

\usepackage[utf8x]{inputenc}

\usepackage[main=english,portuguese]{babel}

\usepackage{iflang}

\usepackage{scrbase}

\usepackage{enumerate}

\usepackage{mathtools, amsmath, amsthm, amssymb, amsfonts}

\usepackage{multirow}
\usepackage{colortbl}
\usepackage{ctable}
\usepackage{booktabs}

\usepackage{graphicx}
\usepackage{subfig}
\usepackage[format=hang,labelfont=bf,font=small]{caption} 
\usepackage[ruled,vlined,algochapter,norelsize,\languagename]{algorithm2e}

\usepackage{listings}
\usepackage[numbered]{./tables_and_code/mcode}

\newcaptionname{portuguese}{\lstlistingname}{Listagem} 
\newcaptionname{portuguese}{\lstlistlistingname}{Listagens} 

\usepackage{xcolor}
\usepackage{color}
\definecolor{forestgreen}{RGB}{34,139,34}

\definecolor{orangered}{RGB}{239,134,64}
\definecolor{lightred}{rgb}{1,0.4,0.5}
\definecolor{orange}{rgb}{1,0.45,0.13}	

\definecolor{darkblue}{rgb}{0.0,0.0,0.6}
\definecolor{lightblue}{rgb}{0.1,0.57,0.7}

\definecolor{gray}{rgb}{0.4,0.4,0.4}
\definecolor{lightgray}{rgb}{0.95, 0.95, 0.95}
\definecolor{darkgray}{rgb}{0.4, 0.4, 0.4}
\definecolor{editorGray}{rgb}{0.95, 0.95, 0.95}
\definecolor{editorOcher}{rgb}{1, 0.5, 0} 
\definecolor{chaptergrey}{rgb}{0.6,0.6,0.6}

\definecolor{editorGreen}{rgb}{0, 0.5, 0} 

\definecolor{olive}{rgb}{0.17,0.59,0.20}
\definecolor{brown}{rgb}{0.69,0.31,0.31}

\definecolor{purple}{rgb}{0.38,0.18,0.81}

\usepackage{setspace}

\usepackage{paralist}

\usepackage{cite}

\usepackage[printonlyused]{acronym}

\usepackage{hyperref}
\hypersetup{ colorlinks=true,
             citecolor=cyan,
             linkcolor=darkgray,
             urlcolor=teal,
             breaklinks=true,
             bookmarksnumbered=true,
             bookmarksopen=true,
             pdftitle=Computational Complexity of Game and Puzzles, 
             pdfauthor=Diogo Costa,  
             pdfcreator=Diogo Costa,   
}

\usepackage{url} 

\usepackage{fancyhdr}
\fancypagestyle{plain}{%
   \fancyhead{} 
   
}
\fancypagestyle{blank}{%
   \fancyhf{} 
   
}
\fancypagestyle{abstract}{%
   \fancyhead{}

}
\fancypagestyle{document}{%
	\fancyhead{}

	\addtolength{\headheight}{2pt} 
}

\usepackage{minitoc}
\renewcommand*{\kernafterminitoc}{\kern0.\baselineskip\kern0.ex}
\mtcselectlanguage{\languagename} 

\def\boxedverbatim{%
  \def\verbatim@processline{%
    {\setbox0=\hbox{\the\verbatim@line}%
    \hsize=\wd0 \the\verbatim@line\par}}%
  \@minipagetrue
  \@tempswatrue
  \setbox0=\vbox\bgroup\vspace*{0.2cm}\footnotesize\verbatim
}
\def\endboxedverbatim{%
  \endverbatim
  \unskip\setbox0=\lastbox 
  \hspace*{0.2cm}
  \vspace*{-0.2cm}
  \egroup
  \fbox{\box0}
}

\newcommand*{\chapnumfont}{%
  \usefont{T1}{pbk}{b}{n}
  \fontsize{150}{130}
  \selectfont
  \color{chaptergrey}
}
\makeatletter
\def\@makechapterhead#1{%
  \vspace*{50\p@}%
  {\parindent \z@ \raggedright \normalfont
    {\chapnumfont\ifnum \c@secnumdepth >\m@ne
        \raggedleft\bfseries \thechapter
        \par\nobreak
        \vskip 20\p@
    \fi}
    \interlinepenalty\@M
    {\raggedleft\Huge \bfseries #1\par\nobreak}
    \vskip 40\p@
  }}
\makeatother

\newcommand{\fancychapter}[1]{\chapter{#1}\vfill\minitoc\pagebreak}

\newcommand{\hlinew}[1]{%
  \noalign{\ifnum0=`}\fi\hrule \@height #1 \futurelet
   \reserved@a\@xhline}
   
\def\Mark#1{\raisebox{0pt}[0pt][0pt]{\textsuperscript{\footnotesize\ensuremath{\ifcase#1\or *\or \dagger\or \ddagger\or%
    \mathsection\or \mathparagraph\or \|\or **\or \dagger\dagger%
    \or \ddagger\ddagger \else\textsuperscript{\expandafter\romannumeral#1}\fi}}}}

\lstdefinestyle{XML} {
	language=XML,
	extendedchars=true, 
	breaklines=true,
	breakatwhitespace=true,
	emph={},
	emphstyle=\color{red},
	basicstyle=\small,
	xleftmargin=17pt,
	columns=fullflexible,
	commentstyle=\color{gray}\upshape,
	morestring=[b][\color{brown}]",
	morecomment=[s]{<?}{?>},
	morecomment=[s][\color{forestgreen}]{<!--}{-->},
	keywordstyle=\color{orangered},
	stringstyle=\ttfamily\color{black},
	tagstyle=\color{blue},
	morekeywords={asn,action,addrType,abilityNAT,audioSampleRate,audiChannels,,bandwidth,bitmapSize,bitRate,connection,codecs,concurrentLinks,dependency,duration,frameRate,from,height,ip,id,lang,mimeType,onlineTime,peerMode,port,priority,peerProtocol,property,release,to,tier,type,transactionID,url,uploadBWlevel,version,width},
	otherkeywords={attribute,xmlns,schemaLocation,PresentationType,availabilityStartTime,availabilityEndTime,minimumUpdatePeriod,minBufferTime,UpdateTime},
}

\lstdefinelanguage{Assembler}{
	morecomment=[l];,
	keywords={ADD,ADDC,SUB,SUBB,CMP,MUL,DIV,MOD,NEG,AND,OR,NOT,XOR,TEST,BIT,SET,EI,EI0,EI1,EI2,EI3,SETC,EDMA,CLR,DI,DI0,DI1,DI2,DI3,CLRC,SHR,SHL,SHRA,SHLA,ROR,ROL,RORC,ROLC,MOV,MOVB,MOVBS,MOVP,MOVL,MOVH,SWAP,PUSH,POP,JZ,JNZ,JN,JNN,JP,JNP,JC,JNC,JV,JNV,JEQ,JNE,JLT,JLE,JGT,JGE,JA,JAE,JB,JBE,JMP,CALL,CALLF,RET,RETF,SWE,RFE,NOP},
	morekeywords={EQU,TABLE,WORD,STRING,PLACE},
} 

\lstdefinestyle{coloredASM}{
	language=Assembler,
	extendedchars=false,
	breaklines=true,
	tabsize=2,
	numberstyle=\tiny,
	numbers=left,
	breakatwhitespace=true,
	emph={},
	emphstyle=\color{red},
	fontadjust=true,
	basicstyle=\small\ttfamily,
	columns=fixed,
	xleftmargin=17pt,
	framexleftmargin=17pt,
	framexrightmargin=5pt,
	framexbottommargin=4pt,
	commentstyle=\color{forestgreen}\upshape,
	morestring=[b][\color{brown}]",
	keywordstyle=\color{darkblue},
	stringstyle=\ttfamily\color{black},
	literate={á}{{\'a}}1 {ã}{{\~a}}1 {â}{{\^a}}1 {é}{{\'e}}1 {É}{{\'E}}1 {ê}{{\^e}}1 {õ}{{\~o}}1 {ó}{{\'o}}1 {í}{{\'i}}1 {ç}{{\c{c}}}1 {Ç}{{\c{C}}}1,
}    

\lstdefinelanguage{CSS}{
	sensitive=true,
	morecomment=[l]{//},
	morecomment=[s]{/*}{*/},
	morestring=[b]',
	morestring=[b]",
	alsoletter={:},
	alsodigit={-},
	keywords={color,background-image:,margin,padding,font,weight,display,position,top,left,right,bottom,list,style,border,size,white,space,min,width, transition:, transform:, transition-property, transition-duration, transition-timing-function}
}

\lstdefinelanguage{JavaScript}{
	morecomment=[s]{/*}{*/},
	morecomment=[l]//,
	morestring=[b]",
	morestring=[b]',
	morekeywords={typeof, new, true, false, catch, function, return, null, catch, switch, var, if, in, while, do, else, case, break}
}

\lstdefinelanguage{HTML5}{
	language=html,
	sensitive=true,	
	alsoletter={<>=-},	
	morecomment=[s]{<!-}{-->},
	tag=[s],
	otherkeywords={
	>,
	<!DOCTYPE,
	</html, <html, <head, <title, </title, <style, </style, <link, </head, <meta, />,
	</body, <body,
	</div, <div, </div>, 
	</p, <p, </p>,
	</script, <script,
	<canvas, /canvas>, <svg, <rect, <animateTransform, </rect>, </svg>, <video, <source, <iframe, </iframe>, </video>, <image, </image>, <header, </header, <article, </article},
	ndkeywords={
	=,
	charset=, src=, id=, width=, height=, style=, type=, rel=, href=,
	fill=, attributeName=, begin=, dur=, from=, to=, poster=, controls=, x=, y=, repeatCount=, xlink:href=,
	margin:, padding:, background-image:, border:, top:, left:, position:, width:, height:, margin-top:, margin-bottom:, font-size:, line-height:,
	transform:, -moz-transform:, -webkit-transform:,
	animation:, -webkit-animation:,
	transition:,  transition-duration:, transition-property:, transition-timing-function:,
	}
}

\lstdefinestyle{htmlcssjs} {%
	backgroundcolor=\color{editorGray},
		fontadjust=true,
	basicstyle=\small\ttfamily,   
	frame=b,
	xleftmargin={0.75cm},
	numbers=left,
	stepnumber=1,
	firstnumber=1,
	numberfirstline=true,	
	identifierstyle=\color{black},
	keywordstyle=\color{blue}\bfseries,
	ndkeywordstyle=\color{editorGreen}\bfseries,
	stringstyle=\color{editorOcher}\ttfamily,
	commentstyle=\color{brown}\ttfamily,
	language=HTML5,
	alsolanguage=JavaScript,
	alsodigit={.:;},	
	tabsize=2,
	showtabs=false,
	showspaces=false,
	showstringspaces=false,
	extendedchars=true,
	breaklines=true,
	literate=%
	{Ö}{{\"O}}1
	{Ä}{{\"A}}1
	{Ü}{{\"U}}1
	{ß}{{\ss}}1
	{ü}{{\"u}}1
	{ä}{{\"a}}1
	{ö}{{\"o}}1
}

\lstdefinestyle{py} {%
	language=python,
	literate=%
	*{0}{{{\color{lightred}0}}}1
	{1}{{{\color{lightred}1}}}1
	{2}{{{\color{lightred}2}}}1
	{3}{{{\color{lightred}3}}}1
	{4}{{{\color{lightred}4}}}1
	{5}{{{\color{lightred}5}}}1
	{6}{{{\color{lightred}6}}}1
	{7}{{{\color{lightred}7}}}1
	{8}{{{\color{lightred}8}}}1
	{9}{{{\color{lightred}9}}}1,
	basicstyle=\small\ttfamily,
	numbers=left,
	numbersep=5pt,
	tabsize=4,
	extendedchars=true,
	breaklines=true,
	keywordstyle=\color{blue}\bfseries,
	frame=b,
	commentstyle=\color{brown}\itshape,
	stringstyle=\color{editorOcher}\ttfamily,
	showspaces=false,
	showtabs=false,
	xleftmargin=17pt,
	framexleftmargin=17pt,
	framexrightmargin=5pt,
	framexbottommargin=4pt,
	backgroundcolor=\color{lightgray},
	showstringspaces=false,
}

\begin{document}

\pdfbookmark[0]{Titlepage}{Title}
%
\univlogo{2cm}{2cm}{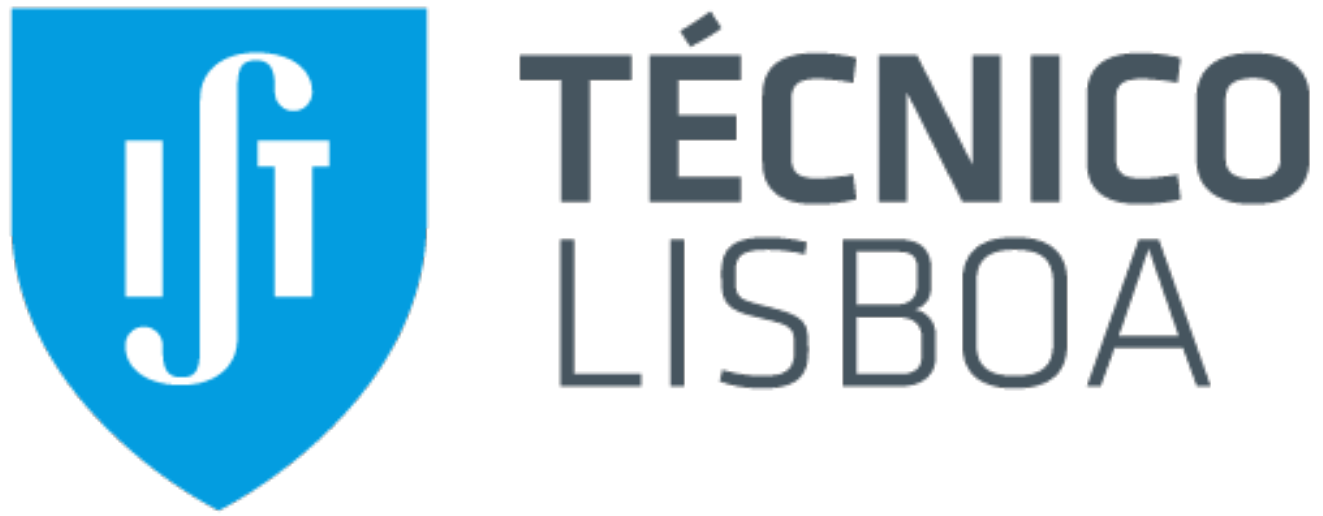}
\thesislogo{2.5cm}{6cm}{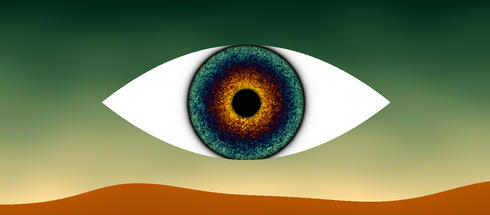}
%
\title{Computational Complexity of Games and Puzzles}
%
\author{Diogo Manuel dos Santos Costa}
%
%
\degree{Information Systems and Computer Engineering}
%
\supervisor{Prof. Luís Manuel Silveira Russo}
\othersupervisor{Prof. Alexandre Paulo Lourenço Francisco}
%
\date{July 2018}
%
\finalthesis{false}
%
\chairperson{Prof. Name of the Chairperson}
\vogalone{Prof. Name of First Committee Member}
\vogaltwo{Dr. Name of Second Committee Member}
\vogalthree{Eng. Name of Third Committee Member}
%
%
\maketitle
\clearpage
\thispagestyle{empty}

\cleardoublepage

\setcounter{page}{1} \pagenumbering{roman}
\baselineskip 18pt 


\pdfbookmark[0]{Abstract}{Abstract}
\begin{abstract}
\noindent In this thesis, we survey techniques and results from the study of Complexity Theory and Games. We then apply these techniques to obtain new results for previously unstudied games. Our contributions in the games Hexiom, Cut the Rope, and Back to Bed may be helpful in further studies by exploiting structure common to several games.
We also highlight some interesting paths for further study, related to uncertainty, that have yet to receive thorough study given their prevalence in today's games.
\end{abstract}
\clearpage
\thispagestyle{empty}
\cleardoublepage

\pdfbookmark[0]{Resumo}{Resumo}
\begin{resumo}
\noindent Nesta tese, estudamos várias técnicas e resultados da relação entre a Teoria da Complexidade e Jogos. Usamos as técnicas estudadas para obter novos resultados. As nossas contribuições para com os jogos Hexiom, Cut the Rope e Back to Bed poderão ser úteis para obter futuros resultados noutros jogos, porque tiram partido de estrutura comum a vários jogos que ainda não foram estudados.
Também salientamos possíveis direcções para estudo adicional, especialmente em relação à incerteza, algo prevalente nos jogos modernos mas ainda pouco estudado. 
\end{resumo}
\begin{palavraschave}
\noindent jogos, puzzles, complexidade, reduções
\end{palavraschave}
\clearpage
\thispagestyle{empty}
\cleardoublepage

\dominitoc
\dominilof
\dominilot

\renewcommand{\baselinestretch}{1}
\pdfbookmark[0]{Contents}{toc}
\tableofcontents
\cleardoublepage
\renewcommand{\baselinestretch}{1.5}

\pdfbookmark[1]{List of Figures}{lof}
\listoffigures
\paragraph*{Notes on the figures:}
Figures of boolean circuits were made using Logisim. Figures of the games Akari, HexCells, Hexiom and Back to Bed were made using custom software. Figures of Offspring Fling and Super Meat Boy were made using the official level editor supplied with those games. All the remaining figures were made using GIMP. Unless noted otherwise in their caption, figures are original to this thesis and its companion article. When taken from a book or paper, we present only a reproduction of the original. 
\cleardoublepage

\pdfbookmark[1]{List of Tables}{lot}
\listoftables




\setcounter{page}{1} \pagenumbering{arabic}
\baselineskip 18pt

\fancychapter{Introduction}
\cleardoublepage
\label{chap:intro}
Games have been at the forefront of computers since their start, and are often associated with breakthroughs in artificial intelligence~\cite{alphago2016,deepblue}. 
Even as early as 1951, one of the first computers built, the Nimrod, was made to play the game Nim, a pub game where players remove beads from different stacks trying to get the last move.
Very recently, AlphaGo~\cite{alphago2016} and AlphaGo Zero have surpassed human Go players.
These two games set the stage for the topic of this report, the Complexity of games, in two opposing extremes.

Although it's not clear at first, Nim is a very simple game to ``solve''. There is a trick to find the optimal move given any state of the game that consists of simply XORing the various stacks. The player can force a victory by finishing a turn while the state of the stacks XOR to 0. Several variations exist, and their complexity has been studied by Fenner and Rogers~\cite{posetgames}.

Go and Chess, on the other hand, are very complex games~\cite{go,chess}, for which we have no easy trick to win. This may be one of the reasons for the difference in popularity and longevity between the games. Chess and Go are still interesting to this day, even with computers performing better than humans.

A formal study of the game's difficulty can be given through Complexity Theory (CT), identifying which complexity classes these problems belong to, showing previously unrecognized relations between many different games. 
Nim\footnote{ By Nim, here, we mean the standard version, where players can take any number of tokens from a single stack. Both \textit{normal} and \textit{mis\`ere} versions are computationally equivalent for this game. In ``normal'' play, the last player to move wins. In ``mis\`ere'', the last player to move loses.} is in the class \textbf{P}, because there is an algorithm that finds the optimal play in a polynomial time in the size of the game.
Go, on the other hand, is in \textbf{EXPTIME}-Complete. This is a class in which we can do no better than to have an algorithm that requires $O(2^{n^k})$ time.

Complexity classes are established on how the difficulty (time to solve) of a problem increases with its size. This doesn't always apply to games directly, which are mostly played on a fixed sized board. 
For example, although Go can be played on boards of different sizes, there is a fixed standard size. 
What looking at a game through a CT lens reveals, however, is the richness of a problem, in a sense. In two player games especially, it hints at the depth of the game and the different kinds of sub problems people will find by playing the game.
A similar argument can be used for puzzle games, since each board configuration is a new puzzle.

If there is a ``trick'' to solve a given game, no matter how many game states will create, we will always be able to find the answer quickly. 
If we can prove that there is no trick, we're likely to be able to find particularly difficult instances of the game that provide interesting puzzles or end-games.

\section{Motivation}
There are several reasons to look at games as more than mere distractions. As will become clearer as we proceed, many games are based on deep computational problems, though often with a lower ``barrier to entry'', since games rarely require a formal educational background to be played.

\subsection{``Serious'' Games}
Recently, especially in the field of Bioinformatics, there's been a wave of puzzles being used to encapsulate real problems~\cite{eterna,foldit}. These are often combinatorial optimization problems, for which we already have algorithms, but that are often sub-optimal (to run in reasonable times). In the game, players' answers are pooled and compared to the current best solution given by the algorithms, and the former replace the latter when an improvement has been found.\footnote{ Some examples are PHYLO \url{http://phylo.cs.mcgill.ca/\#!/EN/About}, Foldit \url{http://fold.it/portal/info/about} and eteRNA \url{http://www.eternagame.org/web/about/} }

Apart from combinatorial algorithms, this method has been applied to correct Image Processing data~\cite{eyewire}.
The algorithms used to find patterns aren't perfect, but they are a good guide. Although there has been significant progress with Machine Learning techniques, humans still have the edge identifying patterns from visual input.
This gives players the chance to correct the results from the Image Processing algorithms in game form.\footnote{ The game Eyewire asks players to map segments of neurons. Players are given a 3D cube from brain imaging, and then work on square cross-sections to map the 3D segment of the neuron. Several players solve the same Cube and the answers are then combined to minimize errors. More information can be found at \url{http://blog.eyewire.org/about/} }

Not yet as common as in Bioinformatics, a similar approach is being taken on Quantum Physics. Although notoriously unintuitive, there is no reason for it to stay that way. The world of quantum is unfamiliar, at least in part, because of the little contact we have with it in our day to day lives. However, no one can say that Chess is intuitive. Games are often made out of arbitrary rules and, by playing them, we gain an intuition of how the rules work and form simpler models to make predictions and evaluations of game states. 
Quantum games are a way to provide that familiarity, introducing rules in the form of games, giving people the chance to play around with the system and create an intuition of how they work.\footnote{One example uses the particle deflection experiments to create puzzles: \url{http://quantumgame.io/}; Another, Decodoku, uses games to develop better Quantum Correction methods to help Quantum Computing: \url{http://decodoku.com/}. Other examples can be found.}

Finally, and more closely related to this thesis, is to use puzzles as a benchmark~\cite{ricochet,ricochet1,ricochet2,hexiomsat} for current algorithms (SAT solvers in particular), and a possible source of heuristics~\cite{ricochetpsy}.
Because we can reduce computational problems (such as SAT and TQBF, defined in Section~\ref{subsec:originalproblems}) to puzzles, we can then ask players to solve them and analyze the way they play to find ways of improving the algorithms we currently have.

\subsection{Game Design}
Another reason to look at games with CT is that it provides concrete ways to analyze a game. It's a useful \textit{tool}.

Puzzles come in a variety of forms. They differ in the length and number of solutions, scoring mechanics based on time, damage taken, etc.
Different game mechanics lead to different types of puzzle. Often, especially in video games, puzzles are not the focus, but used as mini-games to provide variety. 
The kind of challenge each puzzle type provides is likely to be different, depending on the complexity class it belongs in.

Many times, puzzles are hand crafted. A puzzle designer has the task of designing specific puzzles, with certain solutions and paths in mind. Here it helps to know what kind of problems one expects by looking at the mechanics alone.
One of the explored problems regarding puzzle design has to do with the difficulty of counting solutions, or even finding them, given one known solution~\cite{asp}.

Logic also provides a good framework to solve instances of puzzles (optimally), which can be used as a guide or verification of hand crafted puzzles.

More importantly, however, is so called \textit{procedural generation}. Games are increasingly generated by randomized algorithms, whether in map generation, enemy encounters, modeling, or puzzles.
Here, the designers have less control over the puzzles that are generated. Certain properties can be introduced by using different kinds of mechanics and studying the relationship between mechanics and the resulting complexity of a puzzle can be a useful tool to control the random elements.

\section{Objectives}
The objective of this thesis is twofold:
\begin{itemize}
\item 
First, to survey the reduction methods that are commonly used to prove the complexity of games and puzzles, and to understand which techniques fit different types of games by applying them to new games of unknown complexity.

\item
Second, to get an intuition for the type of mechanics that result in the different complexity classes. To find simple mechanisms of getting specific levels of complexity/difficulty.
\end{itemize}

\section{Structure}
The thesis' structure is as follows:
\begin{itemize}
    \item Chapter 1 starts with the motivation and introduces the basic theoretical notions and problems relevant to the work that follows. We also do a brief survey discussing work directly and indirectly related to the topics treated later.
    \item Chapter 2 describes the frameworks used and gives an example proof for each, taken from the respective papers.
    \item Chapter 3 contains all of our results using the frameworks described in Chapter 2.
    \item Finally, Chapter 4 contains possible avenues for further study that we have found over the course of this thesis and concludes this work. 
    \item Appendix A shows an example construction of each of our results. 
\end{itemize}
\cleardoublepage
%
\fancychapter{Background}
\cleardoublepage
\label{chapter:background}
In this chapter, we start Section~\ref{section:basic} by introducing the prerequisite notions of Complexity for the proofs that follow in Chapter~\ref{chapter:results}. In Section~\ref{section:survey} we survey some of the work related to games and complexity in general. Finally, in Section~\ref{section:frameworks}, we detail the results from the papers that serve as a basis for our work. All the proofs presented in Chapter~\ref{chapter:results} start in the frameworks presented here. Section~\ref{section:example} shows three proofs to illustrate the frameworks. The proofs were taken from the respective papers.

\section{Introduction to Complexity Theory and Important Results and Problems}\label{section:basic}
Here we will introduce the important concepts of Complexity Theory that are the foundation of the problems studied in this thesis. We introduce the notion of complexity classes, how a problem is related to them, and some of the field's prototypical problems. 

\subsection{Basic Notions \& Definitions}
\textit{Computational Complexity Theory} is a field of computer science that analyses the difficulty of various algorithmic problems. The following paragraphs briefly describe some central concepts of this field, but we invite the reader to a more rigorous treatment of the topic in~\cite{papadimitriou}, for example.

Whereas \textit{Computability Theory} analysis what kind of problems can be solved by an algorithm, Complexity Theory takes the problems that can, in principle, be solved algorithmically, and tries to establish lower and upper bounds on the resources necessary to solve it. This is important because some problems may be solvable in principle, but require impractical amounts of resources. 

This analysis is typically done by relating unknown problems with other well studied problems. This led to the definition of various \textit{complexity classes} - sets of problems with the same difficulty.
Another approach to Complexity Theory is to study the relations between classes themselves. This has led to the famous \textbf{P}=\textbf{NP} open problem.

To formalize the problem, it's necessary to define the \textit{resources} to measure -- typically time and memory, although others can be used, depending on the context (energy, queries to an external device, etc.) -- and a suitable \textit{model of computation} -- which started with the use of Turing Machines, coming from Computability Theory, to which common digital computers can be reduced.

This thesis deals with \textit{decision problems} - problems with a Yes or No answer. In the context of puzzles, the question is whether or not the given puzzle has a solution. 
The proof of correctness is the solution (sequence of movements, for example) to the puzzle.

The complexity of a problem always deals with a family of problems, rather than specific instances of any one problem.  
We measure how the resources required grow with the size of the problem. As a result, many of the game related proofs require a generalization of the game being analyzed: Chess, for example, must be generalized to a version with arbitrary board sizes (using $N \times N$ instead of $8 \times 8$).

\subsection{Complexity Classes}
Here we briefly define the most relevant complexity classes  to this thesis.

\begin{definition}[The class \textbf{P}]\label{classP}
    \textbf{P} is the class of decision problems that can be decided in polynomial time (some polynomial function on the size of the input), in a deterministic machine.
\end{definition}
This is the class of ``easy'' or tractable.

\begin{definition}[The class \textbf{NP}]\label{classNP}
    \textbf{NP} is the class of decision problems that can be decided in polynomial time, on a \textit{non-deterministic} machine.
\end{definition}
Consider a computation to be a search tree. Where a deterministic machine would need to search every branch individually, a non-deterministic machine searches all branches ``at once'', or ``guesses'' the correct branch.

Another (perhaps more intuitive) way to look at the NP class is as the class of problems whose solutions can be \textit{verified} in polynomial time. Verification means that, given a potential solution, there is a (deterministic) polynomial time algorithm that tells us whether or not the candidate solution is correct.
Many problems of interest (and puzzle games) belong to this class. A popular example is Sudoku~\cite{asp}.

\begin{definition}[The class \textbf{PSPACE}]\label{classPSPACE}
    \textbf{PSPACE} is the class of decision problems that can be decided in polynomial space, on a deterministic machine.
\end{definition}
Space is, in a sense, more powerful than time, because it can be reused. As a result, problems that require a polynomial amount of space may require exponential time (given $n$ binary cells of space, there are $2^n$ different configurations).
It is also an important class for some puzzle games (especially puzzles that require many steps to reach the end goal, like Rush Hour~\cite{ncl}) and some two player games like Hex~\cite{hex,hexthesis} and Othello/Reversi~\cite{ncl}.

\begin{definition}[The class \textbf{NPSPACE}]\label{classNPSPACE}
    \textbf{NPSPACE} is the class of decision problems that can be decided in polynomial space, on a non-deterministic machine. 
\end{definition}
Non-determinism in space doesn't play a big role as in time, however, since we can simulate any non-deterministic machine in a deterministic one, only with a quadratic blowup in space (by Savitch's Theorem~\cite{savitch}). The composition of polynomials is still a polynomial, so \textbf{PSPACE}=\textbf{NPSPACE}.

\begin{definition}[The class \textbf{EXPTIME}]\label{classEXPTIME}
    \textbf{EXPTIME} is the class of decision problems that can be decided in exponential ($O(2^{n^k})$) time, for some $k \in \mathbb{N}$.
\end{definition}
This class includes popular two player board games such as Go~\cite{go}, Chess~\cite{chess} and Checkers~\cite{checkers}. One crucial aspect of these games is that the number of moves in a game is unbounded, increasing its complexity from \textbf{PSPACE} to \textbf{EXPTIME}. This is also the only class that has been proven to be harder than \textbf{P}~\cite{hierarchy}, with \textbf{PSPACE} and \textbf{NP} only being conjectured to be harder than \textbf{P}.

\subsection{Membership, Hardness \& Completeness}
Finally, the concepts of \textit{Membership}, \textit{Hardness} and \textit{Completeness} are introduced. Here we consider considering a generic problem A and class C.

\begin{definition}[Membership]\label{membership}
    A problem A is said to be in C ($A \in C$) if there is an algorithm that solves it using no more resources than the ones specified for C.
\end{definition}

For example, A $\in$ \textbf{P} if there is an algorithm that can solve it using at most polynomial time. If a problem is in \textbf{P}, then it is also in every class that contains \textbf{P}, such as \textbf{EXPTIME}. We can trivially find an algorithm that runs in exponential time for the same problem by running the polynomial time and then padding the problem with redundant operations for an exponential amount of time.
If a problem is in \textbf{EXPTIME}, however, it may not be in \textbf{P}. If a problem always requires exponential time to be solved, then there will not be a polynomially bounded algorithm to solve it, and so A $\notin$ \textbf{P}.
The same applies to space.

\begin{definition}[Hardness]\label{hardness}
    A problem A is C-Hard if it requires \textit{at least as much} time/space as the ones specified in C. 
\end{definition}
This property works in the opposite way as Membership.
Consider \textbf{P} and \textbf{EXPTIME}.
If a problem is \textbf{EXPTIME}-Hard, it is also \textbf{P}-Hard, since, by requiring exponential time, it also requires at least polynomial time.
However, if a problem is \textbf{P}-Hard, it may not be in \textbf{EXPTIME}-Hard, since , if it belongs to \textbf{P}, there is an algorithm that always does better than exponential time.

\begin{definition}[Hardness]\label{completeness} 
    A problem is C-Complete if it is in C \textit{and} it is C-Hard.
\end{definition} 
This is a property of the ``hardest'' problems for the given class C, since solving one amounts to solving every one of them, as will shown in the next section about reductions.

\paragraph*{Problems and Instances.} As mentioned before, complexity studies families of problems rather than single instances. In particular, infinite families. This is because any finite family can, in principle, be precomputed, with every solution stored in a table to be retrieved in constant time (using idealized memory), so the notion of complexity loses its value.

Informally, a problem is the whole set of possible instances. An instance is a particular configuration allowed by the problem's rules. Consider the game of Chess, where the decision problem is ``does Black/White win?'', given a board configuration (size and pieces' positions) and assuming optimal play from the opponent. Any board configuration is a different instance of the general problem, from the starting configuration with every piece on the board, to the end-game configurations with few pieces left. It is important to note that the complexity of the problem is not tied to any particular instance, but the whole set. Taking the Chess example again, it is easy to find a winning sequence of moves (for Black) if White has only a king and Black has a king and 2 towers. The optimal sequence is trivial and with an upper bound of $O(\textit{board size})$ (move the towers alternately, forcing the king to move towards a wall until the edge for a check-mate). The hardest puzzles will require an exponential number of moves.

Referring back to the concept of infinite family introduced above, we can see that the standard game of Chess does not fit the definition. Since the board has a fixed size ($8 \times 8$) and a finite number of pieces, it also has a finite number of possible configurations. This is a very large number, making the game effectively intractable to brute-force, but such is not a concern of Complexity Theory. Most games are finite in this way. To analyze those games, then, we must generalize them. In the case of Chess, we consider $N \times N$ boards and add pieces as naturally possible (for example, each side has $N$ pawns in an $N \times N$ board just as they have 8 pawns in an $8 \times 8$ board). There are multiple possible generalizations and is up to the readers to decide whether or not such a generalization substantially changes the nature of the game in question. Specific generalizations of Chess can be seen in~\cite{chess,chesspspace}.

\paragraph*{Instance Size.} The size of an instance is measured by the space required to write a certain instance as input to the algorithm; a problem's complexity measures how the amount of resources required by the algorithm grow with the instance size. Note that it is not the resources taken for a single instance; instead, the usual approach is to consider the hardest instances of a specific size as the worst case scenarios. It is important to emphasize, again, that the \textit{hardest} instances must still be infinite! If they were finite, we could precompute them, being left with a possibly infinite number of easy instances, and a finite number of hard instances, which became easy by constant time look-up. 

The size of instances is dependent on a chosen \textit{encoding}.
This is especially relevant for number related problems, where numbers are \textit{not} usually considered in unary, but base $2$ (or $10$), reducing the instance's size to a logarithm (the number $n$ only requires $log_k(n)$ bits to represent in base $k$). This leads to the concepts of \textbf{Strong} and \textbf{Weak} \textbf{NP}-Completeness~\cite{strongnp}. Encodings are outside the scope of this thesis, however, and none of the material presented is dependent on any detail of encoding, so we will assume informal encodings such as boolean formulas, graphs, etc. The interested reader can find out more about encodings and number problems in~\cite{papadimitriou} (Chapter 2.2).

To illustrate, an instance of (generalized to $N \times N$) Chess can be given by specifying a board size and the position of every piece on the board. Unspecified pieces can be considered captured, and unspecified positions unoccupied. Note that, using the generalization introduced above, the number of pieces increases only linearly with the board's \textit{side}. As a result, we can consider the input size as simply O($N$) (or O($N \times N$), although a quadratic change is not significant in the context of polynomial or higher complexity) for any instance.

\subsection{Reductions}
A central concept to this thesis is the concept of \textit{reduction}. To find the complexity of a given problem, we compare it to a different problem of known complexity. 
By reducing one problem to another, we can say that the second is at least as hard as the first.

There are two main types of reduction: \textit{Turing reduction} (which uses oracles, “black-boxes” that give the answer to a different problem, to compute the answer of the first problem) and \textit{Many-one reduction}. Only the latter is used in this report.

A Many-one reduction, which we now describe, is a transformation from one problem to another, in a way that if we know how to solve the second, we can use the transformation to solve the first.

Consider two problems, $A$ and $B$. We have an algorithm to solve $A$, but not one to solve $B$. A reduction from $B$ to $A$ will transform an instance of $B$ into an instance of $A$, and a solution to $A$ to a solution to $B$.
Given a reduction from $B$ to $A$, we can solve $B$ as follows:
\begin{enumerate}
\item Given an instance of $B$, transform it into an instance of $A$;
\item Use the algorithm to decide $A$ on the transformed instance;
\item Extract the solution for $B$ from the solution obtained for the instance of $A$.
\end{enumerate}
For a reduction to be useful, however, (1) and (3) must be “easy” to compute. This is especially important when trying to prove the Hardness of problems.
Here, “easy” means \textit{polynomial-time bounded}. The reduction itself runs in O($n^k$) time, for some constant $k$.
Other classes' reductions may require a logarithmic space bounds, for example, or other types of bounds, but we will restrict ourselves to polynomial-time reductions, as only those are relevant for our \textbf{NP} and \textbf{PSPACE} hardness proofs.

Looking at the procedure with this added polynomial time constraint reveals how to prove a problem \textbf{NP}-Hard or \textbf{PSPACE}-Hard.

\paragraph*{NP/\textbf{PSPACE}-Hardness.} 
Consider that $B$ is a problem that is \textbf{NP}-Complete (everything will work the same way replacing \textbf{NP} with \textbf{PSPACE} in this paragraph). We know that, if there is an algorithm that solves $A$ in polynomial time, then, by applying the procedure described above, the same algorithm could solve $B$ in a polynomial amount of time. We say $B$ is \textit{polynomial-time reducible} to $A$, also written as $B\leq_p A$. (1) and (3) take at most polynomial time, since that is a property we're assuming for our reductions. If we have an algorithm to solve $A$ in polynomial time, then (2) also takes at most polynomial time. Because the sum and composition of polynomials is still a polynomial, we can solve any instance of $B$ in polynomial time. Because $B$ in \textbf{NP}-Complete, it is also \textbf{NP}-Hard, and so we can solve every problem in \textbf{NP} in polynomial time. This would prove \textbf{P}=\textbf{NP}.

Given that we do not know whether \textbf{P}=\textbf{NP}, or the complexity of $A$, we can only conclude that $A$ is at least as hard as $B$. Also note that, because $B$ is \textbf{NP}-Complete, every problem in \textbf{NP} can be reduced to it, by definition. So, if we have any problem $C$ $\in$ \textbf{NP}, we can first reduce it to $B$, and then reduce $B$ to $A$. Two reductions is simply the composition of two polynomials, so the overall reductions is still bounded by a polynomial in time. This means we can reduce every problem in \textbf{NP} to $A$, and so $A$ is \textbf{NP}-Hard.

This is the basis for the rest of this thesis. To prove a new game \textbf{NP}-Hard, we start from any \textbf{NP}-Hard problem (or, more generally, C-Hard for the intended class C) and reduce it to the game we're studying. 

Sometimes, we're interested in proving a problem Complete in a given class, and not just Hard. To prove Completeness (after Hardness), we only need to prove its membership in the same class, which tends to be easier.

\paragraph*{Membership in NP/PSPACE.}
One way to do prove membership in a class is to construct a reduction on the opposite direction, reducing $A$ to $B$. Doing so lets us solve $A$ in at most $B$'s time, plus a polynomial reduction, which doesn't affect the overall complexity of the algorithm to solve $A$.

An easier way is to describe a naive algorithm to decide A in the appropriate model of computation. For \textbf{NP}, for example, we typically describe a non-deterministic machine running in polynomial time. The algorithm ``guesses'' a solution (which is limited to a polynomial in the case of \textbf{NP}) and then verifies it in polynomial time, also a property of \textbf{NP} problems. The same technique applies to \textbf{PSPACE}, although the verification needs not take polynomial time, only polynomial space. In both cases, we only need to worry about proving the verification process (although \textbf{PSPACE} is a deterministic model of computation, we know by Savitch's Theorem~\cite{savitch} that \textbf{PSPACE}=\textbf{NPSPACE}, so we can also rely on non-determinism to simplify the proofs).

\subsection{Original Problems}\label{subsec:originalproblems}
Here, we will briefly introduce the computational problems that were used to reduce from to prove each game's complexity. We will describe the problem and point to where its own complexity proof was given.

\paragraph*{SAT (Boolean Formula Satisfiability)}.
SAT is one of the central problems in Complexity Theory, and it was the first problem to be proved \textbf{NP}-Complete, by Stephen Cook in 1971~\cite{cook1}. The following year, 1972, Richard Karp presented a series of reductions of 21 problems, starting by reducing SAT to CLIQUE and 0-1 ILP (Integer Linear Programming) as a stepping stone to the remaining 19 problems~\cite{karp2}.
In 1978, Schaefer published a paper generalizing the SAT problem, giving several sufficient and necessary conditions for each variation belonging to the classes \textbf{P} and \textbf{NP}. Some of the variations include ONE-IN-THREE SAT and NOT-ALL-EQUAL 3-SAT~\cite{Schaefer}.

A boolean formula has a set of boolean variables (their values are $\textit{true}$ or $\textit{false}$) connected by boolean operators ($\textit{AND}$, $\textit{OR}$, $\textit{NOT}$, $\textit{IMPLY}$). The decision problem is whether there is an assignment of variables that satisfies the whole formula (i.e. makes the formula true).

This problem in usually considered in CNF (conjunctive normal form). Any boolean formula with the usual connectives ($\textit{AND}$, $\textit{OR}$, $\textit{NOT}$, $\textit{IMPLY}$) can be translated into CNF with only a linear increase in size~\cite{handbooksat} (see Chapter 2 CNF encodings), so it has no impact in its belonging or hardness in \textbf{NP}.
3-SAT, where clauses have \textit{at most} three variables, is also \textbf{NP}-Complete, as shown by Cook~\cite{cook1}. 3-CNFSAT is used in Section~\ref{proof:mario} to prove the complexity of Super Mario Bros~\cite{nintendopaper}.

\paragraph*{CIRCUIT SAT}
CircuitSAT is the decision problem of whether a boolean circuit has a model that makes it true. It is very similar to SAT, and also considered one of the prototypical \textbf{NP}-Hard problems. It is used to prove Akari's complexity~\cite{akari} in Section~\ref{proof:akari}.

\paragraph*{HAMILTONIAN CIRCUIT}.
Hamiltonian circuit is the problem of whether a given graph has a path that visits every node exactly once (except for the starting node, which matches the ending one and is visited twice).
This problem was also part of Karp's series of reductions~\cite{karp2}, where he shows the problem to be \textbf{NP}-Complete for both directed and undirected graphs. It is used to prove Lode Runner's \textbf{NP}-Hardness~\cite{viglietta}.

\paragraph*{TQBF (True Quantified Boolean Formula)}.
TQBF is the problem of Boolean satisfiability with the addition of universal and existential quantifiers. Given a quantified Boolean formula, the decision problem is whether or not the formula evaluates to true. Unlike the problems listed above, however, it is not \textbf{NP}-Complete, but \textbf{PSPACE}-Complete~\cite{alternation}. It is used to prove The Legend of Zelda: A Link to the Past's \textbf{PSPACE}-Hardness~\cite{nintendopaper} in Section~\ref{proof:zelda}.

\paragraph*{PARTITION}
Partition is the problem of if, given a finite set A and an integer B, there is any subset $A' \subseteq {A}$ such that $\Sigma{} a \in A'= B$. It was also proved \textbf{NP}-Complete by Karp~\cite{karp2}.
\textit{3-PARTITION}, a variation where the set A is partitioned into sets of 3 elements, each with equal sum, has been also proven \textbf{NP}-Complete~\cite{3part}.

\section{Related Work}\label{section:survey}
Here, we survey some of the study of Complexity Theory in the context of games. Each subsection surveys a different category of problems.

\subsection{Partition \& Packing Problems}
The most popular example of a PARTITION reduction in the context of games might be the proof that Tetris is \textbf{NP}-Hard~\cite{tetris}, from 2002.
PARTITION is also used as a starting point to proving Rectangle and Square packing results, then applied to both Polyomino (similar to Tetris' pieces, but not restricted to four units) Packing, Edge-Matching, as well as Jigsaw puzzles~\cite{jigsaw}. They present a loop of reductions, making it possible to easily transform one puzzle type into another without very little blowup in size, revealing how closely related all three games are.

\subsection{Hamiltonicity}
The Hamiltonian paths and cycle problems are very close to games and puzzles. In fact, the name comes from the Icosian game, created by Sir William Rowan Hamilton in 1857.\footnote{\url{http://www.mathematica-journal.com/issue/v11i3/contents/superhamilton/superhamilton.pdf}}

The game is played on a dodecahedral  graph. One player chooses vertices, and the other must figure how a \textit{path} that reaches every vertex without repeating any. It's called a \textit{cycle} when the starting and ending vertices are the same (the only vertex that can be reached more than once).
Many variations of the problem are \textbf{NP}-Complete, and many puzzle games have taken advantage of that property to make appealing puzzles.\footnote{\url{https://rotopo.com/} is about Hamiltonian Cycle in three dimensional ``planar'' (non-crossing) graphs.
\url{http://store.steampowered.com/app/266010/LYNE/} LYNE is based on a variation on the problem, played on a square grid (with Moore neighborhood - adjacent cells include diagonals) and disjointed paths. Generic vertices are added (shared by the otherwise disjointed paths) with an arbitrary (but fixed) number of required passes.
Many other examples can be found.}

The Hamiltonian Cycle problem has also been used to prove other puzzles, not obviously connected to this problem.
Two examples are the Slitherlink~\cite{asp} and Hashiwokakero~\cite{hashiwokakero} games, both pen-and-pencil games published by Nikoli. It is also used to prove study platformer games in ~\cite{viglietta}.

\subsection{CircuitSAT \& Nikoli games}
Circuit SAT is a problem that has been used to proof several pen-and-pencil puzzles to be \textbf{NP}-Hard.
Some examples are:
\begin{itemize}
\item Spiral Galaxies~\cite{spiral} 
\item Akari/Light-Up~\cite{akari} 
\item Minesweeper~\cite{minesweeper} 
\item Fillmat~\cite{fillmat} 
\item Nurikabe~\cite{nurikabe} 
\end{itemize}
Because of CircuitSAT's roots in physical circuits, there are some results that make the proofs easier to work with. Because those results are useful for the construction of new proofs rather than the understanding of specific proofs, they will be explained in detail in Section~\ref{approach:circuitsat}.

The framework for Circuit SAT is to directly translate each wire and gate into game elements. For example, in the example circuit given in Figure~\ref{circuit_conv_before}, one would need to simulate each of the four gates and connect the inputs and outputs correctly given the game's mechanics. An example proof will be given in Section~\ref{proof:akari}.

\paragraph*{Another Solution Problem (ASP)}
In the context of puzzle games, especially from a design standpoint, it can be important to know whether a given solution to a puzzle is unique, or to count the solutions of a puzzle. This has been given a formal treatment in~\cite{asp}. The authors define a new complexity class, ASP, as the class of problems for which finding a solution $s_1$ given solution $s_0$ is \textbf{NP}-Hard. The \textbf{ASP}-Hardness reductions are similar to \textbf{NP}-Hard reductions, with the additional requirement that reductions are parsimonious (i.e. the number of solutions are preserved, which is also used to prove the complexity of \textit{counting} solutions in \textbf{\#P}-Completeness). Circuit SAT proofs, in particular, tend to have that property, since variables and gates typically to have one-to-one correspondence with gadgets (see Sections \ref{proof:akari}, \ref{proof:akaritri} and \ref{proof:akarihex}).

\subsection{Video Games}
Several isolated video games have been independently studied under Complexity Theory, including Minesweeper~\cite{minesweeper}, Tetris~\cite{tetris}, Lemmings~\cite{lemmingscormode,lemmingsviglietta}, and recently Angry Birds~\cite{angrybirds}. 
However, many games also have common or very similar sets of rules, usually classified as genres, such as platformers, first-person shooters, etc.

A series of papers have been dedicated to synthesizing the difficulty of several video games by looking at their common features. 
Although it started with 2D classic platformers~\cite{forisek,viglietta,nintendopaper}, it has since been extended to modern 3D games~\cite{3dgames} using Non-Deterministic Constraint Logic (see Section~\ref{subsec:ncl}).

These papers introduced general frameworks with generic, game independent gadgets. These gadgets are then implemented in the game under consideration. 
One of the frameworks is used to prove \textbf{NP}-Hardness, and the other \textbf{PSPACE}-Hardness.
We will explain two reductions in detail from~\cite{nintendopaper} in Section~\ref{section:approach}.
The first reduction has been used to prove Super Mario Bros. \textbf{NP}-Hard (although it was later shown that it is, in fact, \textbf{PSPACE}-Complete~\cite{hardermario}). We present it in Section~\ref{proof:mario}. 
The second reduction has been used to prove that The Legend of Zelda: A Link to the Past is \textbf{PSPACE}-Hard. We present it in Section~\ref{proof:zelda}.

\subsection{Non-Deterministic Constraint Logic}\label{subsec:ncl}
In 2009, Robert Hearn and Erik Demaine wrote a book~\cite{ncl}, which resulted from Hearn's PhD thesis at MIT.~\footnote{\url{https://dspace.mit.edu/handle/1721.1/37913}}

The main purpose was to provide a unifying framework with which to prove the hardness of several classes of games (from 0 player ``games'' like Conway's Game of Life to 1-player puzzles like Rush Hour to two player board games like Go). The book contains a graph-based theory of logic that aims to simplify many games' hardness proofs by providing an intermediary problem to reduce from.

Not unlike the Circuit SAT framework introduced earlier, Hearn and Demaine's framework eliminates the need for crossover gadgets and requires only the implementation of a few gadgets akin to logic gates. Some of the gadgets are redundant, but can be useful to make the framework more versatile/flexible. The biggest advantage of this framework over Circuit SAT is that it is extended to classes like \textbf{PSPACE} and \textbf{EXPTIME} and also works for multiplayer games (in general, more than two players can be reduced to two players with a ``single'' player against several opponents counting as one alliance).

\begin{table}
\renewcommand{\arraystretch}{3}
\begin{tabular} { r | c | c | c | c | }
    \multicolumn{1}{r}{}
 &  \multicolumn{1}{p{2.5cm}}{Zero player (simulation)}
 &  \multicolumn{1}{p{2.5cm}}{One player (puzzle)} 
 & \multicolumn{1}{p{2.5cm}}{Two player}
 & \multicolumn{1}{p{2.5cm}}{Team imperfect information} \\
    \cline{2-5}
    Unbounded & PSPACE & PSPACE & EXPTIME & RE (undecidable) \\
    \cline{2-5}
    Bounded & P & NP & PSPACE & NEXPTIME \\
    \cline{2-5}
\end{tabular}
\renewcommand{\arraystretch}{1}
\caption{Games' classification from~\cite{ncl}.}
\label{table:ncltable}
\end{table}

As can be seen in Table~\ref{table:ncltable}, they classify game by the number of players and whether or not they're ``bounded'', each leading to a different level of (expected) hardness.
Each of the entries has a variation on Constraint Logic that suits those rules and leads to a different complexity class. A game like Sudoku is an example of a bounded game, because each move cannot be undone in the solution (it can be in the search for a solution, but not the final solution). As a result, each decision reduces the search space significantly.
Another example is a hamiltonian cycle: because each vertex can only be visited once, the number of moves are limited by the number of vertices.

A game like Sokoban, however, is unbounded. This is because a block may need to be moved to let a different block pass through it's previous position, but then moved back to its original spot. 
This can lead to unbounded (or, at least, not polynomially bounded) solutions.

Although they provide a framework for two player Unbounded games in the book, they have not used it to prove any game's hardness (open or otherwise). It is difficult to judge how well the provided framework fits those types of games, where the main difficulty is to create stable wires configurations and force winning or losing conditions.

\subsection{Two Player Games of Perfect Information}
Two player games have been studied in Combinatorial Game Theory~\cite{numbersandgames,winningways} for a long time, from small games like Nim or Dots and Boxes, to large games like Chess and Go.
The large games, however, had proved difficult to analyze with Complexity Theory.

One paper to tackle the problem was published in 1979, but had no results for ``real'' games. Instead, it defined new abstract games that were \textbf{EXPTIME}-Complete~\cite{provdiff}. Until then, there were results proving \textbf{PSPACE}-Hardness for Go and Checkers problems, but not their completeness in any class.

This paper included six games played on propositional formulas, where players would assign the values of variables in turns, and the player to make the formula true was considered the winner. They also showed a game involving sliding tokens on a graph, and a more ``traditional'' interpretation of the formula games as a physical game that they called ``Peek''.

Only in 1981, was a paper was published by Fraenkel and Lichtenstein that proved that Chess was \textbf{EXPTIME}-Complete~\cite{chess}, tightening the previous \textbf{PSPACE}-Hardness result~\cite{chesspspace}. Then, in 1983 and 1984, respectively, Robson proved that both Go~\cite{go} and Checkers~\cite{checkers} were \textbf{EXPTIME}-Complete. The proofs are substantially more complex than the proofs for puzzle games and even two player bounded games.

More results of \textbf{PSPACE}-Complete games, for games with a bounded number of moves. Examples include Hex~\cite{hex,hexthesis}, Othello~\cite{othello}.
Three other games, Amazons, Konane and Cross Purposes were proved \textbf{PSPACE}-Complete using a variant of Non-Deterministic Constraint Logic~\cite{ncl}.

\section{Frameworks}\label{section:frameworks}
In this section, we describe the frameworks we used to obtain our new results in Chapter~\ref{chapter:results}.

\subsection{Gadgets}
All of the proofs in this thesis rely on the use of \textit{gadgets}. A gadget is a piece of the proof that represents some local property or component of the problem we're reducing \textit{from}.

In the case of SAT, for example, there are clause and variable gadgets, which are then implemented in the problem we're reducing \textit{to}. Other gadgets may be required by the reduction, as a bridge between the two problems, although not necessarily part of the original problem. One common example is a \textit{crossover}, usually required when reducing from non-planar to planar problems. 

\subsection{Circuit SAT}\label{framework:circuit}
Circuit SAT's framework is the most natural of the three shown in this section, as the gadgets required follow the components of a circuit very closely, typically without any added gadgets.
The gadgets required are the following:
\begin{itemize}
    \item Wire that propagates two types of signal (true/false) from one end to the other; this gadget may require turns, depending on game in question.
    \item Wire start/end with fixed and variable value (to select variables and have a desired final value for the formula).
    \item FAN-OUT, a way to multiply an input signal into more than one.
    \item A functionally complete set of boolean gates (usually $\textit{NOT}$ and $\textit{OR}$/$\textit{AND}$).
\end{itemize}
By implementing these four elements, a game can be proved \textbf{NP}-Hard by directly simulating a boolean circuit. Technical details and an example are given in Section~\ref{approach:circuitsat}.

For those unfamiliar with circuits and their terminology, a gate has input and output signals. A \textit{boolean} signal has only two possible values, referred to as \textit{true} and \textit{false}. The most common gates are shown in Table \ref{table:gates}. $\textit{NOR}$, $\textit{NAND}$ and $\textit{XNOR}$ are simply compositions (the output of the first is used as an input to the second) of $\textit{OR}$, $\textit{AND}$, $\textit{XOR}$ gates with a $\textit{NOT}$ gate, but may also have an irreducible implementation.
\begin{table}
\center
\subfloat[AND.]{
\begin{tabular}{ |c|c|c| }
 \hline
 Input1 & Input2 & Output  \\ [0.5ex] 
 \hline \hline
 false & false & false \\ 
 false & true & false \\
 true & false & false \\ 
 true & true & true \\
 \hline
\end{tabular}
}\subfloat[OR.]{
\begin{tabular}{ |c|c|c| }
 \hline
 Input1 & Input2 & Output  \\ [0.5ex] 
 \hline \hline
 false & false & false \\ 
 false & true & true \\
 true & false & true \\ 
 true & true & true \\
 \hline
\end{tabular}
}\subfloat[XOR.]{
\begin{tabular}{ |c|c|c| }
 \hline
 Input1 & Input2 & Output  \\ [0.5ex] 
 \hline \hline
 false & false & false \\ 
 false & true & true \\
 true & false & true \\ 
 true & true & false \\
 \hline
\end{tabular}
}\\\subfloat[NOT.]{
\begin{tabular}{ |c|c|c| }
 \hline
 Input1 & Output  \\ [0.5ex] 
 \hline \hline
 true & false \\
 false & true \\
 \hline
\end{tabular}
}
\caption{Common boolean gates.}
\label{table:gates}
\end{table}

\subsection{3-CNFSAT}\label{framework:np}
\begin{figure}[h!]
\centering
\includegraphics[scale=0.8]{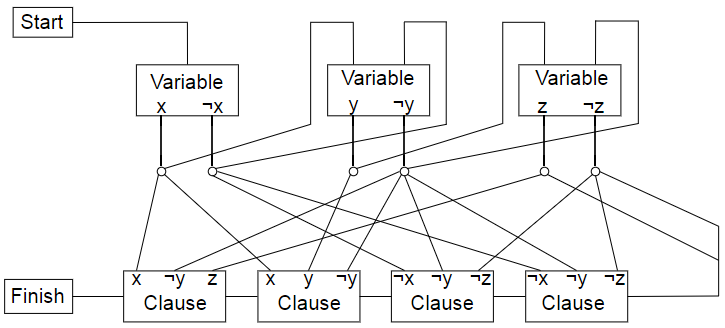}
\caption{\textbf{NP}-Hardness framework from~\cite{nintendopaper}.}
\label{fig:cnf_fw}
\end{figure}

Here we will show the framework used in~\cite{nintendopaper} in order to prove the \textbf{NP}-Hardness of classic Nintendo platformers. An overview is illustrated in Figure~\ref{fig:cnf_fw}.

\paragraph*{Overview.} The player controls an avatar with the objective of traversing a level from Start to Finish. The avatar will visit several gadgets that are linked with wires. A crossover gadget is used to make sure that the player continues on the correct wire when they cross. The framework creates a direct reduction from 3-CNFSAT, where clauses are implemented as locked paths, and choosing a variable's truth value will open (satisfy) the clauses in which it appears. This unlocking is represented with the wires coming out of each variable, into the top of the clauses. The clauses' paths are represented by the wires entering and exiting horizontally.

The gadgets required for the proof are the following:
\begin{itemize}
    \item An avatar that can be controlled;
    \item Locks that can be opened;
    \item Crossover gadget;
    \item One-way paths;
\end{itemize}

\paragraph*{Variables.} In each ``Variable gadget'', the player can take one of two paths. One of the paths sets the variable to true, the other to false. A ``one-way'' gadget is required here, to make sure that the player can't choose both values simultaneously (which would make the reduction inconsistent).

\paragraph*{Clauses.} Each ``Clause gadget'' can be seen as a set of three \textit{locks}, since we're working with 3-CNFSAT. Because a SAT clause is satisfied when \textit{any} of its positive variables is set to true (or negative variables set to false), only one of the locks has to be open for the player to successfully traverse the gadget. Depending on the game, the implementation can vary, the only property that must hold being that \textit{any} satisfying assignment makes the clause traversable, and only satisfying assignments make it traversable.

\paragraph*{One-way paths.} One-way paths are gadgets that have a fixed orientation, and can't be traversed backwards. The most common example are ``long falls'' in games with gravity, which can only be traversed from top to bottom. In fact, a one-way path gadget can be seen as a generalization of the ``long falls'' from~\cite{forisek}. Their purpose, in this framework, is to prevent the player from choosing two values for a single variable by going backward after opening the respective clauses. Depending on the reduction details, ``single-use'' paths may be sufficient (paths that are consumed, and so can only be traversed once), since each variable is only visited once.

\paragraph*{Crossovers.} A crossover gadget is simply a gadget that crosses two paths without letting the avatar change from his initial path. An example can be seen in Figure~\ref{circuit_cross}. The avatar can never go from $x$ to $y$, or $y$ to $x$.

\begin{theorem} The 3-CNFSAT framework shown in Figure~\ref{fig:cnf_fw} is \textbf{NP}-Hard.\end{theorem}
\noindent {\it Proof:} because 3-CNFSAT is \textbf{NP}-Hard, we only need to show how to reduce from it to the framework, in polynomial time, to prove that it too is \textbf{NP}-Hard.
The level is created, schematically, according to Figure~\ref{fig:cnf_fw}. The avatar's initial position is ``Start'', and the only winning condition must be in ``Finish''. The avatar will first visit every variable, choose a value for each one, and open the respective clauses. Note that clauses have separate ``unlock'' and ``traverse'' paths. Choosing a variable's value is done by following one of two paths, with no ability to go back due to the \textit{one-way gadget}. Because the avatar only traverses in wires, and a \textit{crossover} gadget is placed at every intersection, the player must choose an assignment to every variable before reaching the ``Check in'' and start the clauses' traversal. Finally, because each variable only opens the clauses where it appears ($\lnot{x}$ only opens clauses that contain $\lnot{x}$), to traverse every clause, the player must choose a sequence of variable paths that opens every clause. In other words, must choose a variable assignment that satisfies every clause. The player will only be able to finish the level successfully if the original formula is satisfiable, and a solution can easily be extracted from the variable paths taken. If the 3-CNFSAT formula has no satisfying assignment, however, regardless of the assignment that the player chooses, there will always be one clauses that can't be traversed, as all three of its locks are closed and the level will be impossible to complete. As a result, a game that implements this framework is \textbf{NP}-Hard. \qed

\subsection{Hamiltonian Cycle}\label{framework:hamilton}
This framework comes from~\cite{viglietta}, Meta-Theorem 1. It is a reduction from the Hamiltonian Cycle in undirected, 3-regular graphs (each vertex has 3 undirected edges). Although on a restricted type of graph, the problem is still \textbf{NP}-Complete~\cite{simple_npc}.

An Hamiltonian Cycle is a cycle (a walk on a graph that starts and ends on the same vertex) in which every vertex is visited \textit{exactly once}.

The framework relies on \textit{single-use} paths and \textit{location traversal}. Single-use paths, as the name indicates, are simply paths that can be traversed at most once, in any direction. Location traversal is the mechanism that forces the avatar to visit every vertex.

\begin{lemma} Single-use paths are enough to force at most one visit per vertex.\end{lemma}
\noindent {\it Proof:} Recall that the graph is 3-regular, so every vertex has degree exactly three. Consider any vertex $v$ different from the starting/ending vertex. Because edges are implemented as single-use, they can never be used twice. To enter $v$, one of its three edges is consumed, and a second to leave. To reach $v$ a second time, its third edge must be consumed, and then it is impossible to exit the vertex a second time. Since it will be impossible to leave the vertex (and assuming it is not the start/end vertex), it will also be impossible to finish the cycle. As a result, each vertex can only be visited once.

\begin{lemma} Single-use paths and location traversal are \textbf{NP}-Hard.\end{lemma}
\noindent {\it Proof:} The reduction from Hamiltonian Cycle in undirected, planar, 3-regular graphs works as follows: add a single vertex gadget $f$ linked only to the starting/ending vertex gadget $s$ that contains the level's Finish. Each vertex in the graph is replaced by the vertex gadget (which implements location traversal), and every edge with the single-use path gadget. To reach $f$, the player will have to start at  $s$ and visit every other vertex exactly once (at least once due to location traversal and at most once due to the single-use paths), and get back to $s$, which is exactly the path in the Hamiltonian Cycle. If no such solution exists, the level can't be completed either. As a result, games with these two properties can be proved \textbf{NP}-Hard, being reduced to the Hamiltonian Cycle problem, which is \textbf{NP}-Complete. \qed

\subsection{TQBF}\label{framework:pspace}
Also from~\cite{nintendopaper}, we describe the framework used to prove other Nintendo platformers' \textbf{PSPACE}-Hardness, namely The Legend of Zelda: A Link to the Past.

\begin{figure}[h!]
\centering
\includegraphics[scale=0.8]{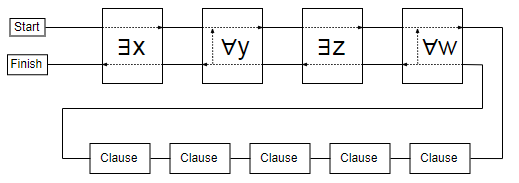}
\caption{\textbf{PSPACE}-Hard framework from~\cite{nintendopaper}.}
\label{fig:qbf_fw}
\end{figure}

The general framework is shown in Figure~\ref{fig:qbf_fw}. This one proves \textbf{PSPACE}-Hardness, reducing directly from True Quantified Boolean Formula (TQBF). It is similar to 3-CNFSAT, but variables now have a universal or existential quantifier associated with them. It is also important to note that the decision problem is no longer ``is there is an assignment that satisfies the formula?'', since the quantifier formulation claims a solution. The decision problem becomes ``is the quantified formula true?''. 

As in the previous proof, a \textit{crossover gadget} is required. A \textit{door} mechanism is also needed, but with special care not to be confused with the \textit{locks} from the 3-CNFSAT framework. Doors need to be \textit{opened} as well as \textit{closed} for the reduction to work with universal quantifiers. Where in the previous framework every lock was ``consumable'', doors alternate between opened and closed an arbitrarily large number of times.

The elements required for the reduction are the following:
\begin{enumerate}
    \item An avatar that can be directed;
    \item Crossovers;
    \item Doors that can be opened and forcefully closed;
\end{enumerate}
Note that single-use paths are no longer required. This property can be implemented with doors. After traversing a path, we can simply force the player to close a door behind them, preventing the backwards traversal. Doors at typically controlled at a distance.

\paragraph*{Overview.} The player starts at ``Start'' and chooses a variable assignment in the existential blocks, like the previous 3-CNF SAT case, opening the respective clauses. The universal blocks, however, require two passes to be traversed - one for each value, true and false. This is enforced by the use of the ``Close'' mechanism of the doors. 

It is important to note that, unlike in the 3-CNFSAT framework, the clauses are traversed up to an exponential number of times, due to the universal quantifiers. Referring to Figure~\ref{fig:qbf_fw}, the clauses must be satisfiable for every combination of $y$ and $w$ values, because they're both universally quantified. As the number of universally quantified variables increases, so does the numbers of clause traversals.

To make the point precise, consider a formula with universally quantified variables $v_0...v_n$ and some current variable $v_k$. $\forall_{i=0...k} v_i \in B$ and $\forall_{i=k...n} v_i \in A$. Each variable in $A$, needs to be assigned \textit{both} values (in separate visits) for each assignment of the variables in $B$. As a result, $v_i$ will need to be visited $2^i$. Each visit will lead to a clause check traversal, so we can see that the length of the solution (given by the sequence of movements or choices), the verification, will no longer be done in polynomial time. 

The answer to the TQBF problem is given by whether or not the player was able to reach "Finish".

\begin{figure}[!h]
\center
 \subfloat[TQBF clause gadget.\label{fig:qbf_clause}]{%
   \includegraphics[scale=0.8]{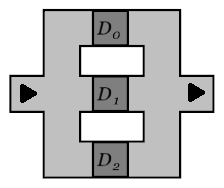}
 }
 \subfloat[TQBF existential quantifier gadget for x.\label{fig:qbf_exi}]{%
   \includegraphics[scale=0.8]{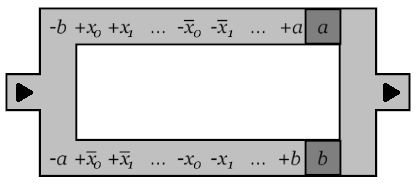}
 }\\
 \subfloat[TQBF universal quantifier gadget for x.\label{fig:qbf_uni}]{%
   \includegraphics[scale=0.8]{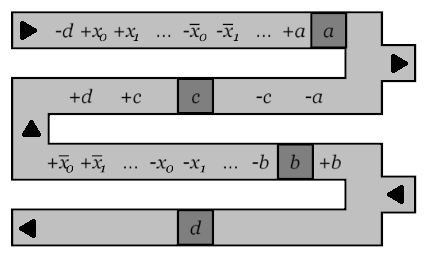}
 }
 \caption{TQBF gadgets from~\cite{nintendopaper}.}
 \label{fig:qbf_gadgets}
\end{figure}

\paragraph*{Door gadget.} In this framework, the door gadget typically has three different paths. One path opens the door, one path closes it, and one paths traverses it. The traverse path can only be successfully traversed if the door is open. These are three distinct paths and need not be close to each other. 
Each box in Figure~\ref{fig:qbf_fw} (quantified variables, clauses) is shown as a composition of door gadgets in Figure~\ref{fig:qbf_gadgets}. 
Bold squares represent the door's traversal path. $-$ (minus) represents a door's closing mechanism, and $+$ (plus) represents its opening mechanism. The closing mechanism is never optional. If the player passes through it, the respective door's state must end closed. The opening mechanism doesn't need such a property.

\paragraph*{Crossover gadget.} This gadget functions exactly as in the previous framework. It crosses two distinct paths, forcing the avatar to continue along its initial path.

\paragraph*{Clause gadget.} The clause gadget shown in Figure~\ref{fig:qbf_clause} represents $D0 \lor{} D1 \lor{} D3$. Variables can be negated. The bold squares are traversal paths that require opening from the respective existential or universal gadget. As in 3-CNFSAT, clauses are satisfied if any of its variables are satisfied, which is represented by the three disjoint doors. The clause is traversable as long as any one of its variables is satisfied.

\paragraph*{Existential quantifier gadget.}
The existential gadget shown corresponds to variable $x$. The upper path assigns the value $\textit{true}$, opening all the clauses that contain $x$ and closing the the ones that contain $\lnot{x}$. Clauses containing both $x$ and $\lnot{x}$ will have one of their doors open and another closed. 
The bottom path does the reverse, opening $\lnot{x}$'s doors and closing $x$'s.
$a$ and $b$ are simply auxiliary doors to make sure that the player can only choose one of the paths. Also note that either path \textit{undoes} whatever the other does. If the player walks halfway through one of the paths and then turns back to choose the other, the result is the same as simply traversing once through the last path.

\paragraph*{Universal quantifier gadget.}
Finally, we have the universal blocks. These are by far the most important, since they're what makes the problem \textbf{PSPACE}-Hard rather than simply \textbf{NP}-Hard.
The variable-clause relationship is exactly as in the existential block. What it needs explaining is the various paths and auxiliary doors $a, b, c, d$.
The purpose of this gadget is to force the assignment of both values to a variable and to check that the formula is satisfied in both cases.
The player starts at the top left. Similarly to the existential block's top path, the first step is to assign $\textit{true}$ to $x$ and open all the respective doors. The player is then forced to leave through the top right (since the door $c$ in the middle is starts closed. The player will then have to complete the whole circuit seen in Figure~\ref{fig:qbf_fw}, assigning other variables and passing through the clauses until he reaches the bottom part of the block (right to left on the framework). Note that if the current assignment does not satisfy every clause, the player will never be able to traverse the clauses.
Assuming it does, the player will be back to the universal block on the bottom right. At this point, it means that the formula was satisfiable given a model where $x$ was $\textit{true}$. Door $d$ is closed, so the player is forced to go back up through the middle path. This path assigns $\textit{false}$ to $x$. This middle path also opens door $d$, but closes $b$ (preventing backtracking), so the player has to traverse the circuit again, with $x$ set to $\textit{false}$. If the formula is still satisfiable with $x$ set to $\textit{false}$, then the player will be back at the bottom right of the gadget. At this point, both values satisfied the formula and door $d$ is open, so it's now possible to leave through the bottom left. This ensures that both value of $x$ make the remainder of the formula $\textit{true}$. Traversing this gadget proves that both values of $x$ satisfy the formula, successfully implementing a universal quantifier.

\begin{theorem}\label{theorem:pspace} The TQBF framework shown in Figures~\ref{fig:qbf_fw} and \ref{fig:qbf_gadgets} is \textbf{PSPACE}-Hard.\end{theorem}
\noindent {\it Proof:} because TQBF is \textbf{PSPACE}-Hard, we only need to show how to reduce from it to the framework, in polynomial time, to prove that it too is \textbf{PSPACE}-Hard.
The level is created, schematically, according to Figure~\ref{fig:qbf_fw}. The avatar's initial position is ``Start'', and the only winning condition must be in ``Finish''. The clauses have to be traversed for every assignment of universal variables, while existential variable only need one value to satisfy the formula for each universal configuration. The resulting level can be complete if and only if the original formula is true. An algorithm that determines whether the level is traversable also determines if the TQBF formula is true, so the framework is \textbf{PSPACE}-Hard. \qed

\section{Example Proofs}\label{section:example}
In this section, we present three proofs as concrete examples of the described Frameworks. None of these proofs is original work and were taken from the respective referenced papers~\cite{akari,nintendopaper}. Figures were taken from the respective papers, with the exception of Akari, which we've remade for coherence with the original proofs in Chapter \ref{chapter:results}. The chosen examples were used as the basis of our Approach (Section~\ref{section:approach}) and the earlier proofs in Chapter \ref{chapter:results}.

\subsection{Akari}\label{proof:akari}
Here, we will use game Akari/LightUp as an example of a CircuitSAT proof. The gadgets are taken from~\cite{akari}, although we reproduced the figures to be consistent with the new proofs presented in Chapter \ref{chapter:results}.

The general idea of the reduction, as said previously, is to simulate boolean circuits by implementing gates and wires (with binary signals) using game/puzzle elements. Each gate and wire of the original circuit is replaced by a gadget implemented with puzzle elements.

Akari/LightUp is a pen-and-pencil game from Nikoli (the japanese company famous for publishing for Sudoku, among many others puzzles, many of them \textbf{NP}-Complete~\cite{akari,asp,hashiwokakero,fillmat}).
The game is played on a square grid. Each cell can be empty or an obstacle (numbered or otherwise). 
The objective is to place lamps on the grid's empty cells such that:
\begin{itemize}
    \item numbered cells (with number n) have \textit{exactly} n edge-adjacent lamps;
    \item no two lamps illuminate each other;
    \item every empty cell on the grid is illuminated;
\end{itemize}
Lamps illuminate every cell in all of the four directions of the grid until their light hits an obstacle, stopping its propagation.

\begin{figure}[!h]
\center
 \subfloat[Empty puzzle.\label{akari_puzzle}]{%
   \includegraphics[scale=0.5]{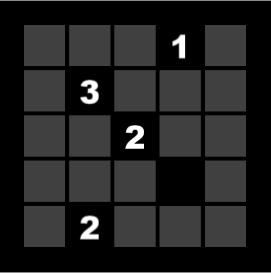}
 }
 \subfloat[Partially solved puzzle.\label{akari_partial}]{%
   \includegraphics[scale=0.5]{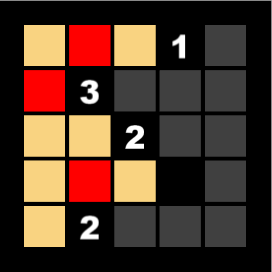}
 }
 \subfloat[Solved puzzle.\label{akari_solved}]{%
   \includegraphics[scale=0.5]{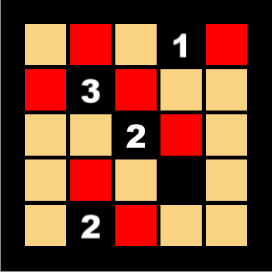}
 }
 \caption{Puzzle example.}
 \label{akari_example}
\end{figure}

Figure~\ref{akari_example} shows a puzzle example in three phases of completion. Gray represents an empty cell, black represents an obstacle cell, red represents a lamp and yellow represents an empty but illuminated cell. Numbered cells, being black, also act as obstacles.

\begin{figure}[!h]
\center
 \subfloat[Variable selection and wire gadgets.\label{akari_square_wire}]{%
   \includegraphics[scale=0.5]{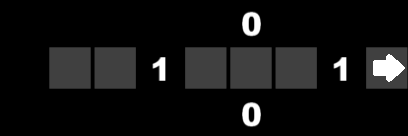}
 }
 \subfloat[Left-Top and Top-Left wire turn.\label{akari_square_turn}]{%
   \includegraphics[scale=0.6]{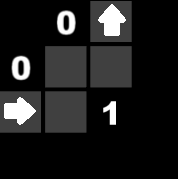}
 }
 \\
 \subfloat[FAN-OUT (top and bottom) and $\textit{NOT}$ (middle).\label{akari_square_fanout}]{%
   \includegraphics[scale=0.45]{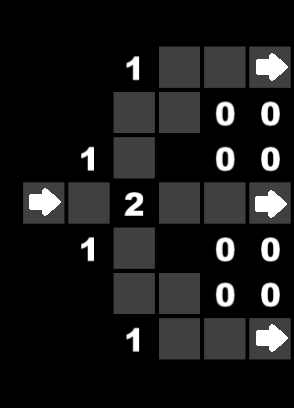}
 }
 \subfloat[OR gadget.\label{akari_square_or}]{%
   \includegraphics[scale=0.5]{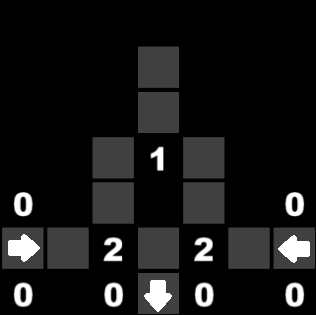}
 }
 \caption{Akari's gadgets \textbf{NP}-Hard.}
 \label{akari_square_gadgets}
\end{figure}

To explain the gadgets, we assume puzzles are always solved by first choosing the variable assignments (by placing the lamps in Figure~\ref{akari_square_wire}'s first or second empty cells) and then using simple propagation to find whether or not the assignment is a solution of the puzzle. This defines an orientation for the signals and gadgets, going outward from the variables.
Although a real solution is a static configuration with no lamp placement order, signal propagation reflects the way players look for solutions to the puzzle. This convention hopefully makes the gadgets more intuitive.

We will now briefly describe the proof. Figure~\ref{akari_square_gadgets} presents every gadget \textit{needed} for the proof.

Although the original paper presented more gates, we omit them here, as they are not required for the correctness of the proof. This is because the set of connectives $\{\textit{NOT}, \textit{OR}\}$ is already functionally complete. Technical details regarding this and the construction of the proofs are given in Section~\ref{section:approach}.

The white arrows define the input to output orientation, consistent with the orientation mentioned above. Every gadget can be rotated and mirrored, transforming the arrows accordingly.

\paragraph*{Variable selection and Wires.}
Figure~\ref{akari_square_wire} shows how a variable's value is selected (first two cells) and how the true/false signal is propagated through the wire. Note that the single 0-cell in the middle can be extended to a sequence of $n$ consecutive 0-cells to make an arbitrarily long wire.
\begin{lemma} The gadget shown in Figure~\ref{akari_square_wire} functions as a variable selection (left) and wire, propagating the chosen value.\end{lemma}
\noindent {\it Proof:} Because every cell on the board must be lit up, either the first or second left-most cells must have a lamp to light the left-most corridor. This determines the variable's value. If the lamp is on the first (left-most cell), there must also be a lamp to the right of the 1-cell by propagation. The corridor will be illuminated (to the right), representing a \textit{true} signal. On the other hand, if the first lamp is placed on the second cell (left of the 1-cell), then the corridor to the right will \textit{not} be illuminated, so it represents a \textit{false} signal. The illumination of this corridor is necessary, achieved by placing a lamp to the left of the right-most 1-cell, by propagation.
As a result, there are two locally viable configurations, each one defining a different signal, which can be connected to different gadgets, behaving as a wire gadget. \qed

\paragraph*{Wire turns.}
Figure~\ref{akari_square_turn} shows how to turn wires. This gadget connects to wires and gates. This gadget can be rotated (by multiples of $90$ degrees) or reflected across the $x$ or $y$ axes.
\begin{lemma} The gadget shown in Figure~\ref{akari_square_turn} propagates a signal, changing the direction of propagation by $90$ degrees.\end{lemma}
\noindent {\it Proof:} Consider the signal coming from the left to the top. If the corridor on the left was illuminated, a lamp will have to be placed on top of the 1-cell (as it requires exactly one adjacent lamp). As a result, the corridor on top will also be illuminated. On the other hand, if the corridor on the left was not illuminated, the player must place a lamp to the left of the 1-cell (because the corridor must be illuminated from either side). Then, there can be no lamp on top of the 1-cell, and the top corridor will not be illuminated. In both cases, the input signal is propagated (whether lit or unlit, true or false), and the direction is incremented by $90$ degrees. \qed

\paragraph*{NOT \& FAN-OUT gadgets.}
Figure~\ref{akari_square_fanout} achieves two purposes in a single gadget. The middle corridor is a \textit{NOT} gate. The top and bottom corridors make a \textit{FAN-OUT} gate. It is easy to place additional obstacles to close off some of the corridors and make the desired gadget out of the possible two.

The $\textit{NOT}$ gadget simply carries the opposite of its input signal. The FAN-OUT gadget duplicates the input signal, carrying it to both outputs. It is used to connect the same variable to several gates, as needs to happen in the clause $(x \lor{y}) \land{(x \lor{z}})$, where a single variable $x$ must connect to two $\textit{OR}$ gates.

\begin{lemma} The gadget shown in Figure~\ref{akari_square_fanout}, middle corridor, implements the constraints of a $\textit{NOT}$ gate, negating the input signal.\end{lemma}
\noindent {\it Proof:}
If the input corridor is illuminated, two lamps need to be placed above and below the 2-cell to satisfy the two 1-cell constraints. Because the 2-cell requires exactly 2 adjacent lamps, no lamp can be placed in the middle corridor, and so the corridor will not be illuminated. A true signal becomes false; consider, on the other hand, that the input corridor is not illuminated. Then, because it needs to be illuminated, a lamp must be placed to the left of the 2-cell, also satisfying the two 1-cell constraints. As a result, the second lamp can only be placed to the right of the 2-cell. This illuminates the corridor, transforming a false signal into a true signal. This shows that the input signal is always negated, implementing a $\textit{NOT}$ gate.\qed

\begin{lemma} The gadget shown in Figure~\ref{akari_square_fanout}, top and bottom corridors, implements the constraints of a FAN-OUT gate, propagating the input signal to both its outputs.\end{lemma}
\noindent {\it Proof:} Consider the input corridor is illuminated. As in the $\textit{NOT}$ case, a lamp will be placed above and below the 2-cell. As a result, to illuminate the empty cell to the left of the 0-cell, we will need to place a lamp above it, to the right of two extremity 1-cells. This illuminates the two output corridors, propagating a true signal in both cases.
Consider the opposite case, where the input corridor is not illuminated. Then lamps will be placed to the left and right of the two cell. To illuminate the short (length 2) corridors above and below the 2-cell, we have to place lamps below the top most 1-cell and above the bottom most 1-cell. As a result, we can't place a lamp on their right, and so both output corridors propagate the same false signal.
This shows that the input signal is always duplicated, implementing a FAN-OUT gate. \qed

\paragraph*{$\textit{OR}$ gadget.}
Finally, we show how the $\textit{OR}$ gate is implemented in a gadget in Figure~\ref{akari_square_or}. The signal comes from the sides of the gadget and is propagated through the bottom. The gadget can be rotated by any multiple of 90 degrees, changing the orientation of the signals (coming in and out) accordingly.
An $\textit{OR}$ gadget outputs true if at least one of its inputs is true. So the output corridor should be illuminated if any of its input corridors is illuminated, and should not be illuminated if neither of its inputs is.
\begin{lemma} The gadget shown in Figure~\ref{akari_square_or} implements the constraints of the $\textit{OR}$ gate.\end{lemma}
\noindent {\it Proof:}
Consider that both inputs are true (so both input corridors are already illuminated). To satisfy the 2-cell constraints, we need to place a lamp above each 2-cell, and one between them. To satisfy the top 1-cell constraint, we place a lamp adjacent to it. The whole gadget is illuminated and so is the output corridor. Both inputs being true, so is the output.

Consider now that both inputs are false (neither input corridor is illuminated). We need to place two lamps to illuminate the input corridors. Now, we are forced to place a lamp above each 2-cell. This may not be immediately obvious, but it can be understood by trying to place a lamp between the 2-cells. Then the two short corridors above the 2-cells could not be both illuminated without breaking the 1-cell's constraint, or one of the 2-cells' constraint. As a result, the output corridor can't be illuminated when both inputs are false.

Finally, we look at the case where only one input being active. The gadget is symmetric, so we look at only one case. Consider that the left input is true, but not the right input. We need to place a lamp above the left most 2-cell, and one between the two 2-cells by direct propagation. Because the right input corridor is not illuminated, we also need to place a lamp to the right of the right most 2-cell. We can't place a lamp above the right most 2-cell as it already has 2 adjacent lamps. As a result, we are forced to place a lamp to the right of the 1-cell to illuminate the corridor above the right most 2-cell. Finally, we only need to illuminate the top 2-cells. Because the 1-cell already has one lamp adjacent to it, we are forced to place a lamp on the top most cell. Recall that the we placed a lamp between the two 2-cells, so the output corridor is illuminated, carrying the signal true.

All four cases are consistent with the $\textit{OR}$ gate, and also the puzzle's constraints, so the gadget correctly implements an $\textit{OR}$ gate. \qed

\paragraph*{Crossing wires.}
Only one detail is left to finish the proof. In some proofs, the problem of crossover arises given overlapping wires. One of the properties of Circuit SAT that makes it an easier problem to reduce from is that no crossover is required. The details are considered in Section~\ref{section:approach}.

\begin{theorem} Akari (played on a square grid) is \textbf{NP}-Complete. \end{theorem}
\noindent {\it Proof:}
Any boolean circuit can be translated into an Akari/LightUp puzzle using the gadgets shown previously, and so the problem is \textbf{NP}-Hard. Because each gadget has a unique configuration for each variable value, the overall puzzle has as many solutions as the base circuit. The solution for the circuit can easily be extracted from the variable selection gadget.

Akari/LightUp $\in{\textbf{NP}}$, because a solution can be verified in polynomial time. Consider a rectangular board of width $n$ and height $m$. The number of lamps are bounded by the number of cells in the board $(n \times m)$. Each lamp also propagates light in to at most $(n + m)$ cells. To verify a solution, all we need is to place all the lamps, propagate their light, and then check each cell for its neighbourhood constraints. There is at most a polynomial number of lights, each propagating light to at most a polynomial number of cells. Finally, the neighborhood constraints are at most four, for every cell, so they too are polynomial. The whole verification is polynomial, so Akari is in \textbf{NP}. \qed

\subsection{Super Mario Bros.}\label{proof:mario}
We will now show an implementation of the 3-CNFSAT framework using Super Mario Bros. Both the proof and the framework come from~\cite{nintendopaper}.
In Super Mario Bros., the objective is to get Mario, the avatar, from the start of a level to the end. For this reduction, we use the following game mechanics:
\begin{enumerate}
    \item Mario can have two states: big and small. Small Mario can pass through gaps with a height of a single unit, while big Mario can only pass through gaps with a height of at least two units; Furthermore, big Mario can break bricks by jumping into them from below, but not small Mario. The transition works as follows: big Mario will become small Mario if he collides with an enemy (unless he jumps on its head); small Mario will become big Mario by consuming a red mushroom, contained in one of the ``?'' blocks.
    \item ``?'' blocks contain a consumable item which can be a mushroom, or a star (for the purposes of this reduction), deterministically (chosen when building the level). These can only activate once.
    \item Mario dies when he collides with fire (or enemy when in his small state), restarting the level.
    \item Star power-ups give Mario invincibility for a few seconds, making him immune to fire.
\end{enumerate}

\begin{figure}[!h]
\center
 \subfloat[Variable gadget for Super Mario Bros..\label{fig:mario_var}]{%
   \includegraphics[scale=0.7]{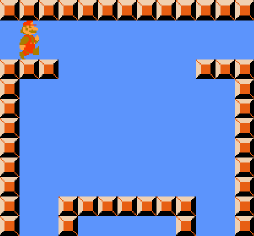}
 }
 \subfloat[Clause gadget for Super Mario Bros..\label{fig:mario_clause}]{%
   \includegraphics[scale=0.7]{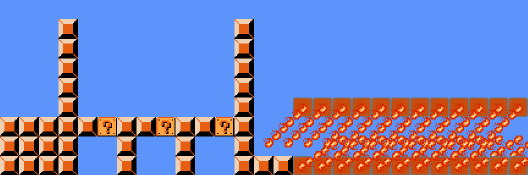}
 }\\
 \subfloat[Crossover gadget for Super Mario Bros..\label{fig:mario_cross}]{%
   \includegraphics[scale=0.7]{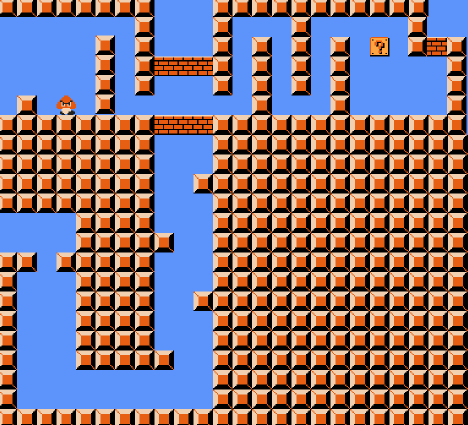}
 }
 \caption{Super Mario Bros.'s gadgets from~\cite{nintendopaper}.}
 \label{fig:smb_gadgets}
\end{figure}

In this proof, fire is used as a lock, preventing Mario from crossing a path. Stars are used as a key, letting him cross only if he managed to place a star close to the locked path. The two Mario states are used to implement a crossover gadget, one which can only be cross of his small state, and the other on his big state. Outside this gadget, Mario is always in his big state. All the gadgets are shown in Figure~\ref{fig:smb_gadgets}.

\paragraph*{Variable gadget.} Figure~\ref{fig:mario_var} shows the variable gadget. There is one of these gadgets for every distinct variable in the original formula. The avatar enters through one of the left or right paths (depending on previous choices, as seen in Figure~\ref{fig:cnf_fw}). The left path down assigns the variable to true, the right path down assigns it to false, or vice-versa. A corridor taller than Mario's jump is necessary to enforce the one-way rule. As shown in the 3-CNFSAT Framework (Section~\ref{framework:np}), the ``true'' path will visit every clause where the variable appears non-negated ($x$), and the ``false'' path will unlock every clause where it appears negated ($\lnot x$).

\paragraph*{Clause gadget} Then, we have the clause gadget in Figure~\ref{fig:mario_clause}. There is one of these for every clause in the original formula. Visiting the clause means that the player has reached the gadget from the bottom, activating one of the ``?'' blocks with each visit (of which only one is necessary). This will place a star in the upper middle section of the level. ``Check in'' and ``Check out'' work from left to right, where Mario needs to cross the fire to successfully traverse the clause.

\begin{lemma} A fire bridge preceded by a star implements a lock mechanism that can be unlocked. \end{lemma}
\noindent {\it Proof:} as mentioned before, Mario loses the level if he touches the fire, unless he has consumed a star, giving him temporary invincibility (enough to cross the bridge). This means that this corridor can't be traversed without a star. This is implements a locked path where the unlocking mecanism is that activation of one of the ``?'' blocks, placing the star to be used to cross the path. Without this, the path remains locked. 

This unlocks the clause by connecting each  variable to one of the ``?'' paths, so that the player can unlock the clause by visiting any of its variables.

One final remark is that the fire bridge must be long enough that a single star's timer is insufficient to traverse two clauses. Otherwise, the player could leave some clauses unsatisfied and still finish the level. \qed

\paragraph*{Crossover gadget.}
Finally, we have the crossover gadget as shown in Figure~\ref{fig:mario_cross}. This gadget has two possible paths: top-left to right, and bottom-left to top. It is used for each wire crossing in Figure~\ref{fig:cnf_fw}.
We assume that Mario is in his ``Big'' state for the whole level. The only state changes occur in the crossover gadget, but are always undone for Mario to successfully leave the gadget. In this state, Mario jumps higher and can break bricks that jumping into them from below. However, he can not fit into the one square sized gaps horizontally (for that, he must be in his ``Small'' state).

\begin{lemma} The gadget shown in Figure~\ref{fig:mario_cross} successfully implements a crossover gadget, forcing the player to exit through the right if he entered on the top left, and to exit through the top if he entered on the bottom left.\end{lemma}
\noindent {\it Proof:}
We consider the two possible traversals:
\begin{itemize}
\item if Mario enters through the top left, he must be attacked by the enemy, changing to his ``Small'' state in order to  fit through the small gap. In this state, he cannot break the bricks, so he's forced to continued all the way to the right, where the ``?'' block has a mushroom. To exit the gadget, he must consume the mushroom to change back into his ``Big'' state and break the brick to the right. He cannot go back to the center area because of the small gaps.
\item if he enters through the bottom left, he must go up through the center shaft, break once of the bottom bricks, then use the others as platforms. Then he breaks one of the top bricks, and uses that others as platforms to leave through the top. He can't go left or right because of the small gaps.
\end{itemize} \qed

\begin{theorem} Super Mario Bros. (SMB) is \textbf{NP}-Hard. \end{theorem}
\noindent {\it Proof:} Using Framework \ref{framework:np}, every 3-CNFSAT instance can be translated into a SMB level. Each framework element is replaced with the respective gadgets presented above. To unlock each variable, the player must direct Mario to the correct variable assignment that visits every clause at least once. This assignment only exists if the formula has a solution, and the solution can be retrieved by taking the sequence of variable choices. Because we have one gadget per variable, one gadget per clause, and a polynomial number of crossing (at most O($n^2$) when every wire crosses every other), we have a polynomial time reduction, proving SMB \textbf{NP}-Hard. \qed

It is unlikely that it is also in \textbf{NP}, as a later proof proved it \textbf{PSPACE}-Complete~\cite{hardermario}. This proof relied on a different enemy type, however, so it is possible that this restricted version is in \textbf{NP} after all.

\subsection{The Legend of Zelda: A Link to the Past}\label{proof:zelda}
We now give an example of the TQBF framework (Section~\ref{framework:pspace}) for The Legend of Zelda: A Link to the Past, implementing each gadget. Both the proof and the framework come from~\cite{nintendopaper}.
As in SMB, the objective is to take an avatar, Link, from a starting to a finishing position. Unlike SMB, however, the game is played from a top-down perspective, where the character doesn't jump. 
The necessary gameplay elements to understand the proof are the following:
\begin{itemize}
    \item There is a player avatar that moves in 4 directions/orientations;
    \item The avatar cannot jump over gaps or obstacles;
    \item There are gates with down/up states, controlled by a special tile that toggles (switches the blocks' state) once it is stepped on.
    \item Although the game is 2D, there is a discrete 3D element with two height levels that make bridges and pits possible;
    \item There are one-way teleport tiles (in pairs).
\end{itemize}

\begin{figure}[h!]
    \centering
    \includegraphics[scale=0.7]{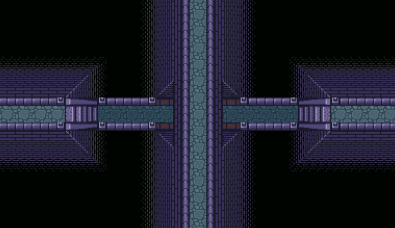}
    \caption{Zelda's crossover gadget from~\cite{nintendopaper}.}
    \label{fig:zelda_cross}
\end{figure}

\paragraph*{Crossover}
The crossover gadget is shown in Figure~\ref{fig:zelda_cross}, which follows directly from the bridge mechanic. This will make it possible to cross the paths without a more complicated construction.

\begin{figure}[h!]
    \centering
    \includegraphics[scale=0.9]{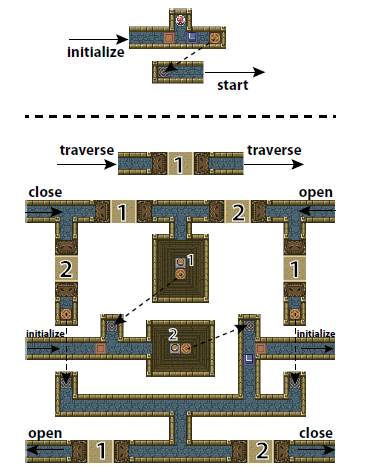}
    \caption{Zelda's door gadget from~\cite{nintendopaper}.}
    \label{fig:zelda_door}
\end{figure}

\paragraph*{Door}. As said in Framework \ref{framework:pspace}, doors are gadgets that need to open and close a path at will. This gadget is shown in  Figure~\ref{fig:zelda_door}. It uses 1-to-many toggles (each toggle can switch the state of multiple doors) to construct a door gadget with a path that opens it, one that closes it, and one that traverses it.
The buttons on the bottom of the pits change their respective door's state (given by the number label). After pressing the button, the player is forced to move into the one-way teleport, unable to press the button twice to avoid switching the door's state.

On the path labeled ``traverse'' is the actual door that can be opened and closed by traversing the two other paths. Gates labeled 1 and 2 are always in opposite states, so if gate 1 is open, gate 2 is closed and vice-versa, determining the door's state.

\begin{lemma} The gadget shown in Figure~\ref{fig:zelda_door} implements a Door gadget conforming to the constraints described in Framework \ref{framework:pspace}.\end{lemma}
\noindent {\it Proof:}
The gadget has 2 entry points: ``open'' and ``close''. The Door's state is a function of the 2 gates' states: 
open : (open, closed); closed : (closed, open).
We can follow the player's steps to make sure the gadget works correctly.
\begin{itemize}
\item Assume the player enters through the ``Close'' path while the door is \textit{closed} (gates' states are (closed, open)). The player is forced to go downwards all the way to the bottom, and then turn right. The gates do not change state, so the door doesn't either. We can conclude the same for the ``Open'' path with the door open, by symmetric reasoning.

\item Now assume the player enters through the ``Close'' path while the door is \textit{open}. The player is forced to go right and down, falling into the pit, toggling gate 1. At this point, the gates' states are (closed, closed). The player is then forced to step into the teleport tile. Now, once again, he has to go right and fall into the second pit, thus toggling gate 2. The gate's states are now (closed, open). 
He then follows the second teleport and can then travel all the way to the bottom of the gadget and leave through the bottom right, exiting the ``Close'' path. The door's state has changed, since both gates' did.
Symmetric reason shows that the ``Open''path is also correct.
\end{itemize}
As a result, the gadget does have three disjoint paths. To be able to traverse the ``Closed'' path, the player is forced to traverse the gadget in a way that results in the door having the state ``Closed'' upon exit. Similarly, ``Open'' results in the door having the state ``Open''. The ``Traverse'' path does not change the gadget's state, and can only be traversed if the door is in its ``Open'' state. \qed

\begin{theorem}The Legend of Zelda: A Link to the Past (Zelda) is \textbf{PSPACE}-Hard.\end{theorem}
\noindent {\it Proof:}
 Following framework \ref{framework:pspace}, we show that TQBF $\le _p$ Zelda by replacing the doors with Door-Traverse gadget, Open door to the Close path of the door gadget and Open door to the Open path of the door gadget. Any crossing paths are replaced with a Crossover gadget. The player will only be able to reach the end of the levels if the original quantified formula is true, proving Zelda \textbf{PSPACE}-Hard. \qed

\cleardoublepage
%
\fancychapter{Results}
\cleardoublepage
\label{chapter:results}
In this chapter, we present our approach in Section~\ref{section:approach}, taking care of some technical details relating to each framework used. We then dedicate Sections~\ref{proof:cuttherope} to~\ref{proof:backtobed} to presenting eight new proofs using the methods presented in the previous chapter.
3-CNFSAT was used to prove Cut the Rope \textbf{NP}-Hard (Section~\ref{proof:cuttherope}).
CircuitSAT was used to prove \textbf{NP}-Hardness of Hexagonal Akari (Section~\ref{proof:akarihex}), Triangular Akari (Section~\ref{proof:akaritri}), HexCells (Section~\ref{proof:hexcells}) and Hexiom (Section ~\ref{proof:hexiom}).
Hamiltonian Cycle was used used to prove Super Meat Boy \textbf{NP}-Hard (Section~\ref{proof:supermeatboy}).
Finally, TQBF was used to prove \textbf{PSPACE}-Hardness of Offspring Fling (Section~\ref{proof:offspringfling}) and Back to Bed (Section~\ref{proof:backtobed}).

\section{Approach}\label{section:approach}
Here, we will detail the frameworks shown previously, taking care of the technical details to facilitate the construction of new proofs with those frameworks.

\subsection{Circuit SAT}\label{approach:circuitsat}
We will use the Circuit SAT framework presented in Section~\ref{framework:circuit}, page~\pageref{framework:circuit}.
For this problem, there are 2 important features to mention. The first concerns \textit{functional completeness}, which determines the set of gates required for the proof.
The second concerns \textit{circuit planarity}, which is solved simply by any functional complete set of gates, as we will explain.

The figures in this subsection were done using Logisim~\footnote{http://www.cburch.com/logisim/}.

\paragraph*{Functional Completeness} refers to how many and which logical connectives (gates) we need to express every boolean expression. Although there are many connectives ($\textit{AND, OR, XOR, IMPLY}$, etc.), many of them can be represented by a composition of others.
There is also the notion of \textit{minimal functionally complete set}, which is a set of gates that is \textit{irreducible} - removing any of the gates from the set would remove its functional completeness property.

As a result of this property, there are multiple sets of gates that are \textit{sufficient} for a proof. This multiplicity provides some flexibility, as some gates may follow more or less naturally than others from each game's mechanics. It may also provide more compact proofs, with the possibility of creating real puzzles from simple formulas.

Some of the functionally complete sets are very small. Some include a single a gate, like $\textit{NAND}$ or $\textit{NOR}$. In practice, it's easier to provide $\textit{AND}/\textit{OR}$ and $\textit{NOT}$ separately.

\begin{figure}[!h]
\center
 \subfloat[$\textit{AND}$ implementation with $\textit{NAND}$ gates.\label{nand_and}]{%
   \includegraphics[scale=0.8]{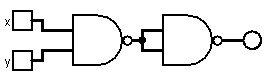}
 }
 \subfloat[$\textit{OR}$ implementation with $\textit{NAND}$ gates.\label{nand_or}]{%
   \includegraphics[scale=0.7]{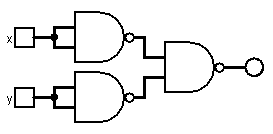}
 }\\
 \subfloat[$\textit{XOR}$ implementation with $\textit{NAND}$ gates.\label{nand_xor}]{%
   \includegraphics[scale=0.6]{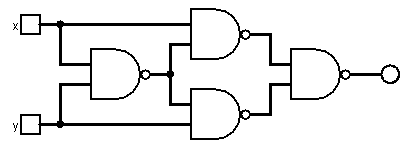}
 }
 \subfloat[$\textit{NOT}$ implementation with $\textit{NAND}$ gates.\label{nand_not}]{%
   \includegraphics[scale=1]{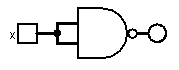}
 }
 \hfill
 \caption{Individual gate conversion to $\textit{NAND}$ gates only.}
 \label{fig:fc_nand}
\end{figure}

\begin{figure}[!h]
 \subfloat[Circuit before $\textit{NAND}$ conversion.\label{circuit_conv_before}]{%
   \includegraphics[scale=0.5]{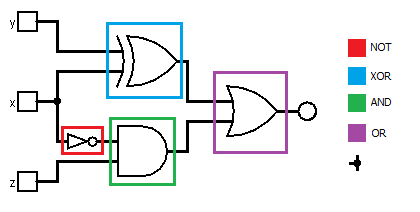}
 }
 \hfill
 \subfloat[Circuit after $\textit{NAND}$ conversion.\label{circuit_conv_after}]{%
   \includegraphics[scale=0.4]{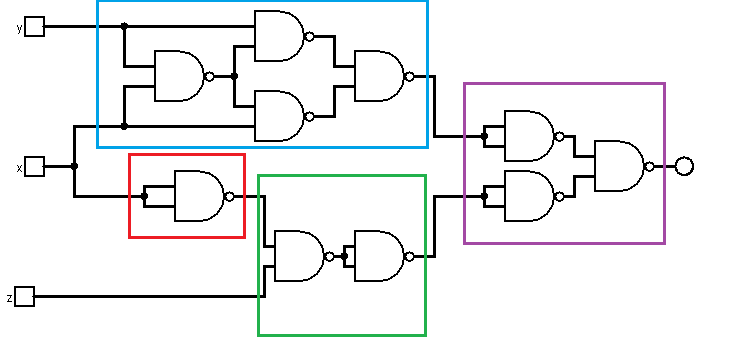}
 }
 \hfill
 \caption{$(x \oplus y) \lor (\lnot x \land z)$ before and after $\textit{NAND}$ conversion.}
 \label{fig:functional_completeness}
\end{figure}

Figures \ref{fig:fc_nand} and \ref{fig:functional_completeness} illustrate functional completeness by showing how to convert each outlined gate as a circuit of $\textit{NAND}$ gates. Figure~\ref{circuit_conv_before} shows the formula $(x \oplus y) \lor (\lnot x \land z) $ before the $\textit{NAND}$ conversion. Figure~\ref{circuit_conv_after} shows the same formula after the conversion. More about functional completeness can be found in ~\cite{funccomp}.

\begin{figure}[h!]
\centering
\includegraphics[scale=0.7]{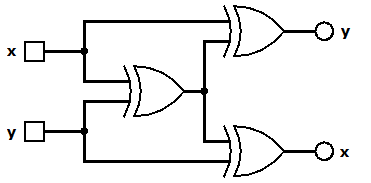}
\caption{Crossover gadget using X$\textit{OR}$ gates.}
\label{circuit_cross}
\end{figure}

\paragraph*{Planarity.} Planarity is not a problem for CircuitSAT, since it is easy to implement a crossover gadget using X$\textit{OR}$ gates, as shown in Figure~\ref{circuit_cross}.

Every time we have crossing wires in the circuit, we can add these 3 gates and remove the crossing, making it planar. Because the proof already relies on having a functionally complete set of boolean gates, we are guaranteed to have an X$\textit{OR}$ gate (or equivalent composition of other gates), solving the crossing problem without new gadgets.

\paragraph*{Parity.} Sometimes, CircuitSAT reductions' wires will have a length multiple of some $k$. This can lead to parity issues when trying to fit all the gadgets together. This can be solved by an additional ``Parity Shift'' gadget that introduces a delay or an advance to wires. This can be seen in Minesweeper~\cite{minesweeper}'s proof, designated as a ``Phase-changer''. We solve a similar problem using two $\textit{NOT}$ gates in HexCells' and Hexiom's proofs (see Sections \ref{proof:hexcells} and \ref{proof:hexiom}).

\subsection{3-CNFSAT}

\paragraph*{Conversion to CNF.}
The framework shown in Section \ref{framework:np} assumes that formulas are in the Conjunctive Normal Form (CNF).

The naive approach, through application of De Morgan's rules and distributivity can lead to an exponential blow-up in size. Using the ``Tseytin transformation'', however, it is possible to achieve the linear bound, and get a logically equivalent formula. It works by first identifying subformulas, and then assigning them to a new auxiliary variable, through equivalency ($aux\_var <=> subformula$). This can be seen as 2 implications.  Because the number of sub-formulas can't exceed the formula's size, we have a $2*|formula|$ = $O(|formula|)$ blowup. The interested reader can learn more about encodings in~\cite{handbooksat}.

\paragraph*{Lock mechanism.} In this framework, the most important gadget is the lock, and it should be the first element to be searched for when applying this framework. 
A game is typically very easy to prove in the presence of tokens or keys. Several variations are given in~\cite{forisek,viglietta} of generic game mechanics that can serve as a lock mechanism. 

Our proof for Cut the Rope shows a different example in which a lock mechanism was built out of multiple game mechanics in a less direct manner. 

\paragraph*{Teleporting.} Several games have teleporting mechanics that link two arbitrary points in space. One example is Cut the Rope's magic hats (see Section~\ref{proof:cuttherope}). This simplifies the framework considerably, since crossovers are no longer needed.

\subsection{TQBF}
\paragraph*{Prenex Form.}
Although this framework assumes that the quantified formulas are in \textit{prenex form} (first a sequence of quantifiers, followed by an unquantified boolean formula), this is not true for general quantified boolean formulas. To build levels out of generic examples, then the conversion must be done. This conversion is outside the scope of this report, as the reductions are not as elegant as CircuitSAT's, with which we can build fairly small puzzles, but the interested reader can find more in~\cite{mathlogic,prenex2016}.

\paragraph*{Door gadget.}
The main difference between the \textbf{PSPACE} and \textbf{NP} frameworks is precisely the lock and door mechanisms. When looking at a game (specifically platformers or games with avatars that must traverse a level) it's the most helpful to check whether the game's mechanics (the candidates for doors and locks) are consumable/single-use or more permanent, and can change between states indefinitely. This is similar to the differences in Non-Deterministic Constraint Logic frameworks given for bounded and unbounded games~\cite{ncl}, where the variation lies in whether or not moves can be undone.

\subsection{\textbf{NP} and \textbf{PSPACE}}
It is important to note that every \textbf{PSPACE}-Hard problem is also \textbf{NP}-Hard. As a result, when tackling problems and games of unknown complexity, it may be helpful to consider the weaker notion of \textbf{NP}-Hardness before \textbf{PSPACE}-Hardness, since the latter would be a dead-end in the case of \textbf{NP}-Complete problems. This approach was taken for Offspring Fling (Section~\ref{proof:offspringfling}); the initial \textbf{NP}-Hardness proof is not presented as it is made redundant by the \textbf{PSPACE}-Hardness proof. Super Mario Bros.'s hardness followed a similar path, first proved \textbf{NP}-Hard in ~\cite{nintendopaper} and later \textbf{PSPACE}-Hard in ~\cite{hardermario}.

\textbf{NP} membership can also be tested after a \textbf{NP}-Hardness proof to avoid trying to prove the stronger \textbf{PSPACE}-Hardness result. For example, the game of Akari is easily seen to be in \textbf{NP}, the argument given in Section ~\ref{proof:akari}.

\section{Cut the Rope (3-CNFSAT)}\label{proof:cuttherope}
Cut the Rope is a very popular physics based mobile game developed by Zepto Lab and first released in 2010. It has since spawned many sequels and variations.

Here, we prove its \textbf{NP}-Hardness using the 3-CNFSAT (Section~\ref{framework:np}, page~\pageref{framework:np}) used earlier to prove the \textbf{NP}-Hardness of Super Mario Bros.

\begin{figure}[!h]
    \center
     \subfloat[Rope pivot.\label{fig:cuttherope_rope}]{%
       \includegraphics[scale=0.35]{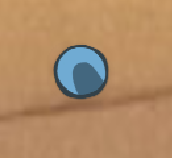}
     }
     \subfloat[Movable rope.\label{fig:cuttherope_moverope}]{%
       \includegraphics[scale=0.3]{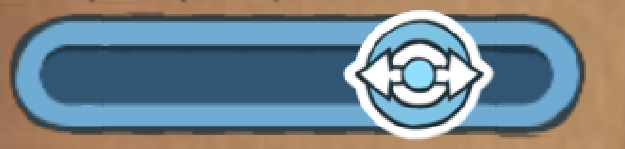}
     }
     \subfloat[Teleport hat.\label{fig:cuttherope_hat}]{%
       \includegraphics[scale=0.4]{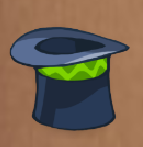}
     }
     \\
     \subfloat[Spikes.\label{fig:cuttherope_spike}]{%
       \includegraphics[scale=0.5]{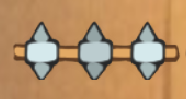}
     }
     \subfloat[Balloon.\label{fig:cuttherope_balloon}]{%
       \includegraphics[scale=0.3]{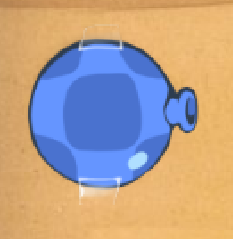}
     }
     \subfloat[Bubble.\label{fig:cuttherope_bubble}]{%
       \includegraphics[scale=0.4]{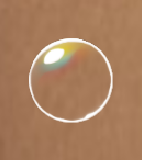}
     }
     \caption{Cut the Rope elements required for the proof}
     \label{fig:cuttherope_mechanics}
    \end{figure}

The objective of the game is to carry a candy (avatar) to Om Nom's mouth (Finish). The player does not control the candy directly, however. The levels have a variety of physics-based objects that the player interacts with to change the candy's path. The relevant mechanics for this proof, illustrated in Figure~\ref{fig:cuttherope_mechanics}, are the following:
\begin{itemize}
    \item Ropes (Figure~\ref{fig:cuttherope_rope}) trigger when the candy enters their radius, then acting as a pendulum. The player can swipe his finger on the rope to cut it. Ropes only trigger once, and are no longer usable once cut. A rope's length is the same as its trigger radius.
    \item Movable (Figure~\ref{fig:cuttherope_moverope}) ropes are ropes that the player can move around within the visible limits. 
    \item Teleporting hats (Figure~\ref{fig:cuttherope_hat}) come in pairs. The candy enters through one, it leaves through the other. The candy's velocity is conserved after teleporting, but the movement's direction changes with the hats'.
    \item Spikes (Figure~\ref{fig:cuttherope_spike}) break the candy on contact, making the player to loose the level.
    \item Balloons (Figure~\ref{fig:cuttherope_balloon}) can be pressed by the player to blow air. This applies force to the candy.
    \item Bubbles (Figure~\ref{fig:cuttherope_bubble}) activate on contact and make the candy float. Once pressed (by the player), the bubbles pop and lose their effect for the remainder of the level.
\end{itemize}

\begin{figure}[!h]
\center
 \subfloat[Variable choice.\label{fig:cuttherope_variableselect}]{%
   \includegraphics[scale=0.52]{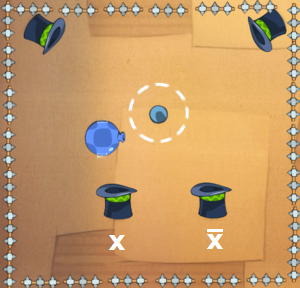}
 }
 \subfloat[Lock.\label{fig:cuttherope_lock}]{%
   \includegraphics[scale=0.4]{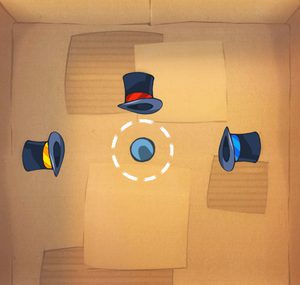}
 }
 \subfloat[Speed adjuster.\label{fig:cuttherope_speed}]{%
   \includegraphics[scale=0.34]{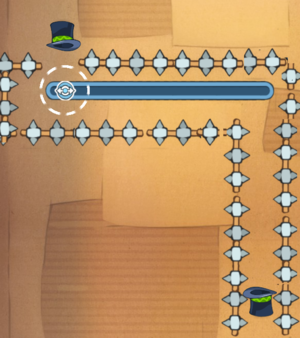}
 }\\
 \begin{center}
 \subfloat[Clause unlock.\label{fig:cuttherope_clausevisit}]{%
   \includegraphics[scale=0.7]{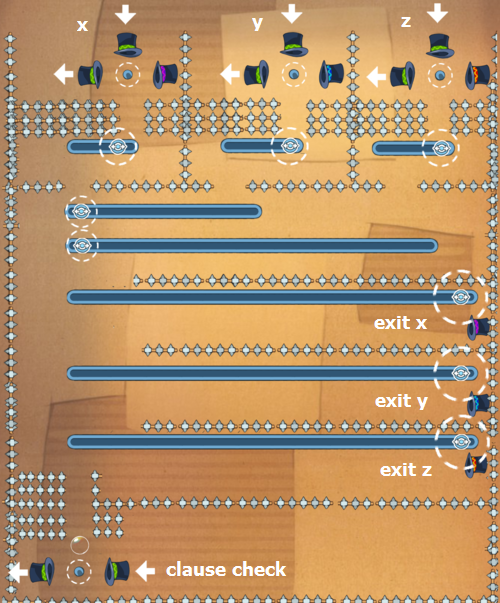}
 }
 \end{center}
 \hfill
 \caption{Cut the Rope gadgets.}
 \label{fig:cuttherope_gadgets}
\end{figure}

Figure~\ref{fig:cuttherope_gadgets} shows the gadgets required for the proof. Because of the physics' nature of the game, an additional gadget is required to make the proof work, independent of the ones mentioned in Framework \ref{framework:np}, page~\pageref{framework:np}, to adjust the candy's speed.

\paragraph*{Variable assignment.} The variable assignment gadget (Figure~\ref{fig:cuttherope_variableselect}) starts with the candy triggering the rope at the top, after entering from one of the two teleport hats at the top. The player enters through the left hat if he has chosen the value ``true'' to the previous variable, and the right hat if he has chosen the value ``false''. The player activates the balloon to swing the candy, cutting the rope at the appropriate moment to let the candy fall into either hat.

One of the entry hats corresponds to the previous variable's assignment being true, and the other false, as seen in Section~\ref{framework:np}, page~\pageref{framework:np}. 

Following the framework in Figure~\ref{fig:cnf_fw}, page~\pageref{fig:cnf_fw}, consider a variable $v$. One hat represents \textit{true} and will lead to the ``clause visit'' gadget for $v$, unlocking every clause where $v$ appears. The other hat represents \textit{false}, unlocking every clause where $\lnot{v}$ appears.

\paragraph*{Lock mechanism.}
Figure~\ref{fig:cuttherope_lock} shows how we implemented a lock in Cut the Rope. This ``lock'' is used in 2 ways: as the clause check (the clause must be unlocked before it is traversed, see Figure~\ref{fig:cuttherope_clausevisit} and the bottom left) and also to let the player enter the same gadget from multiple paths, yet forcing him to leave through the same path from which he entered (also Figure~\ref{fig:cuttherope_clausevisit}).

\begin{lemma}The gadget shown in Figure~\ref{fig:cuttherope_lock} behaves as a lock. The horizontal path can only be traversed if the top of the gadget has been used to unlock it.\end{lemma} 
\noindent {\it Proof:}
The gadget is traversed horizontally, and the top hat (or top entry, if not a hat) is used as the unlocking mechanism. To cross the horizontal path, the player must \textit{first} trigger the rope at the center. Otherwise, the rope will stop (or misdirect) the candy's movement. To this end, we add a separate entry from the top that is used to trigger the rope and float back up with the help of a bubble. Later, the player may traverse the gadget horizontally.

To make this gadget work, we assume that the entry from the top has no horizontal movement (this can be guaranteed with the gadget shown in Figure~\ref{fig:cuttherope_speed}) so that the player can't enter from the top and leave through the left/right by swinging. We also assume that the rope cannot be cut before it affects movement. These assumptions make it impossible to cross the \textit{locked} gadget horizontally.

Attempting to traverse the horizontal path without unlocking it first will invariably lead to the candy being destroyed (or stuck), so the player never has the opportunity to unlock it horizontally. To traverse the path, then, the player must first visit it from the top, successfully implementing a lock mechanism. \qed

\paragraph*{Clause gadget.}
Figure~\ref{fig:cuttherope_clausevisit} shows the most important gadget of this reduction, the clause visit gadget. This includes both the unlocking and the ``Clause Check'' traversal. The ``Clause Check'' is implemented by the two hats at the bottom left of the gadget. The rest of the gadget implements the clause unlocking mechanism. 

Because this represents a clause in 3-CNFSAT, there are 3 entries and exits corresponding to each variable in the clause. Consider the three sets of three hats at the top. Each set corresponds to one variable. The top hat is the entry point. The left hat is the exit point. The right hat is used to traverse inside the gadget, with the pairs being color coded (red, blue and purple) to indicate which hat connects to which. Note that between each set of 3 hats is a rope. This is one use of the lock mechanism.

The hats at the bottom represent the ``Clause Lock''. The left hat connects to the next clause and the right hat to the previous one. The bubble is placed so that the candy will trigger the clause's rope, cut it, and then float back up to exit the gadget. If the bubble has already been used, this step will be skipped and the candy will be caught by the respective movable rope on its way down to exit the gadget  before reaching the ``Clause Lock''.

The function of the gadget is \textit{to force the candy to leave through the variable from which it came}. For this to be correct, three properties must hold.

\begin{lemma}Property 1: The player must unlock the respective lock (consume the rope between the 3 hats on the top) to use the exit. \end{lemma}
\noindent {\it Proof:} This follows from how the lock mechanism works. To traverse the path horizontally, the rope must have been previously consumed, or the candy will get stuck and must then fall. \qed
    
\begin{lemma}Property 2: The player can't reach the lock from below and go back down.
\end{lemma}
\noindent {\it Proof:} Unlocking from below can only happen with the use of the bubble, floating all the way up to the top. However, to avoid the spikes and align the candy with the lock, the top most movable rope must be used. Note that the rope's range is short enough so that, to float up and trigger the lock, the player must cut the movable rope. After triggering the lock, however, the player must burst the bubble and cut the rope to move back down. Doing so would let the candy fall onto the spikes, however, since the movable rope has already been consumed. \qed
    
\begin{lemma}Property 3: The player can't reach the lock from the same colored hat twice.
\end{lemma}
\noindent {\it Proof:} This follows from a similar principle to property 2's. To reach the colored hat, the candy must consume the colored hat's movable rope. Doing this, the player could open an incorrect lock and move back down. However, since the rope is consume, there is no way to reach the same colored hat again, and so the player will not be able to leave through the incorrect entry point. \qed

From the three properties, we can conclude that the only way to traverse the clause safely is to unlock the lock by entering through the top, and then using the single-use movable rope to exit the gadget. Note that the Exit movable ropes can intercept the candy in its fall, so the clause does not \textit{need} to be unlock every time (or even any time). The player should then open it on the first clause visit, and simply traverse the gadget for the subsequent variables. 

To finish the reduction, we'll simply consider some details to do with velocity and speed.

\paragraph*{Forcing exact speed when leaving hats.}
Every time the candy moves from one gadget to the other, it is through teleport hats. To make sure the candy is always thrown to the correct place, we may need to make sure that it does not have too little or too much velocity. To overcome this problem, we have an additional gadget (Figure~\ref{fig:cuttherope_speed} that can be added between any teleport hat pair.

\begin{lemma}The gadget shown in Figure~\ref{fig:cuttherope_speed} can force the candy to leave the gadget with arbitrary exit velocity, given an arbitrary entry velocity.\end{lemma}

\noindent {\it Proof:}
The candy enters through the top hat triggering the rope (thus removing vertical speed) and moves to the right. To descend without colliding with the spikes, the player must wait for the candy to stop swinging (minimizing horizontal speed). The player then cuts the rope and lets the candy fall under gravity. When the candy reaches the hat, it will have negligible (if any) horizontal speed, and any desired horizontal speed, depending on the length of the fall, which can be adjusted to the game physics' parameters.\qed

The proof should now be robust for different settings of gravity and air friction and every detail is taken care of. Although the framework is not as clean for a physics' based game (instead of a more rigid platformer like Super Mario Bros.), it can still be successfully applied.

\begin{theorem}Cut the Rope is \textbf{NP}-Hard.\end{theorem}
\noindent {\it Proof:} We show that 3-CNFSAT $\le _p$ CutTheRope by implementing the variable assignment and clause gadgets from Framework~\ref{framework:np}, page~\pageref{framework:np} with the respective CutTheRope gadgets. Wire endpoint are replaced with teleport hats, removing the need for crossover gadgets. Between any two teleport hats, we add the additional gadget in Figure~\ref{fig:cuttherope_speed} to account for speed gain/loss during the gadgets' traversal. 
For every 3-CNFSAT formula, we create a CutTheRope level that can be successfully finished if and only if the original formula is satisfiable. The variables' values between the level and the formula have a one-to-one correspondence, given by the path traversed in Figure~\ref{fig:cuttherope_variableselect}, where the left hat represents true and the right hat represents false. \qed

\paragraph*{Example.} To see how everything fits together, we present a schematic example with numbered hats in Figure~\ref{fig:cuttherope_example}, in the Appendix A, Section~\ref{example:cuttherope}. 

\section{Hexagonal Akari (CircuitSAT)}\label{proof:akarihex}
Here we present a CircuitSAT proof for Akari, he game already proved \textbf{NP}-Hard by Brandon McPhail~\cite{akari}. However, we changed the game's grid to an hexagonal one. A proof for a triangular grid is also presented in Section~\ref{proof:akaritri}.

\begin{figure}[h!]
\centering
\includegraphics[scale=0.5]{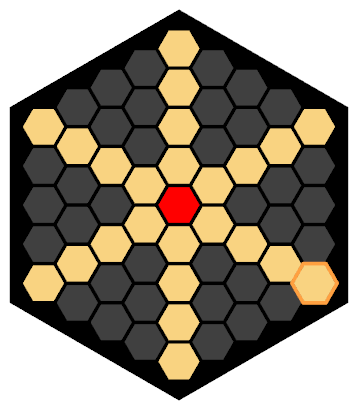}
\caption{Lamp (center) propagating light.}
\label{fig:akarihex_prop}
\end{figure}

\paragraph*{Hexagonal Grid.}
The game's rules aren't significantly changed by using a different grid, the only change being how light propagates. We generalized the propagation rule in the most natural way: instead of propagating in only 2 directions as in the square grid, it now propagates in 3, as seen in Figure~\ref{fig:akarihex_prop}.

\begin{figure}[!h]
 \subfloat[Wire.\label{fig:akarihex_wire}]{%
   \includegraphics[scale=0.35]{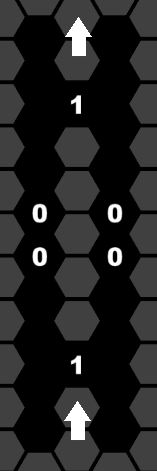}
 }
 \hfill
 \subfloat[60 degree turn.\label{fig:akarihex_turn60}]{%
   \includegraphics[scale=0.85]{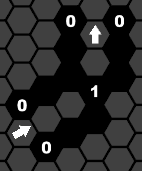}
 }
 \hfill
 \subfloat[120 degree turn.\label{fig:akarihex_turn120}]{%
   \includegraphics[scale=0.35]{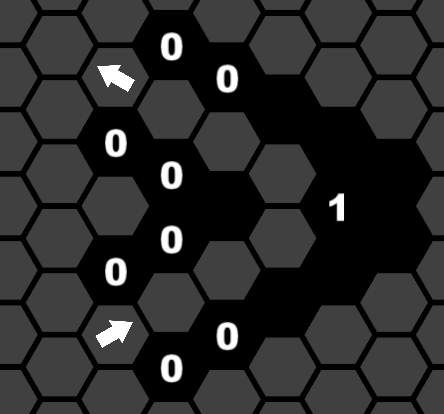}
 }
 \\
 \subfloat[Negated Fan-Out gate.\label{fig:akarihex_fanout}]{%
   \includegraphics[scale=0.3]{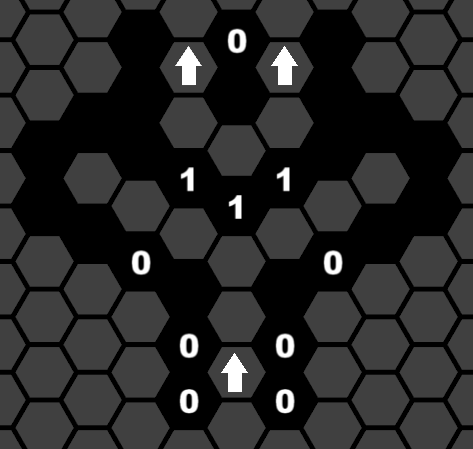}
 }
 \hfill
 \subfloat[$\textit{NOT}$ gate.\label{fig:akarihex_not}]{%
   \includegraphics[scale=0.5]{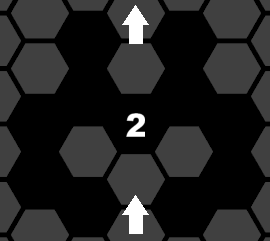}
 }
 \hfill
 \subfloat[$\textit{OR}$ gate.\label{fig:akarihex_or}]{%
   \includegraphics[scale=0.4]{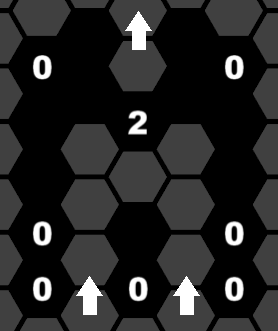}
 }
 \hfill
 \caption{Hexagonal Akari's gadgets.}
 \label{fig:akarihex_gadgets}
\end{figure}

\paragraph*{Gadgets.}
Figure~\ref{fig:akarihex_gadgets} shows the gadgets required. As in the proof explained in Section~\ref{proof:akari}, we define the puzzle's orientation as a circuit's, going from the input (variables) to the output. The true signal is represented by the corridor being forward illumination (from the input to the output). Unilluminated corridors will force backward illumination (from the output to the input), representing the false signal.
Every gadget is Figure~\ref{fig:akarihex_gadgets} is orientated from bottom to top (with the exception of the turn gadget), as illustrated by the white arrows. The turn gadgets are symmetric and can be turned by any angle $mod$ 60 degrees, so we don't need to define any orientation other than that they need to be aligned with the input wire. Gray cells outside the gadgets are not part of the puzzle/reduction and can be assumed to be obstacles.

\paragraph*{Wires and Turns.} These work exactly as in the square grid, so we refer back to Section~\ref{proof:akari}.

\paragraph*{$\textit{NOT}$ gadget.}
Figure~\ref{fig:akarihex_not} shows the $\textit{NOT}$ gadget. The gadget works symmetrically, but we can assume the signal is propagated from bottom to top. 

\begin{lemma}The gadget shown in Figure~\ref{fig:akarihex_not} implements a $\textit{NOT}$ gate, negating the input signal.\end{lemma}
\noindent {\it Proof:}
If the input corridor (bottom) is already illuminated, the lamp cannot be placed below the 2-cell, since that cell is already illuminated. To illuminate the two cells to the side (left and right), both of them must have a lamp placed on them. The 2-cell then has two adjacent lamps, and no lamp must be placed at the top. As a result, if the input signal is true (corridor illuminated), the output signal is false (the top corridor will not be illuminated). 

On the other hand, if the corridor is \textit{not} illuminated, a lamp must be placed below the 2-cell, to illuminate the input corridor. A lamp below the 2-cell also illuminates the two cells on the side, so the second lamp must be placed above the 2-cell, illuminating the output corridor.

Those are the only two valid configurations, both negating the input signal, successfully implementing a $\textit{NOT}$ gate. \qed

\paragraph*{Negated FAN-OUT.}
Figure~\ref{fig:akarihex_fanout} shows a negated FAN-OUT gadget, used to multiply the input signal. Usually, the aim is for a gadget that simulates a FAN-OUT gate, not its negation. This is not problematic, because of the $\textit{NOT}$ gadget; we can always apply it to negate the input, leading to a normal FAN-OUT gate.

\begin{lemma} The gadget shown in Figure~\ref{fig:akarihex_fanout} implements a negated FAN-OUT gate by propagating two signals as its output, with the opposite value of its input.\end{lemma}
\noindent {\it Proof:}
There are only two valid configurations for this gadget. If the input is true, no lamp may be placed below the center 1-cell, because that cell is illuminated from the input. As a result, the lamp must be placed above it, which is also adjacent to the side 1-cells. This prevents any lamp from illuminating any of the output corridors from this gadget, propagating a false signal from both outputs.

If the input is false, on the other hand, a lamp must be placed below the center 1-cell to illuminate the input corridor. Then, the cell above it must not have a lamp, forcing one lamp to be placed above each of the side 1-cells. This illuminates both output corridors, thus propagating a true signal from both outputs.

The input signal uniquely determines the outputs, which are its negation, successfully implementing a negated FAN-OUT gate. \qed

\paragraph*{$\textit{OR}$ gadget.}
Figure~\ref{fig:akarihex_or} shows the $\textit{OR}$ gadget. It is easy to see that if any input corridor is illuminated, then a cell must be placed above the 2-cell, propagating a true signal.

\begin{lemma} The gadget shown in Figure~\ref{fig:akarihex_or} implements an $\textit{OR}$ gate, propagating a false signal if and only if both of its inputs are false. \end{lemma}

\noindent {\it Proof:} If both inputs are true, then there are only two unilluminated cells where lamps can be placed: below and above the 2-cell. This propagates a true signal.

If only one of the inputs is true, then there is one unilluminated cell above the 2-cell and two below it. However, the two cells below are adjacent, so they can't both have lamps. We can conclude that one lamp must be placed above the 2-cell (and the other to the side, to illuminate the false corridor), also propagating a true signal.

Finally, if none of the inputs are true, then two lamps must be placed below the 2-cell in order to illuminate the two corridors. This prevents a lamp from being above the 2-cell, thus propagating a false signal. \qed

\begin{theorem} Hexagonal Akari is \textbf{NP}-Complete. \end{theorem}
\noindent {\it Proof:} We show that CircuitSAT $\le _P$ Hexagonal Akari by referring to Section~\ref{framework:circuit}, page~\pageref{framework:circuit}. We start by converting every boolean gate into a composition of the gates from the functionally complete set $\{OR,NOT\}$. We can then replace every WIRE (straight or turned) and FAN-OUT gate with the respective gadgets from Figure~\ref{fig:akarihex_gadgets}. This reduces any CircuitSAT instance to a Hexagonal Akari puzzle, proving it is \textbf{NP}-Hard.
We show that Hexagonal Akari $\in \textbf{NP}$ due to the existence of a polynomial time verifier. A solution is given by each lamp's position. The number of lamps is limited by the board's size, so it is linear. Light propagation is also limited by the board's size, since it propagates on straight lines. To verify each cell's constraints, we only need to keep a track of a bit that states whether each cell is illuminated or not, and then checking the adjacency constraints, which can be done in constant time (always 6 neighbours).

Hexagonal Akari, then, is both \textbf{NP}-Hard and $\in$ \textbf{NP}, making it \textbf{NP}-Complete. \qed

\paragraph*{Example.} To make sure every gadget fits together, we present a solvable puzzle (from a boolean formula) in Figure~\ref{fig:hexakari_example}, in the Appendix A, Section~\ref{example:akarihex}.

\section{Offspring Fling (TQBF)} \label{proof:offspringfling}
Offspring Fling is a puzzle-platformer game by independent developer Kyle Pulver, released in 2012. Here we prove it \textbf{PSPACE}-Hard using the framework presented in Section~\ref{framework:pspace}, page~\pageref{framework:pspace}.

\begin{figure}[!h]

 \subfloat[Offspring (left) and avatar (right).\label{fig:of_avatar}]{%
   \includegraphics[scale=0.9]{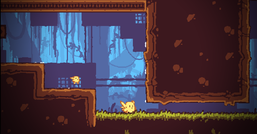}
 }
 \hfill
 \subfloat[Carrying offspring increases avatar's height.\label{fig:of_carry}]{%
   \includegraphics[scale=1]{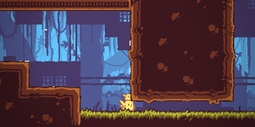}
 }\\
 \center
 \subfloat[Offspring throw.\label{fig:of_throw}]{%
   \includegraphics[scale=1]{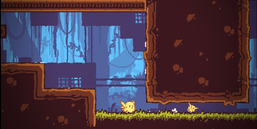}
 }
 \hfill
 \caption{Offspring Fling's movement and carrying constraints.}
 \label{fig:of_mechanics}
\end{figure}

\paragraph*{Game Rules.}
The player controls an avatar and has to collect its offspring and get them to an exit location. Carrying offspring adds height to the character, however, so the player must fling (throw) them to pass through narrow passages. Figure~\ref{fig:of_mechanics} illustrates this mechanic. Once thrown, the offspring's movement only stops once it hits an obstacle (or is caught in mid-air by the player avatar).

\begin{figure}[!h]
 \subfloat[Active blocks.\label{fig:of_toggle0}]{%
   \includegraphics[scale=0.4]{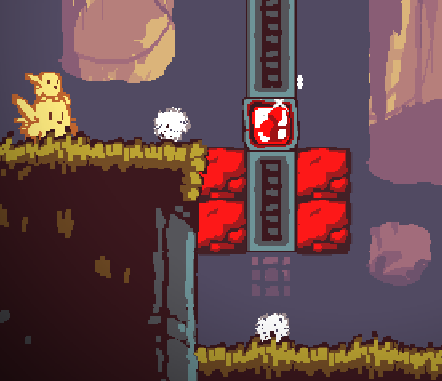}
 }
 \hfill
 \subfloat[Offspring thrown.\label{fig:of_toggle1}]{%
   \includegraphics[scale=0.4]{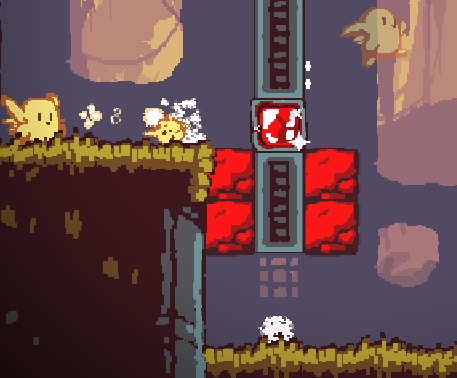}
 }
 \\
 \subfloat[Offspring hitting the switch.\label{fig:of_toggle2}]{%
   \includegraphics[scale=0.4]{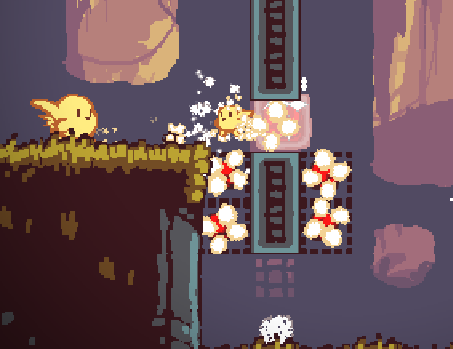}
 }
 \hfill
 \subfloat[Inactive blocks.\label{fig:of_toggle3}]{%
   \includegraphics[scale=0.45]{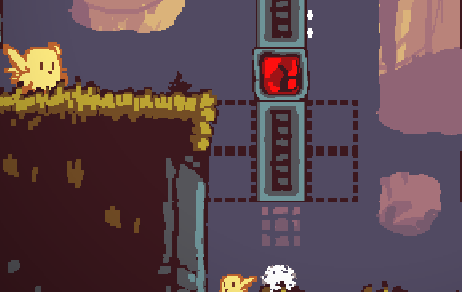}
 }
 \caption{Offspring Fling's toggle switch.}
 \label{fig:of_toggle}
\end{figure}

The game also features \textit{toggle switches} (crystals) that activate/deactivate certain blocks when hit with the offspring. Figure~\ref{fig:of_toggle} illustrates how to toggle a switch. These will be used to implement the Door mechanism, similar to the door in Zelda's proof (Section~\ref{proof:zelda}). A block can only be activated/deactivated by a single switch, but each switch may activate/deactivate multiple blocks.

Now that we've explained the necessary mechanics, we can move on to the proof.
\paragraph*{Proof overview.}
Recall from Section~\ref{framework:pspace}, page~\pageref{framework:pspace} that the only gadgets necessary to prove \textbf{PSPACE}-hardness are a Crossover and a Door. In this game, a crossover can be achieved with the door gadget, so only one gadget is effectively required. This gadget is shown in Figure \label{fig:of_gadgets}.

\begin{figure}[!h]
\center
 \subfloat[Door gadget - Traverse path.\label{fig:of_doortraverse}]{%
   \includegraphics[scale=0.7]{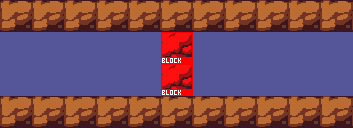}
 }
 \vfill
 \subfloat[Door gadget - Open and Close paths. Also works as a crossover.\label{fig:of_dooropen}]{%
   \includegraphics[scale=0.32]{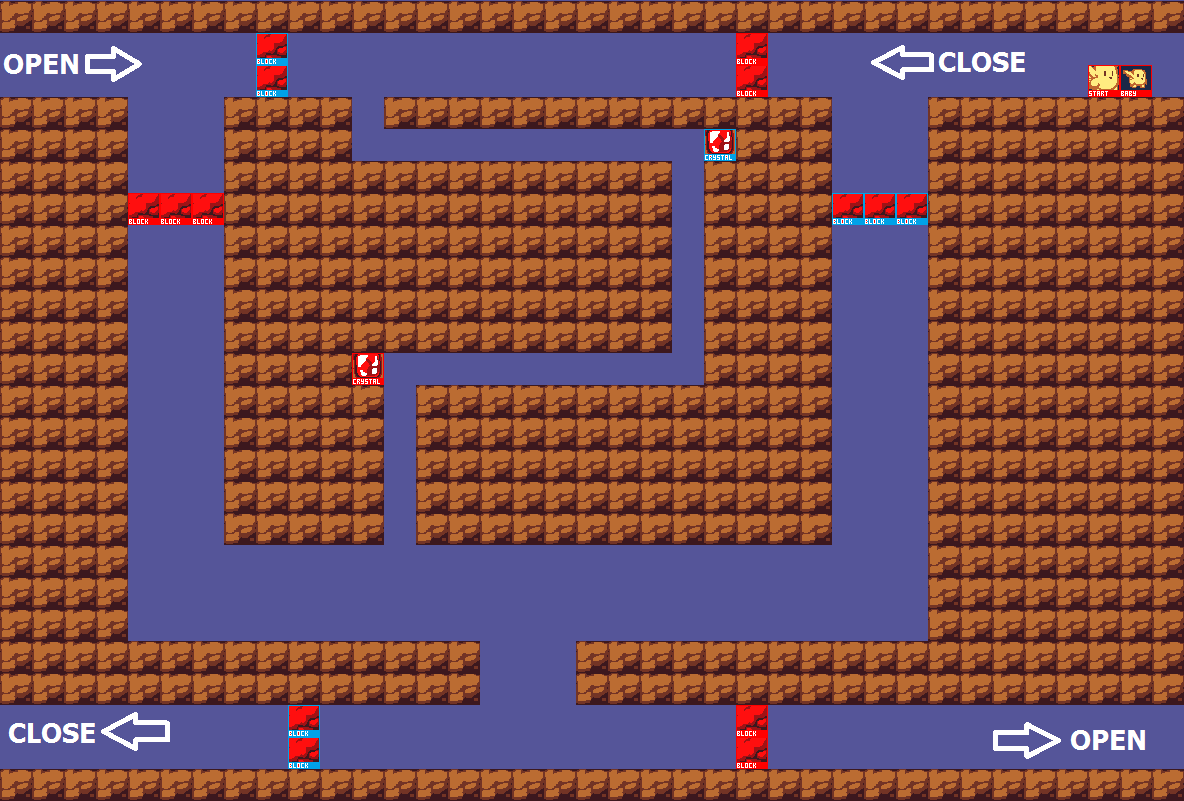}
 }
 \hfill
 \caption{Offspring Fling's gadgets.}
\end{figure}

\paragraph*{Door gadget.}
Figure~\ref{fig:of_doortraverse} and \ref{fig:of_dooropen} show the door gadget. The first thing to note is that Red and Blue blocks start in different states. If the red blocks are active, the blue blocks are not, and vice-versa. Although this property may not be maintained \textit{while} the player is opening or closing the door (while it is in the Open/Close section of the gadget), it is always restored once the player leaves gadget with the offspring.
Because the ``Traverse'' path uses red blocks, the Door's state is Open whenever the red blocks are inactive and the blue blocks are active. It is closed when the red blocks are active and the blue blocks are inactive. The avatar always enters at the top of the gadget and leaves through the bottom.

\begin{lemma}\label{lemma:ofdoor} The gadget shown in Figure~\ref{fig:of_dooropen} implements the Open and Close mechanisms to the door in Figure~\ref{fig:of_doortraverse}.\end{lemma}
\noindent {\it Proof:} We'll look at the possible cases separately:
\begin{itemize}
    \item Consider that the player enters through the ``Open'' path and the Door is closed. This means the red blocks are active and the blue blocks are inactive. As a result, the player is forced to go through the blue blocks at the top and traverse the middle section. To pass the narrow corridors, the offspring must be thrown twice, toggling the switch each time. Each switch changes both colored blocks' states, so the Door is indeed open once the player reaches the bottom. The blue blocks are now active and the red blocks inactive, so the player has to leave through the bottom right.
    \item Now consider that the player enters through the ``Open'' path but the door is already open. The player then has to go down by falling through the left chasm. Again, the player leaves through the bottom right. We can conclude the ``Open'' path always leaves the door open when traversed.
    \item The gadget is symmetric, so the ``Close'' path is similarly correct, simply by changing sides and colors.
\end{itemize}
\qed

\paragraph*{Crossover gadget.}
One interesting aspect of this proof is the fact that the Door open/close gadget (Figure~\ref{fig:of_dooropen}) also works as a crossover, if we remove or ignore its traverse gadget.

\begin{lemma} The gadget shown in Figure~\ref{fig:of_dooropen} implements a crossover gadget. \end{lemma}
\noindent {\it Proof:} We refer to the proof of Lemma \ref{lemma:ofdoor} for the details. Entering from the top left forces the player to leave through the bottom right; conversely, entering from the top right forces the player to leave through the bottom left, as described for the door gadget. The result is two paths crossing, with the player never being able to change from one to the other, which is the requirement of a crossover gadget. \qed

\begin{theorem} Offspring Fling is \textbf{PSPACE}-Hard. \end{theorem}
\noindent {\it Proof:} Following framework \ref{framework:pspace}, we show that TQBF $\le _p$ Offspring Fling by replacing the doors with Door-Traverse gadget, Open door to the Close path of the door gadget and Open door to the Open path of the door gadget. Any crossing paths are replaced with a Crossover gadget. The player will only be able to reach the end of the levels if the original quantified formula is true, proving Offspring Fling \textbf{PSPACE}-Hard, as shown in Theorem \ref{theorem:pspace}. \qed

\section{Super Meat Boy}\label{proof:supermeatboy}
Super Meat Boy is a 2D platformer game with a focus on speed and dexterity rather than puzzle solving.

Despite this focus, it features several of the key elements sufficient to prove \textbf{NP}-Hardness. Although a perhaps simpler proof could be obtained using 3-CNFSAT's framework, we use this proof to illustrate the Hamiltonian Cycle's framework.

\begin{figure}[!h]
\begin{center}
 \subfloat[Deadly saws.\label{fig:supermeatboy_saws}]{%
   \includegraphics[scale=0.75]{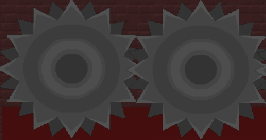}
 }
 \subfloat[Breakable tiles.\label{fig:supermeatboy_breakable}]{%
   \includegraphics[scale=1]{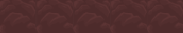}
 }
 \\
 \subfloat[Lock and Key.\label{fig:supermeatboy_key}]{%
   \includegraphics[scale=0.9]{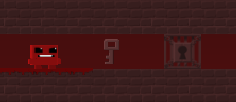}
 }
 \subfloat[Lava.\label{fig:supermeatboy_lava}]{%
   \includegraphics[scale=0.9]{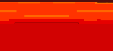}
 }
 \end{center}
 \caption{Super Meat Boy mechanics.}
 \label{fig:supermeatboy_mechanics}
\end{figure}

\paragraph*{Game Rules.}
The game has a player-controlled avatar, Meat Boy, that must traverse a level from a starting to a finishing location. The avatar is affected by gravity but can also jump on walls, climbing arbitrarily high walls.
Touching lava or saws will kill the avatar, forcing the player to restart the level.
The key elements for the proof, illustrated in Figure~\ref{fig:supermeatboy_mechanics}, are: 
\begin{enumerate}
    \item breakable tiles (tiles that disappear permanently a short time after the avatar has touched them);
    \item lava and saws (both killing meat boy if touched);
    \item keys that open locks.
\end{enumerate}

\begin{figure}[!h]
\center
 \subfloat[Single-use path.\label{fig:supermeatboy_singleuse}]{%
   \includegraphics[scale=0.5]{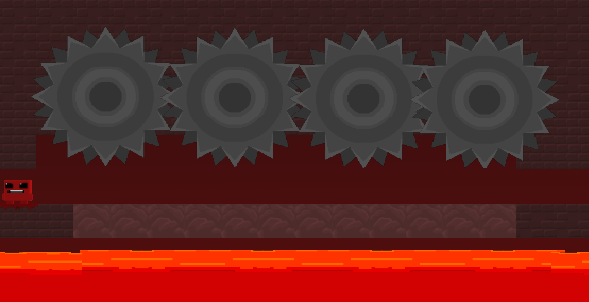}
 }
 \subfloat[Vertex.\label{fig:supermeatboy_vertex}]{%
   \includegraphics[scale=0.5]{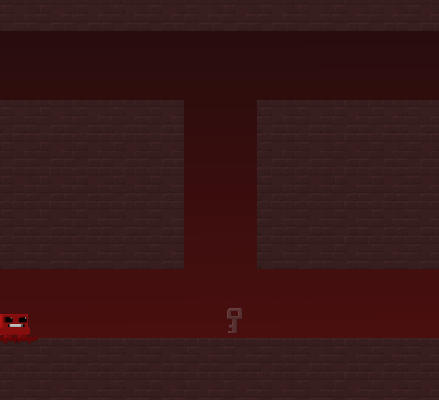}
 }
 \\
 \begin{center}
 \subfloat[Finish.\label{fig:supermeatboy_finish}]{%
   \includegraphics[scale=0.6]{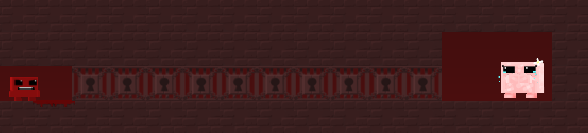}
 }
 \end{center}
 \caption{Super Meat Boy gadgets.}
 \label{fig:supermeatboy_gadgets}
\end{figure}

Following Metatheorem 1 from~\cite{viglietta}, we construct gadgets for single-use paths and location traversal to reduce from undirected, 3-regular Hamiltonian Path. Figure~\ref{fig:supermeatboy_gadgets} illustrates the gadgets required for the proof. Location traversal is trivially implemented using scattered keys that open the locks placed at the end of the level.

\begin{lemma} The gadget in Figure~\ref{fig:supermeatboy_singleuse} implements a single-use path.\end{lemma}
\noindent {\it Proof:} The gadget can be symmetrically traversed from the left or from the right. The blocks at the bottom are breakable, so they will disappear soon after being walked on. One traversal of the gadget is guaranteed to break every block since jumping would cause Meat Boy to touch the saws and die, restarting the level. Once the first traversal is complete, the floor disappears and it is impossible to traverse it again, conforming to the single-use requirement. \qed

\begin{lemma} The gadget in Figure~\ref{fig:supermeatboy_vertex} implements the vertex with location traversal. \end{lemma}
\noindent {\it Proof:} The only requirements for vertices, because of the 3-regularity of the graph, is that they must be visited once. This is implemented with the use of a key in each vertex, and a sequence of locks at the end of the level, as shown in Figure~\ref{fig:supermeatboy_finish}. Meat Boy will only reach the exit if he manages to reach every vertex, unlocking every lock. \qed

Although the figure shows a degree-4 vertex, it is easy to block one of the paths with a wall. Recall that Meat Boy can wall-jump to reach the top of the gadget from the bottom. This gadget connects with the Single-use paths that implement edges.

\begin{theorem}SuperMeatBoy is \textbf{NP}-Hard.\end{theorem}
\noindent {\it Proof:} SuperMeatBoy $\le _p$ HamiltonianCycle in a undirected, planar, 3-regular graph, by replacing each vertex with a the gadget of Figure~\ref{fig:supermeatboy_vertex} and edges with the single-use paths of Figure~\ref{fig:supermeatboy_singleuse}.
These are all the gadgets required to reduce from a Hamiltonian Cycle on a 3-regular, undirected graph (see Section~\ref{framework:hamilton}) to Super Meat Boy, thus proving the game \textbf{NP}-Hard. A sequence of vertices that forms a valid solution to the Hamiltonian Cycle in a graph has a one-to-one correspondence with sequence of vertex gadgets in the corresponding Super Meat Boy level, so solving the graph problem gives us a solution for the corresponding Super Meat Boy level and vice-versa. \qed

\paragraph*{Example instance.} An example with a 3-regular planar graph and the respective gadget configuration is shown in Figure~\ref{fig:supermeatboy_instance}, page~\pageref{fig:supermeatboy_instance}, of the Appendix.

\paragraph*{\textbf{PSPACE}-Hardness.} We conjecture that Super Meat Boy is \textbf{PSPACE}-Hard, using game elements not present in this proof. The required elements are: 
\begin{itemize}
    \item Burnt Boys, enemies that chase Meat Boy once it gets within their range;
    \item Buttons that temporarily deactivate some blocks;
\end{itemize}
The reduction would be similar in spirit to Back to Bed's (Section~\ref{proof:backtobed}), where an enemy is used to block one of two paths (either the Traverse or Close path of the Door gadget). The state transition would be implemented with the buttons, letting the enemy chase Meat Boy to change the door's state. We leave the details as an open problem.

\section{Triangular Akari (CircuitSAT)}\label{proof:akaritri}
Here we present an additional extension to the Akari proof, using a triangular instead of hexagonal grid.

\begin{figure}[!h]
 \center
 \includegraphics[scale=0.5]{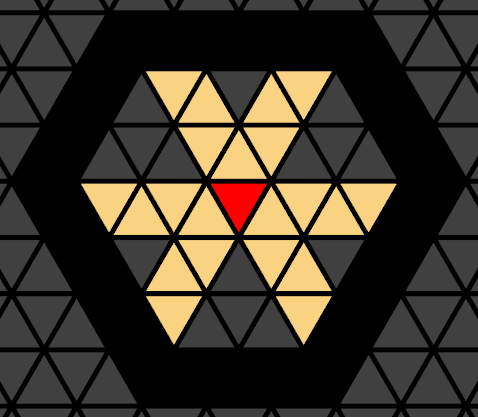}
 \caption{Light propagation for triangular grids.}
 \label{fig:akaritri_lightprop}
\end{figure}

Similarly to the hexagonal grid, light propagates in three directions, as seen in Figure~\ref{fig:akaritri_lightprop}. Lamp adjacency is only considered as \textit{edge adjacency} and not vertex adjacency. We maintain the convention of lit (unlit) corridors as a representation of true (false) signals oriented from input to output.

\begin{figure}[!h]
\center
 \subfloat[Variable selection and Wire.\label{fig:akaritri_wire}]{%
   \includegraphics[scale=0.4]{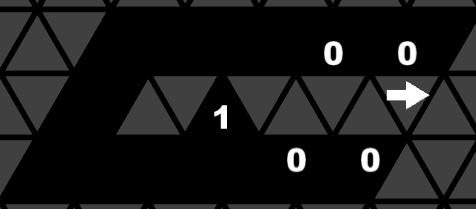}
 }
 \subfloat[Turn.\label{fig:akaritri_turn}]{%
   \includegraphics[scale=0.4]{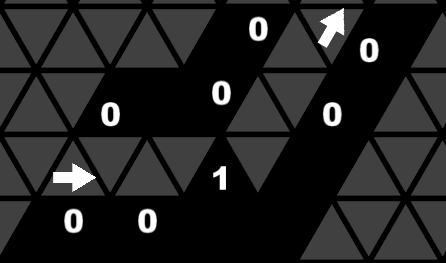}
 }
 \subfloat[Double turn.\label{fig:akaritri_doubleturn}]{%
   \includegraphics[scale=0.4]{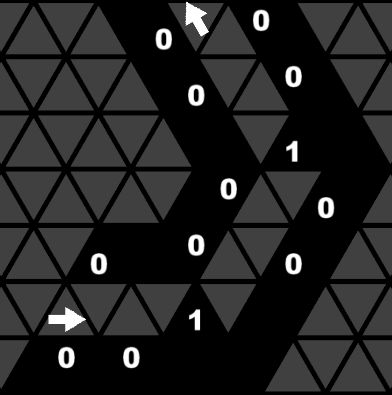}
 }
 \\
 \subfloat[FAN-OUT gadget.\label{fig:akaritri_fanout}]{%
   \includegraphics[scale=0.5]{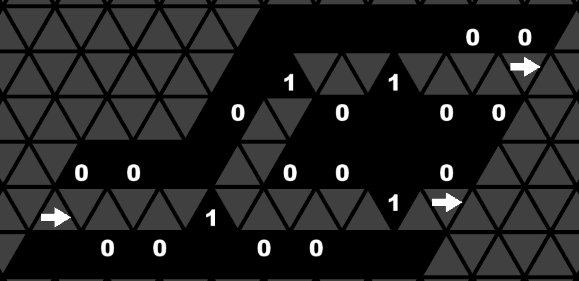}
 }
 \subfloat[$\textit{NOT}$ gadget.\label{fig:akaritri_not}]{%
   \includegraphics[scale=0.5]{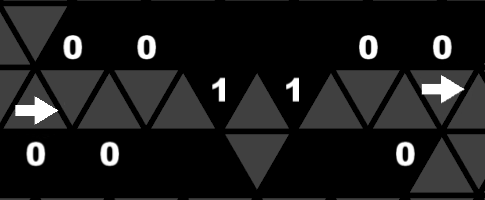}
 }
 \\
 \subfloat[$\textit{OR}$ gadget.\label{fig:akaritri_or}]{%
   \includegraphics[scale=0.6]{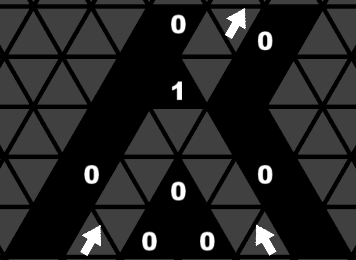}
 }
 \caption{Triangular Akari's gadgets.}
 \label{fig:akaritri_gadgets}
\end{figure}

\paragraph*{Gadgets.} Figure~\ref{fig:akaritri_gadgets} shows the gadgets used in the proof. They are essentially the same as in the Hexagonal Akari (Section~\ref{proof:akarihex} proof. Gray cells outside the gadgets are not part of the puzzle/reduction and can be assumed to be obstacles.

\paragraph*{Wires and Turns.} The basic wire follows exactly the same principles as the wires in the two other Akari proofs. The signal is determined at the extremities of the wire with an empty corridor (outer cells filled with zeroes) of arbitrary length. Here, only sixty degrees' turns exist, as shown in Figure~\ref{fig:akaritri_turn}. Two sixty degrees' turns can easily be composed into a one hundred and twenty degrees' turn, as shown in Figure~\ref{fig:akaritri_doubleturn}.

\paragraph*{$\textit{NOT}$ gadget.} This gadget can be introduced into a wire seamlessly, as shown in Figure~\ref{fig:akaritri_not}. The signal is inverted by the two 1-cells. The gadget is symmetric, but we will consider the that it is oriented from left to right, below.
\begin{lemma} The gadget shown in Figure~\ref{fig:akaritri_not} implements a $\textit{NOT}$ gate, inverting the signal.\end{lemma}
\noindent {\it Proof:}
If the signal is true, there is no lamp to the left of the left-most 1-cell. As a result, we must place a lamp between the two 1-cells, and none to the right of the right-most 1-cell. Because the output corridor is not illuminated, the signal is false.

If a signal is false, then a lamp must be placed to the left of the left-most 1-cell to illuminate the input corridor. This means that the second lamp in the center must be placed on the cell bellow, and the third one adjacent to the right-most 1-cell. The output corridor is illuminated, so the signal is true.

Both input signals are inverted, resulting in a $\textit{NOT}$ gate. \qed

\paragraph*{FAN-OUT gadget.} This gadget is built as a merge of a simple wire and a turn, as seen in Figure~\ref{fig:akaritri_fanout}. One of the signals will keep the direction of the wire, while the other is turned. A second turn can be added to the turned signal to turn the signal back into its original direction. The propagation works in the same manner as normal wires.

\paragraph*{$\textit{OR}$ gadget.} To finish the proof, we have an $\textit{OR}$ gate, as shown in Figure~\ref{fig:akaritri_or}. The inputs enter at the bottom of the gadget and the output is propagated at the top. The 1-cell only has two empty adjacent cells. The important idea in this gadget is to see that any true input will prevent a lamp to be placed below the 1-cell. Similarly, a lamp placed below the 1-cell illuminates both corridors.

\begin{lemma}The gadget shown in Figure~\ref{fig:akaritri_or} implements an $\textit{OR}$ gate, having a false output only when both inputs are false.\end{lemma}

\noindent {\it Proof:}
The first thing to note is that if \textit{any} of the inputs are true (lit from below), their light propagates all that way to the empty cell below the 1-cell. This immediately forces the lamp to be placed above the 1-cell, propagating a true signal. If both inputs are true, the bottom part of the gadget is fully illuminated; if only one is true, there is a single valid lamp placement to illuminate the false input's corridor.

Finally, if none of the signals are true, then a single lamp must be placed below the 1-cell to illuminate both input corridors. This prevents a lamp from being placed above the 1-cell. As a result, the signal from two false inputs will also be false. \qed

\begin{theorem}Triangular Akari is \textbf{NP}-Complete.\end{theorem}
\noindent {\it Proof:} We show that CircuitSAT $\le _P$ Triangular Akari by referring to Section~\ref{framework:circuit}, page~\pageref{framework:circuit}. We start by converting every boolean gate into a composition of the gates from the functionally complete set $\{OR,NOT\}$. We can then replace every WIRE (straight or turned), FAN-OUT with the respective gadgets from Figure~\ref{fig:akaritri_gadgets}. This reduces any CircuitSAT instance to a Triangular Akari puzzle, proving in \textbf{NP}-Hard.
We show that Triangular Akari $\in \textbf{NP}$ due to the existence of a polynomial time verifier. A solution is given by each lamp's position. The number of lamps is limited by the board's size, so it is linear. Light propagation is also limited by the board's size. To verify each cell's constraints, we only need to keep a track of a bit that states whether each cell is illuminated or not, and then checking the adjacency constraints, which can be done in constant time (always 3 neighbours).

Being \textbf{NP}-Hard and in \textbf{NP}, Triangular Akari is \textbf{NP}-Complete. \qed

\paragraph*{Example.} To conclude, and making sure every gadget fits together, we present a solvable example in Figure~\ref{fig:triakari_instance}, in the Appendix A, Section~\ref{example:akaritri}, page~\pageref{example:akaritri}.

\section{HexCells (CircuitSAT)}\label{proof:hexcells}
HexCells is a puzzle game based on Minesweeper (whose consistency hardness proof we use as a reference\cite{minesweeper}), but played on a hexagonal grid. It also has some additional rules which we can use to simplify the proof.

In Minesweeper, the game board is given by a set of cells in a square grid. These cells may be have a bomb, or numbered. Numbered cells tell the player exactly how many adjacent bombs to it there are. Additionally, each cell may be revealed or unrevealed. The objective of the game is to reveal every cell without ever trying to reveal a bomb. Each time a cell is revealed, it will become numbered.

In HexCells, the rules are similar, with the following modifications: 

\begin{enumerate}
    \item the cells are hexagonal; 
    \item clear cells may not reveal any information about the number of adjacent bombs;
    \item the number of bombs is fixed and known to the player. 
\end{enumerate}
There are extra rules, but since the ones above are enough to prove \textbf{NP}-Hardness, we will not address them further.

Following Kayes' proof~\cite{minesweeper}, we prove that HexCells' \textit{Consistency} is \textbf{NP}-Hard (Minesweeper itself being co-\textbf{NP}-Hard~\cite{minesweeperconp} where they consider a different, stronger, decision problem). The decision problem is whether a game board with a subset of revealed cells is consistent - that is, if there is \textit{any} placement of bombs that satisfies all of the revealed cells. This mimics how some people play the game, tentatively considering one possible move and committing only if the move doesn't lead to an inconsistent board - a sure sign of a bomb.

It is interesting to note that our proof was a direct consequence of Hexiom's proof (see Section~\ref{proof:hexiom}), as Hexiom can be seen as HexCells' Consistency where bombs are also numbered, an additional constraint. This numbering of the bombs makes the proof more difficult, so we present this one first. Both use the CircuitSAT framework (see Section~\ref{framework:circuit}, page~\pageref{framework:circuit}).

\begin{figure}[!h]
\center
 \subfloat[Variable selection and Wire.\label{fig:hexcells_wire}]{%
   \includegraphics[scale=0.35]{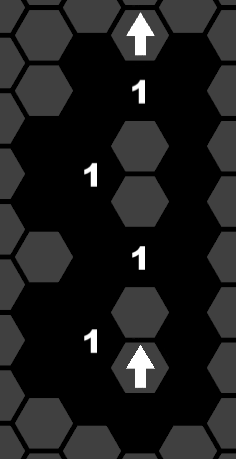}
 }
 \subfloat[Turn.\label{fig:hexcells_turn}]{%
   \includegraphics[scale=0.37]{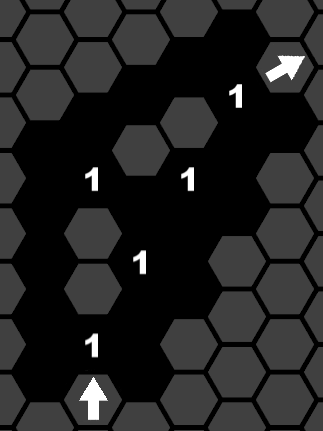}
 }
 \subfloat[Negated FAN-OUT gadget.\label{fig:hexcells_fanout}]{%
   \includegraphics[scale=0.36]{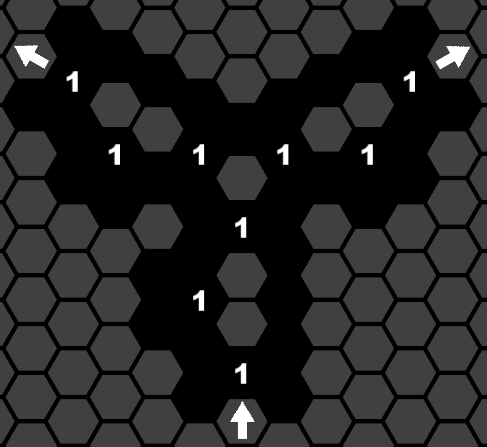}
 }
 \\
 \subfloat[$\textit{NOT}$ gadget.\label{fig:hexcells_not}]{%
   \includegraphics[scale=0.825]{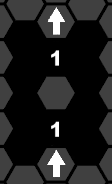}
 }
 \subfloat[N$\textit{OR}$ gadget.\label{fig:hexcells_nand}]{%
      \includegraphics[scale=0.5]{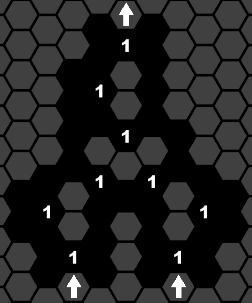}
     }
 \caption{HexCells gadgets.}
 \label{fig:hexcells_gadgets}
\end{figure}

\paragraph*{Gadgets.} The gadgets for this proof are shown in Figure~\ref{fig:hexcells_gadgets}. The gate used is a $\textit{NOR}$ gate which is functionally complete by itself, making a $\textit{NOT}$ gate seemingly redundant. However, the wires have a fixed length of three cells, which could create a parity problem when fitting things together; the $\textit{NOT}$ gate, with a length of 2 cells, solves this problem. It also leads to smaller reductions (since we don't need to create an equivalent $\textit{NOT}$ gate with a composition of $\textit{NOR}$ gates).

Gadgets may be consistent with a variable number of bombs. We assume each adds the maximum number of bombs that make it consistent to the number of bombs on the solution. The excess bombs can easily be placed at the end, in an isolated part of the board completely surrounded by cells that reveal no information.

\paragraph*{WIRE and Variable selection.} The wire, together with the variable selector at the bottom, is shown in Figure~\ref{fig:hexcells_wire}. Figure~\ref{fig:hexcells_turn} shows how to create turns. 
Each wire has two valid configurations: it must have a single bomb, either at the top or at the bottom. By convention, we use a bomb at the bottom to represent a true signal and a bomb at the top to represent a false signal.

\begin{lemma}The gadget shown in Figure~\ref{fig:hexcells_wire} successfully propagates a true/false signal.\end{lemma}

\begin{figure}[!h]
 \center
 \includegraphics[scale=0.4]{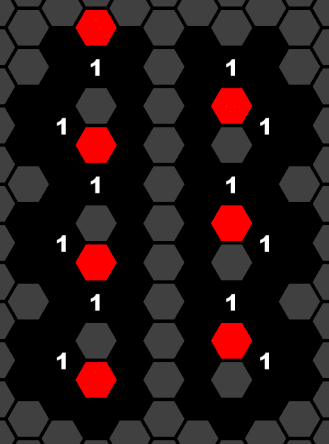}
 \caption{True (left) and False (right) configurations of the WIRE shown in Figure~\ref{fig:hexcells_wire}.}
 \label{fig:hexcells_wire_example}
\end{figure}

\noindent {\it Proof:} Consider signal propagation from bottom to top. The only two consistent configurations are shown in Figure~\ref{fig:hexcells_wire_example}, as the reader can easily verify. The 1-cell on the side, in combination with the 1-cell on the center, force each WIRE segment to have a bomb, and the relative position of the bomb is always forced to be the same. This is all we need to carry a signal to any gadget, as we guarantee the presence of a bomb in one of two positions. \qed

\paragraph*{$\textit{NOT}$ gate.} The $\textit{NOT}$ gate is shown in Figure~\ref{fig:hexcells_not}. This gadget will also be used as a parity-shift gadget, making sure that wires can be made arbitrarily long to fit with every gadget.

\begin{lemma} The gadget shown in Figure~\ref{fig:hexcells_not}, inserted in a wire, will invert the input signal.\end{lemma}

\noindent {\it Proof:} The gadget has two valid configurations: a single bomb between the two 1-cells, or one bomb at each extremity. We can see that this gadget inverts the position of the bombs relative to the 1-cell when introduced, since its two 1-cells have a bomb on the opposite side. This inverts the signal, implementing a $\textit{NOT}$ gate. \qed

\paragraph*{Negated FAN-OUT gate.} Figure~\ref{fig:hexcells_fanout} show how to split the signal. The propagation works similarly to the two previous gadgets, propagating throughout 1-cells. The output is the negation of its input, so an additional $\textit{NOT}$ gate before the gadget (or two after it) is required to make a proper FAN-OUT. The outputs can be easily directed up, which leads to more compact instances, as shown in the example reduction (see Figure~\ref{fig:hexcells_instance}, page~\pageref{fig:hexcells_instance}).

\paragraph*{N$\textit{OR}$ gate.} This gadget is illustrated in Figure~\ref{fig:hexcells_nand}. The figure shows the gate with wire segments on its inputs and output.

\begin{lemma}The gadget shown in Figure~\ref{fig:hexcells_nand} implements a $\textit{NOR}$ gate, having a false output if and only if both of its inputs are true.\end{lemma}

\noindent {\it Proof:} Consider only the triangular arrangement of 1-cells and obstacles (ignoring the wire segments) and propagation from bottom to top. The top-most 1-cell determines the signal's value, as in the FAN-OUT gadget. Because both of the inputs are adjacent to this cell, it can only have a bomb above (signalling true) if both of the inputs are false (have bombs below their respective 1-cells), which is the first property of a $\textit{NOR}$ gate. To conclude, we only need to make sure that the two other configurations are valid. If only one input signal is true, we can easily place the bomb on one of the extremities. If both are true, we place the bomb in the middle cell, as it will satisfy all three 1-cells. The gadget outputs true when both inputs are false, and false otherwise. This satisfies the properties of a $\textit{NOR}$ gate, completing the bulk of this reduction. \qed

\paragraph*{Length Parity.} It is important to note that the wire shown in Figure~\ref{fig:hexcells_wire} has an length multiple of 3 (one 1-cell, followed by two unrevealed cells). The gates also have a fixed length (see Figures \ref{fig:hexcells_fanout} and \ref{fig:hexcells_nand}). This could lead to a parity problem where gadgets wouldn't connect as expected to each other, but one that can be solved by using two consecutive $\textit{NOT}$ gates to align any two wires.

\begin{lemma}WIRE and $\textit{NOT}$ gadgets can be used to align wires arbitrarily.\end{lemma}

\noindent {\it Proof:} First, note that WIRE gadget have length 3 and $\textit{NOT}$ gadgets have length 2. Then, note that using two $\textit{NOT}$ gadgets, we can propagate the signal unchanged (due to double negation). This has length 4.

\begin{figure}[!h]
 \center
 \includegraphics[scale=0.5]{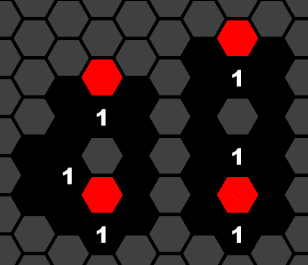}
 \caption{Shifting wires' parity (true configuration).}
 \label{fig:hexcells_wire_align}
\end{figure}

Now, we can shift the alignment of wires by one unit propagating one signal with a WIRE segment (length 3) and the other with two $\textit{NOT}$ gadgets (length 4). Their alignment shifts by 1 unit, as illustrated in Figure~\ref{fig:hexcells_wire_align} for a true signal. Repeating this process we can change the alignment to an arbitrary number, which allows us to connect gadgets freely. \qed

\paragraph*{Fixed number of bombs.} Another problem that requires attention is the fact that the number of bombs is constant, but the gadgets' number of bombs isn't (for example, the FAN-OUT gate may have one bomb at the center, or none at all). To complete the puzzle's solution, we add an empty row of unrevealed cells to accommodate the number of bombs that aren't placed in the circuit.

\begin{theorem} HexCells' Consistency is \textbf{NP}-Complete.\end{theorem}

\noindent {\it Proof:} using CircuitSAT's framework (Section~\ref{framework:circuit}, page~\pageref{framework:circuit}), we can replace every circuit element with one of the gadgets of Figure~\ref{fig:hexcells_gadgets}. If any of the gates doesn't have a gadget, we can easily compose it as a function of $\textit{NOR}$ gates (Section~\ref{approach:circuitsat}), thus showing that CircuitSAT $\le _P$ HexCells' Consistency, thus proving it is \textbf{NP}-Hard.

We show that HexCells $\in \textbf{NP}$ by the existence of a polynomial time verification algorithm. Given a board and every piece’s position, we iterate through every piece,  checking  the  local  constraint~\footnote{The full HexCells game (rather than the restricted version considered here) has additional rules that we do not include in this analysis. Their evaluation still takes polynomial time, since they're all bounded by the size of the board. We leave the details as an open problem.}.  Each  piece  only  needs  to  evaluate  its  six neighbours which is polynomial. Because HexCells is \textbf{NP}-Hard and belongs to \textbf{NP}, HexCells is \textbf{NP}-Complete. \qed

\paragraph*{Example.} To conclude, and making sure every gadget fits together, we present a solvable example in Figure~\ref{fig:hexcells_instance}, in the Appendix A, Section~\ref{example:hexcells}, page~\pageref{example:hexcells}.

\section{Hexiom (CircuitSAT)}\label{proof:hexiom}
Hexiom is a free puzzle game released on the website Kongregate in 2007. The game can be played for free at \url{https://www.kongregate.com/games/Moonkey/hexiom}. Although no complexity results existed, to our knowledge, it has received some independent study in the context of SAT solvers~\cite{hexiomsolve,hexiomsat}. The game was encoded in SAT, relying on SAT solvers to find the solutions for the levels.

\begin{figure}[!h]
\center
 \subfloat[Empty puzzle.\label{fig:hexiom_example_empty}]{%
   \includegraphics[scale=0.5]{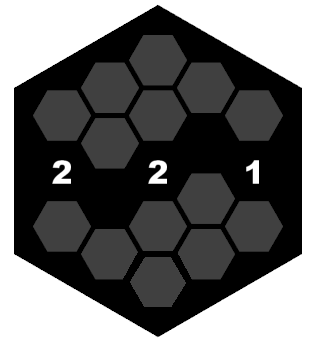}
 }
 \subfloat[Complete puzzle\label{fig:hexiom_example_filled}]{%
   \includegraphics[scale=0.5]{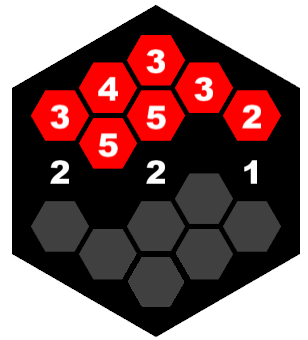}
 }
 \caption{Hexiom puzzle example.}
 \label{fig:hexiom_example}
\end{figure}

\paragraph*{Game Rules.}
The game consists of two groups of hexagonal pieces: on group is the board, which has fixed pieces (numbered or not not) on the hexagonal grid; the other group is the player's ``hand'', which consists of exclusively numbered cells that must be placed on the empty cells of the board. Numbered pieces are locally constrained such that a piece numbered with $n$ has exactly $n$ \textit{numbered} cells adjacent to it. An example of puzzle is shown in Figure~\ref{fig:hexiom_example}. Black hexagons represent fixed pieces that can't be moved; gray hexagons are the empty board cells; red hexagons are the pieces from the player's hand. The red pieces start randomly placed on the board.

\paragraph*{Proof overview.} We point the reader to the proof of HexCells (Section~\ref{proof:hexcells}). In fact, the fixed component of the gadgets is the same for both games, and so is their valid configurations. The main difference is that in Hexiom, instead of placing bombs, we are placing numbered cells; the main construction is kept from HexCells', but additional care must be taken to place the excess pieces. We will describe each gadget by its valid configurations, and in the end we will describe how to place the excess n-cells to finish the proof. 

\begin{figure}[!h]
\center
 \subfloat[\{3\} or \{1,2\} gadget.\label{fig:hexiom_3_1}]{%
   \includegraphics[scale=0.6]{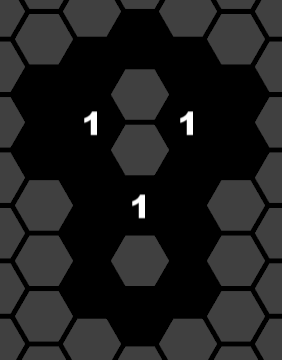}
 }
 \subfloat[\{2,1,1\} or \{1,1\} gadget.\label{fig:hexiom_2}]{%
      \includegraphics[scale=0.52]{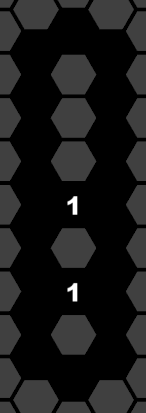}
     }
 \caption{Gadgets to solve Hexiom's excess cells.}
 \label{fig:hexiom_excess}
\end{figure}

This proof has three distinct parts:
\begin{enumerate}
    \item The first part of the proof is composed by the main gadgets, which form the conversion from a circuit as all the preceding CircuitSAT proofs. The main gadgets are shown in Figure~\ref{fig:hexcells_gadgets}.

    \item The second part of the proof will reduce the excess piece problem to a simpler parity problem, by converting every 3 and 2-cell into a 1-cell. The excess correction gadgets are shown in Figure~\ref{fig:hexiom_excess}.
    
    \item The third part solves the parity problem by showing a global property of every valid solution.
\end{enumerate}

We will consider each puzzle to have two sets of pieces, set $F$ that contains every fixed piece (pieces that can't be moved) and set $M$ that contains every mobile piece (pieces that the player must place on the board to solve the puzzle). Each gadget adds pieces to both sets, when used. In Figure~\ref{fig:hexiom_example}, set $F$ is represented by the numbered black cells and set $M$ is represented by the numbered red cells.

\paragraph*{Wire and Variable selection.} The wire and variable selection are shown in Figure~\ref{fig:hexcells_wire}. The two bottom empty cells represent the variable selection. The gadget is extended indefinitely as shown in the figure to form a wire. 

\paragraph*{Signals.} The value of the signal is given by the position of the movable cell relative to the center 1-cell. Placement above the 1-cell represents true, while placement below the 1-cell represents false. 

\begin{lemma}The gadget shown in Figure~\ref{fig:hexcells_wire} carries its input signal forward, like a wire.\end{lemma}
\noindent {\it Proof:} Because of the 1-cell on the side of the wire, one movable cell must be placed in one of the two empty spaces (and only one). Because of the center 1-cells, the movable 1-cell will be placed in the same relative position throughout the whole wire. Because the signal is determined by this relative position, and it is conserved, the signal is correctly propagated. \qed

Some observations must be made regarding the types of cells placed. In the variable selection (bottom portion of the wire), the bottom-most empty space has a single adjacent numbered cell. To select a true value for the variable, then, a 1-cell must be placed. The empty space above it, however, has two adjacent numbered cells. To select a false value for the variable, then, a 2-cell must be placed. Note that this doesn't happen in the wire \textit{extensions}, where a 2-cell is placed to propagate the signal regardless of the signal.

As such, variable selection gadgets, when used in the construction, add one 1-cell and one 2-cell to the set $M$. Wire extensions add only a single 2-cell to $M$.
The former will always have one excess piece (either a 1-cell or a 2-cell), with which we deal after we describe the circuit gadgets. The latter has no excess piece.

\paragraph*{Turns.} Turns can be seen as wire extensions, only propagating the signal in a different direction (see Figure~\ref{fig:hexcells_turn}). Like the wire extension, turns add one 2-cell to set $M$ and has no excess.

\paragraph*{$\textit{NOT}$ gadget.} The $\textit{NOT}$ gadget, shown in Figure~\ref{fig:hexcells_not}, is introduced before/after any wire extension to negate the signal. Aside from being used to negate a signal, it can also be used to shift the wire's position. Because the wires' length is always a multiple of $3$, alignment problems could arise to connect the different gadgets, solved by the composition of two $\textit{NOT}$ gates, equivalent to a wire segment of length $4$, disrupting the modulo $3$ length of the wires.

\begin{lemma}The gadget shown in Figure~\ref{fig:hexcells_not} behaves like a $\textit{NOT}$ gate, reversing its input signal.\end{lemma}
\noindent {\it Proof:} If the input signal is true (there is a numbered cell below the bottom-most 1-cell), the output signal will be false (there will be another cell above the top-most 1-cell). If the input signal is true, there will be a 2-cell between the two 1-cells, and so no cell above the top-most 1-cell, so the output signal is false. \qed 

The gadget will then require a single 2-cell (placed at the center), or no cell at all (since the two 2-cells can be considered part of the wire segments), depending on the incoming signal. The possible excess 2-cell will be treated after the $\textit{NOR}$ gate is described. Each $\textit{NOT}$ gate adds one 2-cell to set $M$.

\begin{lemma}Two $\textit{NOT}$ gates can be combined into a wire segment of length $4$.\end{lemma}
\noindent {\it Proof:} The signal passing through the two $\textit{NOT}$ gadgets will be negated twice, becoming the original signal again. Each $\textit{NOT}$ gate has a length of $2$, so the overall length is $4$. \qed

This can be used to shift the wires by a single cell, because $4 - 3 = 1$. If the wires are misaligned, we can extend one of them with a single wire segment, by 3, and the other with two $\textit{NOT}$ gates, by 4. This will bring them a single unit closer to alignment. Repeating the process will align them completely.

\paragraph*{Negated FAN-OUT gadget.} The negated FAN-OUT gadget is just the triangular configuration of 1-cells (see Figure~\ref{fig:hexcells_fanout}. The remainder of the gadget simply illustrates how it connects to wire extensions. 
\begin{lemma} The gadget in Figure~\ref{fig:hexcells_fanout} behaves like negated a FAN-OUT gate, having two output wires carrying the opposite signal of its input wire.
\end{lemma}
\noindent {\it Proof:} The gadget has two valid configurations - either a 3-cell in the center, or no cell at all. If there is no cell, the 1-cell constraints are satisfied in the wire extensions with 2-cells. \qed

A single 3-cell is added to set $M$. When the input is true, the 3-cell is in excess (can't be placed in the gadget); this will be taken care of after describing the $\textit{NOR}$ gate.

\paragraph*{N$\textit{OR}$ gadget.} The $\textit{NOR}$ gadget (Figure~\ref{fig:hexcells_nand}) is similar to the negated FAN-OUT, in the sense the only a small portion of the gadget, composed of a triangular set of 1-cells is required to make the gadget. The two extra empty cells suffice to create the $\textit{NOR}$ behaviour, and the remainder is a straightforward connection to the wire extensions. 
\begin{lemma} The gadget shown in Figure~\ref{fig:hexcells_nand} behaves like a $\textit{NOR}$ gate, having an output of true when both inputs are false and false otherwise.
\end{lemma}
\noindent {\it Proof:}
There are four valid configurations: a 3-cell in the center when both inputs are true, a 2-cell on the side of the false input when only one input is true (symmetrically for the two inputs counts as two distinct configurations), or no cell at all in the center portion when both inputs are false. \qed

The gadget adds one 3-cell and one 2-cell to set $M$ to cover all four cases. Only one of the pieces is used for a given configuration, the other being in excess. We now proceed to the second part of the proof, explaining how to deal with all the excess cells.

\paragraph*{Excess and Parity.} In HexCells, the bombs that were placed were all identical and unconstrained, so extra empty space sufficed to place additional bombs. This is not the case in Hexiom, since the excess cells have constraints themselves, requiring additional machinery.
We approach this problem by first reducing the excess pieces' problem to a \textit{parity} problem (a single 1-cell in excess), and then show that it can easily be solved.

It is important to note that this reduction uses only 1-cells, 2-cells, and 3-cells (aside from obstacles), so those are the only cells we need to take care of. We start by showing how to take care of the 2-cell problem by incorporating them into another gadget if required. Then, we show how to replace a 3-cell with one 1-cell and another 2-cell. These two excess gadgets are shown in Figure~\ref{fig:hexiom_excess}. Finally, we solve the parity problem itself.

\begin{lemma}
The 2-or-none gadget shown in Figure~\ref{fig:hexiom_2} can be used to consume an excess 2-cell.
\end{lemma}
\noindent {\it Proof:} The gadget can be satisfied by placing a 1-cell below the bottom 1-cell and another 1-cell above the top 1-cell. Alternative, a single 2-cell can be placed between the two 1-cells, and two movable 1-cells are placed adjacent to each other at the top of the gadget. \qed

Two 1-cells are added to the set $M$. We add one of these gadget for each $\textit{NOT}$ gate, $\textit{NOR}$ gate, and 3-or-1 gadget described below. 

\begin{lemma}
The 3-or-1 gadget shown in Figure~\ref{fig:hexiom_3_1} can be used to replace one 3-cell with one 1-cell and one 2-cell.
\end{lemma}
\noindent {\it Proof:} A 2-cell may only be placed on the top-most free cell of the gadget, adjacent to two 1-cells. Then, the bottom cell must be satisfied by placing another 1-cell at the bottom of the gadget. Alternatively, a single 3-cell may be placed in the center of the gadget, satisfying the three adjacent 1-cells.

To make use of this gadget, we add a 1-cell and a 2-cell to $M$. We add one gadget for each $\textit{NOR}$ and FAN-OUT gadgets. In this way, given the previous gadget, we can replace a 3-cell with a 1-cell whenever we get a surplus 3-cell either of the two gates. Because this gadget may also have an excess 2-cell, we add a copy of the previous gadget (Figure~\ref{fig:hexiom_2}) to consume the excess 2-cell. \qed

\paragraph*{Solving parity.} So far, we have managed to place every 3-cell and 2-cell into valid configuration in the board, and are left with some number of excess 1-cells. The first thing to note, then, is that two 1-cells placed adjacent to each other form a stable configuration. As a result, we can take care of every excess 1-cell \textit{pair} given enough free space. For this, we can add, for example, $2\times n$, with $n$ being the number of cells of the construction. Having taken care of every pair, all we're left with is a parity problem - in the end, are we left with a single extra 1-cell?

\begin{figure}[!h]
 \center
 \includegraphics[scale=0.3]{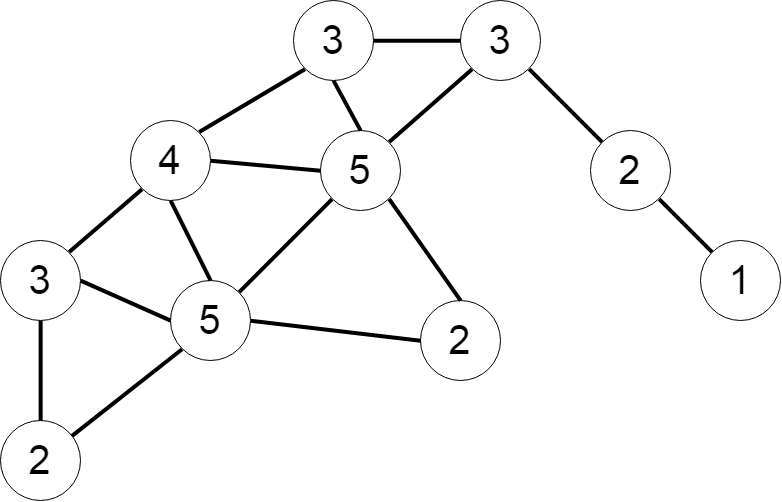}
 \caption{Graph of example in Figure~\ref{fig:hexiom_example}.}
 \label{fig:hexiom_graph}
\end{figure}

Here we show that there is a global parity property \textit{independent of the variable assignment}, for valid configurations, letting us solve the parity problem. Our argument rests on the following observation:
\begin{lemma}\textit{Every connected valid Hexiom configuration is an instance of a graph $G$ (with a set of vertices $V$ and a set of edges $E$) where the number of each cell corresponds to the degree of its respective vertex.}
\end{lemma}
\noindent {\it Proof:} This observation is illustrated in Figure~\ref{fig:hexiom_graph} for the previously shown example in Figure~\ref{fig:hexiom_example}. Each numbered cell becomes a vertex, and every adjacent numbered cell shares an edge. In a valid Hexiom configuration, each n-cell is satisfied if and only if it has exactly n adjacent cells, leading to n adjacent vertices on the graph, which is its \textit{degree}. This is relevant because we know that: $$\sum_{v \in{V}}^{} \deg(v) = 2 \times \left\vert{E}\right\vert$$ Consider starting with a graph with isolated vertices, all of them with degree $0$, and then adding each edge one by one. Each edge added will increase the degree of its two vertices by $1$, thus increasing the overall sum of degree by $2$. From this, we can conclude that the sum of every cell's number, in valid configuration, must be even. \qed

\begin{corollary}Every valid Hexiom configuration is an instance of a graph where the number of each cell corresponds to the degree of its respective vertex.
\end{corollary}
This follows simply because a valid configuration consists of the sum of valid connected configurations. The overall sum of the cells' numbers will be a sum of even numbers, which is also an even number.

The whole construction must then have an even sum of degree.
We then have a global property for the puzzle that, for some $k \in{\mathbb{N}}$: $$\sum_{c \in{F}\cup{M}}^{} num(c) = 2 \times k$$

Each gadget, as described, has a constant sum of the numbered cells in $F \cup M$. This means the the final construction, too, will have a constant sum, as all the elements of $M$ must be placed. This solves our parity problem, because we can simply sum the degree of every gadget and check whether the result is even or odd. If the number is odd, we know that we will always end up with an extra 1-cell once we find an otherwise valid configuration (up to the placement of the last 1-cell), and so we can add one additional 1-cell to offset this parity. Conversely, if it sums to even, we will always end up with an even number of 1-cells, so no additional 1-cell is needed. 

The following table shows the balance of each gadget in terms of fixed and free pieces to make the final accounting easier. In the end, the overall sum must be even.
\begin{center}
\begin{tabular}{ |c|c|c|c| }
\hline
\label{table:hexiom_balance}
 Gadget & Fixed & Free & Total  \\ [0.5ex] 
 \hline \hline
 SELECT & \{1,1\} & \{1,2\} & 5 \\ 
 WIRE & \{1,1\} & \{2\} & 4 \\
 $\textit{NOT}$ & \{1\} & \{2\} & 3 \\ 
 FAN-OUT & \{1,1\} & \{3\} & 5 \\
 $\textit{NOR}$ & \{1\} & \{2, 3\} & 6 \\
 2-EXCESS & \{1,1\} & \{1,1\} & 4 \\
 3-or-\{2,1\} & \{1,1,1\} & \{2,1\} & 6 \\
 \hline
\end{tabular}
\end{center}

\begin{theorem}
    Hexiom is \textbf{NP}-Complete.
\end{theorem} 
\noindent {\it Proof:} We show that CircuitSAT $\le _P$ Hexiom by replacing every circuit wire and gate by its respective gadget. Each gadget will also add its corresponding excess gadgets to the board. Every circuit solution will have at least one Hexiom puzzle solution and every Hexiom puzzle solution has exactly one puzzle solution. The solution from the puzzle to the circuit is taken from the variable assignments which can be extracted in linear time in the number of variables, achieving a polynomial-time reduction. As a result, Hexiom is \textbf{NP}-Hard.

We show that Hexiom $\in$ \textbf{NP} by the existence of a polynomial time verification algorithm. Given a board and every piece's position, we iterate through every piece, checking the local counting constraint. Each piece only needs to evaluate its six neighbours, for a total complexity of $O(|F| \cup |M| \times 6)$, which is polynomial in the size of the original circuit.

Because Hexiom is \textbf{NP}-Hard and belongs to \textbf{NP}, Hexiom is \textbf{NP}-Complete. \qed

\paragraph*{Example.} To conclude, and making sure every gadget fits together, we present a solvable example in Figure~\ref{fig:hexiom_instance}, in the Appendix A, Section~\ref{example:hexiom}, page~\pageref{example:hexiom}.

\section{Back To Bed (TQBF)}\label{proof:backtobed}
Back to Bed is an M.C. Escher and Salvador Dalí inspired puzzle game released in 2014 for computers and smart phones.
The game's objective is to indirectly control a sleep-walker character back to bed by placing obstacles in its way and bridging gaps while avoiding patrolling dogs. The players controls their own character, placing the obstacles down while being constrained by the same platforms as the sleep-walker.

We will show the game to be \textbf{PSPACE}-Hard using the same framework as Offspring Fling's (see Sections~\ref{framework:pspace} and~\ref{proof:offspringfling}).

\begin{figure}[!h]
\center
 \subfloat[Fish bridge.\label{fig:backtobed_bridge}]{%
   \includegraphics[scale=0.85]{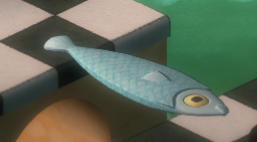}
 }
 \subfloat[Teleport Mirror.\label{fig:backtobed_teleport}]{%
   \includegraphics[scale=0.7]{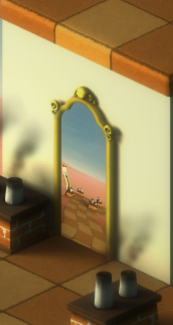}
 }
 \subfloat[Dog.\label{fig:backtobed_dog}]{%
   \includegraphics[scale=0.7]{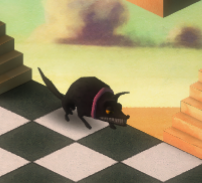}
 }\\
 \subfloat[Player character.\label{fig:backtobed_player}]{%
   \includegraphics[scale=1.1]{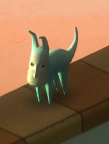}
 }
 \subfloat[Sleep-walker character.\label{fig:backtobed_npc}]{%
   \includegraphics[scale=1]{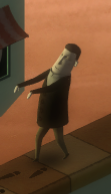}
 }
 \subfloat[Apple.\label{fig:backtobed_apple}]{%
   \includegraphics[scale=1.2]{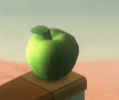}
 }
 \caption{Back to Bed mechanics.}
 \label{fig:backtobed_mechanics}
\end{figure}

\paragraph*{Game rules.}
The game elements required for the proof, shown in Figure~\ref{fig:backtobed_mechanics}, are the following:
\begin{itemize}
    \item $4\times 1$ fishes that act as bridges (Figure~\ref{fig:backtobed_bridge});
    \item Magic mirrors that link to arbitrarily far locations (Figure~\ref{fig:backtobed_teleport});
    \item Dogs that always turn clock-wise when colliding with an obstacles or a gap (Figure~\ref{fig:backtobed_dog});
    \item $1\times 1$ apples that makes the character turn clock-wise when colliding (Figure~\ref{fig:backtobed_apple});
    \item Player avatar that can carry a fish or apple (Figure~\ref{fig:backtobed_player});
    \item Avatar that must be directed to the exit (Figure~\ref{fig:backtobed_npc});
    
\end{itemize}

The player controls a different character that can carry one object at a time (an apple or a fish) and place it on clear tiles. If the sleep-walker (sleep-walker) falls off the edge or collides with a patrolling dog, the level is reset.

The player character never collides with the sleep-walker or with the patrolling dogs, so it can walk freely. However, it cannot fly, the bridge is required to traverse the level. Bridges can be placed as long as its two endpoints are on solid ground and can be picked up by the player from either end.

Note that teleporting mirrors solve any crossover problem that could arise, so we only need to show how to construct a Door that can be opened and closed (closing needs to be forced, opening does not).

\paragraph*{Reduction overview.}
The reduction works from any TQBF formula to a Back to Bed level. Existentially quantified variables give the player a choice of value, while universally quantified variables must verify both values.

The proof relies on the existence of a single fish-bridge on the whole level and the path being punctuated with gaps. This forces the player to escort the sleep-walker closely, preventing the player from changing the state of the doors independently of the sleep-walker (which would lead to inconsistencies between the level and the formula).

Each variable (both existential and universal) gadget also contains a single apple. This apple \textit{must not be carried to other gadgets} to guarantee consistency. The clause gadget works on a similar principle, to direct the player towards the correct door out of the three in each clause.

Doors are implemented with patrolling dogs. These dogs have two possible paths, in which they're locked. Changing the door's state corresponds to changing the dog's path. One of the paths blocks the ``Traverse'' path of the Door, while the other blocks the ``Close'' path of the Door.

The use of small segments of solid path punctuated by gaps is our mechanism to force the player to escort the sleep-walker and make it impossible for it to go back and change the level's state independently of the sleep-walker. The gaps force the player to be within a fixed range of the sleep-walker, who otherwise falls in one of the gaps, forcing a level restart.

\begin{figure}[!h]
\center
 \subfloat[Existential gadget.\label{fig:backtobed_existential}]{%
   \includegraphics[scale=0.4]{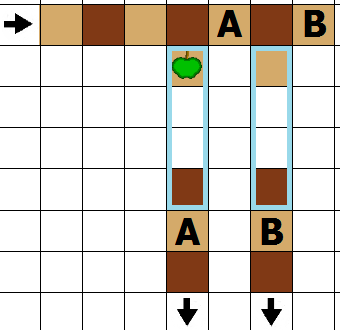}
 }\hfill
 \subfloat[Clause gadget.\label{fig:backtobed_clause}]{%
   \includegraphics[scale=0.4]{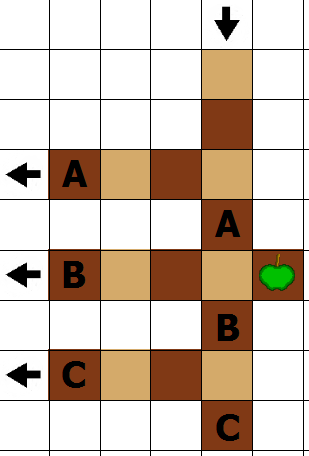}
 }\\
 \subfloat[Path-Segment gadget.\label{fig:backtobed_pathseg}]{%
   \includegraphics[scale=0.4]{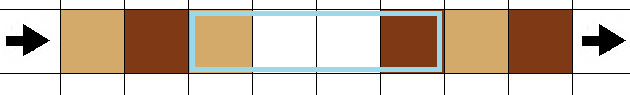}
 }\\
 \subfloat[Extending Path-Segment gadgets.\label{fig:backtobed_pathseg2}]{%
   \includegraphics[scale=0.4]{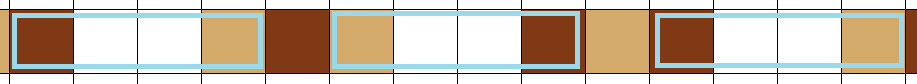}
 }\\
 \subfloat[Door gadget.\label{fig:backtobed_door}]{%
   \includegraphics[scale=0.5]{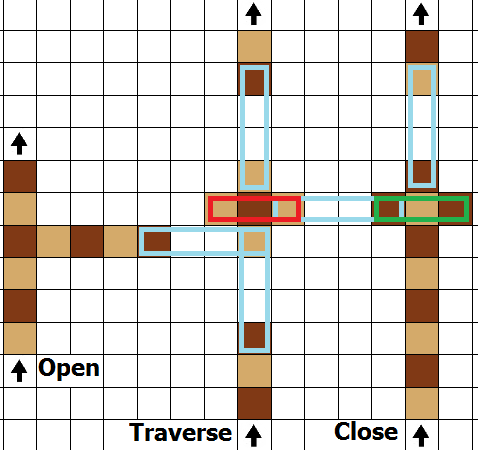}
 }
 \caption{Back to Bed gadgets.}
 \label{fig:backtobed_gadgets}
\end{figure}

\paragraph*{Gadgets.} Figure~\ref{fig:backtobed_gadgets} shows the gadgets used in this proof. We refer back to Section~\ref{framework:pspace}, page~\pageref{framework:pspace} to show how they fit together to form a full reduction.

White cells represent gaps/falls, black cells represent walls/obstacles, and colored cells represent solid ground. The grid lines are merely to ease understanding the gadget in terms of traversal time and where bridges can be placed. The arrows show to path's orientation. It can be expected that a teleporting mirror is placed at every gadget's entry and exit to simplify connectivity.

Light blue rectangles show where the bridges can relevantly be placed. Recall that there is only a single bridge in each level, so only one of the rectangle may be active at any given time; re-positioning a bridge doesn't happen instantly and requires the player avatar to traverse between the two starting point for a bridge to be re-positioned.

\paragraph*{Existential Quantifier (EQ) gadget.} As previously stated, an apple is required to make a choice in the existential gadget. When the sleep-walker collides with an apple, it will turn clock-wise. 

\begin{lemma}The gadget shown in Figure~\ref{fig:backtobed_existential} implements an existential quantifier gadget, letting the player choose the value (true/false) for the current variable.\end{lemma}

\noindent {\it Proof:} The gadget has only two possible exit paths. The apple is placed only of the upper part of the gadget. Placing the apple on A or B will turn the sleep-walker to the corresponding path. We can arbitrarily choose A to be True and B to be False.
If the apple is placed anywhere else in the sleep-walker's path, it will turn clockwise into a gap, falling of the edge and restarting the level; if the apple is not placed at all, it will similarly fall off the right most edge. This fulfills the criteria for an existential gadget. \qed

To prevent the apple from being carried to other gadgets, we use the Path-Segment gadget described next, which is impossible to traverse while carrying an apple.

\paragraph*{Path-Segment gadget.} This gadget is shown in Figure~\ref{fig:backtobed_pathseg}. Note that it can be arbitrarily extended by having alternating a $3\times 1$ solid path and a $2\times 1$ gap, as shown in Figure~\ref{fig:backtobed_pathseg2}. The gadget is used in two ways: firstly, to prevent apples from being taken from one gadget to another; secondly, to prevent the player character from straying far from the sleep-walker and changing the state of the doors in unintended sections of the level.

\begin{lemma} The gadget shown in Figure~\ref{fig:backtobed_pathseg} prevents apples from being carried while the sleep-walker is traversing it.\end{lemma}

\noindent {\it Proof:} Fish-bridges must have one solid square on each extremity; each must also be clear. Bridges must also be clear to be picked back up, meaning that they can't be picked while the sleep-walker is traversing them. Because the Path-Segment has only a length of 3, the bridge can only be picked up and placed down while the sleep-walker is walking on the middle square, letting the player change the bridge's position from one end to the other. This leaves no clear square on which to place the apple, so it can't be carried forward while the sleep-walker is traversing the gadget.

It can't be propagated by keeping it in the Path-Segment behind the NPC either, because the bridge must always move forward to keep up with the sleep-walker before it reaches the third square of the path-segment. As a result, the apple will always be left behind.

Finally, we only need to note that the apple must be used in every traversal of the existential gadget, otherwise the sleep-walker will walk into a gap. This ensures that the apple can't be carried in small amounts in each traversal either, as each traversal effectively resets its position back to the gadget. \qed

\begin{lemma}The path-segment gadget prevents the player from straying too far from changing the state of any gadget without the sleep-walker.\end{lemma}

\noindent {\it Proof:} Here, we only need a timing argument. Each path-segment requires $X$ time for the player to traverse, with $X$ being some multiple of the time that the sleep-walker takes to traverse a single square. To isolated one gadget from the next, we only need to check the maximum number of squares that the sleep-walker will traverse, $S$, and add $S/X$ path-segments between each. Because each gadget has a constant size, this doesn't affect the polynomial size of the reduction. \qed

\paragraph*{Connectivity.} Although the TQBF framework requires crossovers, in general, to connect all the paths, the existence of teleporting mirrors solves any connectivity problems for this reduction. We can assume that there is a mirror at the entry and exit of each gadget's path.

\paragraph*{Door gadget.} This is the most important gadget in this reduction, as it is the feature that makes the game \textbf{PSPACE}-Hard, instead of merely \textbf{NP}-Hard or even polynomial. It is shown in Figure~\ref{fig:backtobed_door}.

The door's state is given by the dog's patrolling area. As stated before, dogs always walk forward, until they encounter an obstacle or gap, at which point they turn 90 degrees clock-wise. In our reduction, each door has one dog, which may be on the green or red paths. The green path represents an Open door (as it leaves the Traverse path unobstructed); a red path represents a Closed door (as its Traverse path is obstructed).

Because no apples may be carried into this gadget, the only way to interact with the dog is through bridges. A bridge between the red and green paths will allow the player to change the door's state by making the dog cross the bridges, switching from a red to a green patrolling path, or vice-versa.

\begin{lemma}The gadget in Figure~\ref{fig:backtobed_door} implements a Door gadget, which can be opened and closed; the traverse path can only be traversed when the door is opened.\end{lemma}

\noindent {\it Proof:} There are three paths in the gadget. Open, Close and Traverse. In the \textit{Open path}, the player may \textit{choose} to change the door's state. The sleep-walker will simply traverse in a straight line, bottom to top. This gives the player time to take the horizontal path and use the bridge to change the dog's state.

In the \textit{Close path}, traversal \textit{always} leaves the door \textit{closed}. For the sleep-walker to walk through the middle section, the dog cannot be patrolling the green area, or it would collide with the dog, resulting in a level restart. This forces the player to close the door (if it is open, with the dog patrolling the green path) by bridging the gap until the dog crosses it and patrols the red path. Note that the continuous path gives the player enough time to change the dog's state (using the bridge), if needed.

The \textit{Traverse path} does not change the door's state, but requires it to be open (dog patrolling the green zone) to be traversed. The gap in the path prevents the player from using the bridge to change the door's state, since it is needed to cross the gap. \qed

To complete this gadget, we only need to make sure that the door's state can't be changed after traversal, as it would make the state inconsistent with the framework.

\begin{lemma} A door's state can't be changed immediately after being "closed" or "traversed". \end{lemma} 

\noindent {\it Proof:} The two paths end in path-segments (\ref{fig:backtobed_pathseg}), which forces the player to escort the sleep-walker closely. Even ignoring the time it takes for the player to move and place the bridges, changing a door's state requires the dog to cross a bridge (which takes at least $4 \times T_d$ time, where $T_d$ is the time the dog requires to traverse a single square), which is longer than the time it takes for the player to traverse a single square. 

From this, we can conclude that a door's state can't be changed after the traversing the Close or Traverse paths. \qed

\paragraph*{Clause gadget.} Because a clause may be satisfied by any of its three variables (as we're working with 3CNF), the player must be able to direct the NPC to one specific door. This is easily done with the use of one apple, as shown in Figure~\ref{fig:backtobed_clause}. Also note that it can't be traversed backwards.

\begin{lemma}The gadget in Figure~\ref{fig:backtobed_clause} implements a clause gadget.\end{lemma}
\noindent {\it Proof:} Each of the paths (A, B, C) leads to a "traverse" path of a door, each corresponding to its respective variable (see Section~\ref{framework:pspace}, page~\pageref{framework:pspace}). Coming from the top, the player wants to direct the sleep-walker to an open variable. To do this, he must placed the apple in either of the A,B,C positions on the vertical path. Placing it anywhere else, the sleep-walker will be forced onto a gap, resetting the level. This ensures the gadget can't be traversed backwards.

The same arguments used for the existential quantifier applies here, with the introduction of path-segment gadgets to prevent the apple from being carried from one gadget to another. \qed

\begin{theorem}Back to Bed is \textbf{PSPACE}-Hard.\end{theorem}
\noindent {\it Proof:} Using the framework described in Section~\ref{framework:pspace}, page~\pageref{framework:pspace} and the gadgets in Figure~\ref{fig:backtobed_gadgets}, we show that TQBF $\le _P$ Back to Bed, thus proving the game to be \textbf{PSPACE}-Hard. The level can be finished if and only if the original quantified boolean formula is true. Finishing the level is proof of the formula's truth, giving the solution to the decision problem.\qed

\paragraph*{Example.} To show how every gadget fits together in a full construction, an example is given in Appendix A, Section~\ref{example:backtobed}.

\cleardoublepage
%
\fancychapter{Final Remarks}
\cleardoublepage
\label{chapter:finalremarks}
In this thesis, we've studied the relation between complexity classes and puzzle games. Although other kinds of games have also received study in relation to complexity, such as parametrized complexity, two player games and even economic games, we finish the thesis by mentioning other paths that have not, to our knowledge, been studied. 

We start by showing a specific game, Bloxorz, whose main mechanic is ``easy'' (solvable in polynomial time), but still leads to compelling puzzles, although aided by elements that also make it \textbf{PSPACE}-Hard. Then, we briefly point to the concepts of guessing and randomness in games which may lead to interesting complexity results. Finally, we also extend the CNF-SAT framework used to support simple approximation problems.

\section{Bloxorz - a Puzzle Game based on an ``Easy'' Mechanic}
When studying the relation between games and complexity, it may be fruitful to look at examples where the interest of the game lies in "easy" mechanics, solvable in polynomial time.

Bloxorz was a popular puzzle game freely available online~\footnote{ \url{https://www.miniclip.com/games/bloxorz/en/}}. The game is essentially a top-down maze where the player wants to reach the Finish position with an avatar. The twist is that, instead of having an avatar like Mario who could run and jump, the avatar was a $2\times\!1$ parallelepiped that can only roll.

The interest in the game lies on the fact that moving the $2\times\!1$ block does not obey the laws we're used to with characters in most games. Instead, moving it left \textit{rolls} it to the left. This means that the position involves also the state of the avatar, depending of whether it is standing still or laying down. The state of the avatar is illustrated in Figure~\ref{fig:bloxorz_asymmetry}.

The game also features breakable tiles and buttons that can toggle the state of other blocks. These are very similar to the elements used in our own NP and PSPACE proofs, and play a similar role in the complexity of Bloxorz. They are, however, secondary to the movement and the general appeal of the game. The challenge is not so much in which doors to open, but in rolling into the correct positions.

\subsection{Movement Complexity}
\begin{figure}[!ht]
 \center
 \includegraphics[scale=0.3]{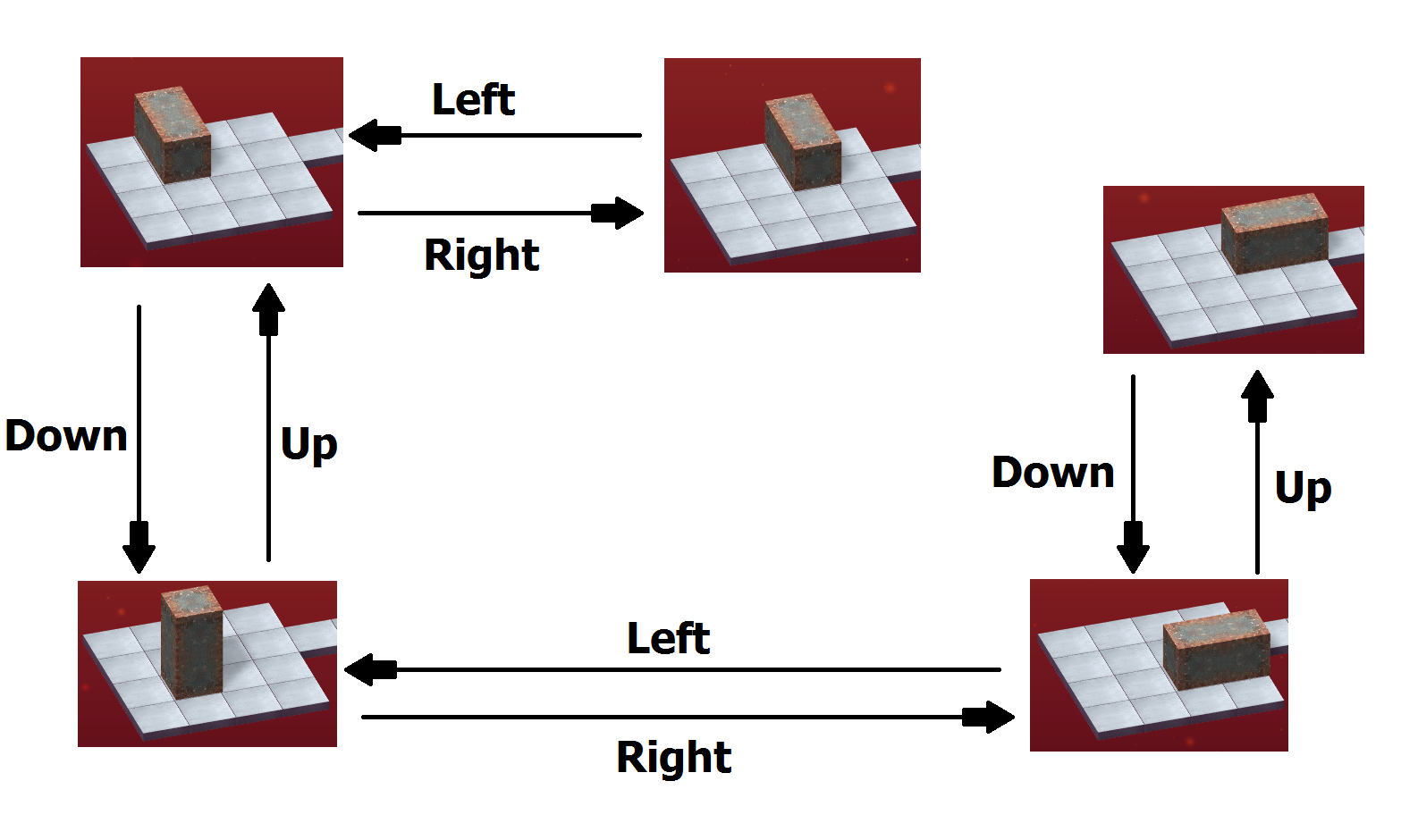}
 \caption{Movement in Bloxorz.}
 \label{fig:bloxorz_asymmetry}
\end{figure}

The feature that makes Bloxorz interesting is its unorthodox movement system. The fact that the character is $2\times\!1$ creates an asymmetry in its movement (see Figure~\ref{fig:bloxorz_asymmetry}), which often conflicts with the greedy approach of always moving towards the goal.
If the block is ``standing'', any movement will knock it down, lying with the same orientation of the movement. If the block is lying, moving it perpendicularly to its orientation simply rolls it, while moving it along its orientation will stand it back up.

Here we show that, despite initially unintuitive, the movement is easily seen to be in $P$, an ``easy'' problem in the context of this thesis. First, we must translate a board instance of Bloxorz to a Bloxorz graph.

\begin{definition}A Bloxorz graph is a graph with set of vertices $V$, where each vertex $v$ has the board position and the state of the avatar, and a set of edges $E$, with an edge $e$ between two vertices, $v_1$, $v_2$, annotated with the action that moves the avatar from vertex $v_1$ to $v_2$.\end{definition}

When the block is lying (vertically or horizontally), there are two units in contact with the ground. We conventionalize its position to be the position of its south-western-most unit. This will make every position unique with no overlap in the Bloxorz graph. 
Any movement that leads to the block falling off the level connects to the starting position's vertex.

\begin{lemma} The translation between a Bloxorz board and a Bloxorz graph only requires a polynomial blowup in size.\end{lemma}
\noindent {\it Proof:} Note that, given a board with size $|B|$ ($|B|$ cells), the graph has exactly $3\times{|B|}$ vertices. This is because the avatar still has a single position in the board, but can be in one of three states. 
At each vertex, the player only has four possible actions: move up, move down, move left, move right. As a result, the number of edges is at most $4\times{3\times{|B|}}$. The whole reduction is then polynomial. \qed

\begin{theorem} Finding a path from $A$ to $B$ in a Bloxorz level has a polynomial time algorithm. \end{theorem}
\noindent {\it Proof:} Given two arbitrary positions in a Bloxorz level, we can find the path between the two (or prove that one doesn't exist) in polynomial path by using a simple Breadth First Search (or Dijkstra's algorithm with uniform weights) on a Bloxorz graph. 
The whole algorithm takes $4\times{3\times{|B|}}$ to translate from a level instance to a graph, and then $O(V+E)$ to find the path. We can get the sequence of movements by keeping the edge annotations. As a result, finding a path between $A$ and $B$ takes polynomial time. \qed

\subsection{General Complexity}
Bloxorz was proved \textbf{PSPACE}-Complete~\cite{bloxorz} in a paper by Tom C. van der Zanden and Hans L. Bodlaender in 2014. The paper is freely available on Arxiv.

We will not detail the proof here. It uses the NCL framework, but, perhaps more importantly, it uses buttons that can toggle bridges, a similar mechanism to the one used in our Offspring Fling proof (Section~\ref{proof:offspringfling}, page~\pageref{proof:offspringfling}). 

\subsection{Conclusion}
We have now looked at Bloxorz from two distinct viewpoints. The first uses the game's main mechanic, which is polynomial and has no dependence whatsoever on buttons. The second features the generic use of buttons, mostly independently of the game's unique movement system, that suffices to prove the \textbf{PSPACE}-hardness.

This is merely to suggest the fact that complexity is not the \textit{only} factor to consider when analysing not only games in general, but also puzzle games. Some games are enjoyable to play for simply confronting players with unfamiliar and unintuitive situations, even if that isn't what we consider a computationally hard problem. Trying to parse those situations and figure out a good method to solve them may be sufficient to make a good game, though perhaps of limited longevity.

\section{Guessing}
Guessing in an important aspect of all of the games that are provably hard. Having no sub-exponential algorithm to solve any of these problems means that, ultimately, solving one puzzle in each of these games relies on a systematic search of all possible solutions, or guessing.

However, in most cases, the idea of solving a good puzzle is not that of trying every possibility, but of learning certain patterns to reduce the search space significantly. 

Further along this line of reasoning, is that of propagation and no guessing. The HexCells series (see Section~\ref{proof:hexcells}, page~\pageref{proof:hexcells}) has become popular due to its level generation which never requires guessing. At every step, there is at least one move that can be propagated from the previous ones. 

Similarly, but not quite the same, guessing can be related to having a unique solution. This has been studied in relation to Minesweeper (the game on which HexCells is based, only on a different grid). Although we've proved that HexCells Consistency is \textbf{NP}-Hard, solving the game is more likely to be co-\textbf{NP}-Complete, just as Minesweeper is~\cite{minesweeperconp}. This complexity is related to arriving at a \textit{single and unique} solution when starting on an already consistent board.

We then have two ways of looking at guessing:
\begin{enumerate}
    \item each move leads to the next with a single induction step (propagation);
    \item regardless of the number of induction steps, there is ultimately a single valid move.
\end{enumerate}

\subsection{Guessing and Propagation}
This kind of guessing results in mostly easy, polynomially solvable games. This is because it relies on forced moves. If each move induces a succeeding move, the puzzles' completion move in a single direction until they're solved. In the games we've studied in \textbf{NP}, we know that the number of moves is bounded by a polynomial. If we bound the number of induction steps required for a single move, induction takes polynomial time, and thus the whole game becomes solvable in polynomial time as well.

\paragraph*{Constraint Satisfaction Problems (CSP) and Domain Propagation.} One way to find each induced move (which may not be immediately apparent) is to use domain propagation, popular in CSP solvers. In a game like Sudoku, for example, each empty square starts with domain $\{1,\ldots,9\}$. Each move then fixes one value to an empty square, and removes it from every other square in the same row, column and 3x3 square. No guessing, in this context, means that, after completing the propagation step, there is one empty square with a single value left in its domain somewhere on the board. We fix this value and repeat the propagation process. A similar approach would work for Minesweeper and HexCells, with much simpler binary domains.

\subsection{Guessing and Co-NP}
This kind of guessing is more specific to games like Minesweeper and HexCells. It happens specifically when, due to hidden information, there are different consistent moves that could lead to a complete solution, but only one of them is an actual solution. This can inevitably prevent ``perfect play'', since choosing the wrong move may reveal a mine and the loss of the game, a well known problem of minesweeper.

The co-\textbf{NP}-Completeness of Minesweeper is studied in~\cite{minesweeperconp}. It relies on a decision problem different from consistency. Instead, the decision question is ``Is there any square that can be determined \textit{with certainty} to contain (or not) a bomb?''. This removes the guessing element from the game, since each move will be ``safe''.

\subsection{Conclusion} Although the second notion of guessing is more interesting from a complexity standpoint, it is more difficult to apply to games, as it relies on an uncommon characteristic of Minesweeper that relies on a fixed board with hidden information, unlike Akari or Hexiom. It could be interesting to find or even create more games that are co-\textbf{NP}-Hard instead of \textbf{NP}-Hard (assuming \textbf{NP}$\neq$co-NP).

\section{Puzzle games with Hidden Information - PSPACE and co-NP}
In the previous section, we've pointed at the relation between co-\textbf{NP}-Hardness and hidden information. This is not a sufficient condition, however, and is a topic that has not received as much study as it deserves given its prominence in many modern games.

\subsection{Games Against Nature}
Papadimitriou has addressed the hidden information problem in~\cite{gamesagainstnature}. In this paper, the problem analyzed is a game where one player plays randomly, while the other wants to maximize the probability of winning. The decision question is ``can the second player win with probability $p \leq 50\%?$''. This results in a new characterization of \textbf{PSPACE}, with the stochasticity lifting some problems from \textbf{NP}-Complete to \textbf{PSPACE}-Complete. With this new notion, he studies the complexity of Stochastic SAT and Dynamic Markov Processes, among others.

\subsection{Revealing Hidden Information.}
The complexity of some games with hidden information have also been studied by revealing the hidden information. One example is Tetris~\cite{tetris}, where the full sequence of pieces is known in advance (revealing the game's hidden information), making the game \textbf{NP}-Hard. Several solitaire card games have been studied in the same way, like Klondike~\cite{klondike}, Spider~\cite{spider} and Mahjong~\cite{mahjongnp}. One especially interesting case is that of Mahjong, that has also been studied without revealing its hidden information, resulting in a \textbf{PSPACE}-Complete proof~\cite{mahjongpspace}. 

\subsection{PSPACE or Co-NP?}
We can see that hidden information can lead to the classes PSPACE or co-NP. The most important aspect to distinguish the two might be in the kind of hidden information in the game. In the solitaire card games mentioned, the game can still be played when the information is revealed; they can even be \textbf{NP}-hard. In the case of Minesweeper or HexCells, however, that is not an option: the hidden information is the solution itself.

\subsection{Conclusion}
Although the foundational work of single player games with hidden information is done in~\cite{gamesagainstnature}, its application is not nearly as common as the techniques used in this thesis for deterministic games. Perhaps not as rich when the original study was done, this area has become more important in recent years with the rising popularity of  randomly generated levels and card games, making it a promising possibility for further study of the role of randomness in games and their respective complexity.

\section{MaxSAT \& Optimization}
Simple modifications of the 3-CNFSAT Framework can be done to allow reducing from optimization problems like MaxSAT (including partial, and sometimes weighted MaxSAT). 
We present them here.

\subsection{MaxSAT}
MaxSAT arises from overconstrained SAT problems, that is, boolean formulas that are unsatisfiable due to too many conflicting constraints. In these cases, however, it might still be useful to obtain an answer, even if it does not satisfy every clause. 
In the case of SAT, the approach is typically to satisfy the maximum number of clauses instead of all of them, the MaxSAT problem. There are two important variations: \textit{partial} and \textit{weighted} MaxSAT. These two approaches can also be combined.

\paragraph*{Partial MaxSAT} is a problem in which one subset of the clauses \textit{must} be satisfied (usually referred to as \textit{hard clauses}), as in SAT, and the other subset is subject to the maximization of MaxSAT (usually referred to as \textit{soft clauses}).

\paragraph*{Weighted MaxSAT} is like MaxSAT, but a weight is assigned to each clause so that the objective is not necessarily to maximize the \textit{number} of satisfied clauses, but the \textit{sum of their weights}.

The combination of the two approaches applies weights to soft clauses, while keeping the hard clauses unchanged.

\subsection{Application to Games} These optimization notions can be applied to games in two respects that we haven't dealt with earlier. The first is \textit{scoring}. Many games have collectibles that are not mandatory to finish a level, but that award the players with more points. Related to scoring is \textit{speed-running}, in which players try to finish the level as fast as possible. Both of these can serve as a basis for optimization in games, in which the notion of MaxSAT is applicable beyond simply SAT and \textbf{NP}-Hardness.

\paragraph*{Modification to 3-CNFSAT Framework.} To reduce MaxSAT to a game, using this framework, we only need to change the ``Clause'' gadget (see Figure~\ref{fig:cnf_fw}, page~\pageref{fig:cnf_fw}). Instead of having a ``lock'' mechanism preventing the player from traversing the clause at the end, we let the player traverse every clause, and apply the lock mechanism to gate a collectible that awards points. 
Upon reaching the end of the level, the question asked is ``Does the player have more than $X$ points?'', where each point corresponds to one clause satisfied. In the case of partial MaxSAT, we keep the old gadget for hard clauses (these clauses can't be traversed unless they're satisfied) and the new gadget for soft clauses. The case of weighted MaxSAT leads to problems in the form of collectibles, as we describe next.

\paragraph*{Exponential Weights.}
Because numbers are represented as a binary string (of size maintaining a $O(log(n))$ for number $n$), weights may be exponentially large. This makes the amount of collectibles to be collected exponential, preventing a polynomial-time reduction and even solution. The same applies with time taken instead of score, because the length of the levels would have to increase exponentially. The Super Mario Bros. (SMB) games are an interesting exception in scoring, because a sequence of enemies hit by a shell will increase the score exponentially (first enemy awards one point, the next two points, then four, etc.), making this kind of reduction possible. The scores of SMB do not follow a perfect exponential and do have a cap due to console limitations, but it would be a fair generalization to consider.

\paragraph*{Time.} If we intend to use time instead of score, we create a gadget that requires $X$ time to traverse in place of the clause (forcing the player to take a longer path). We can consider that satisfied clauses take $X$ time to traverse, while unsatisfied ones take $k\times X$ for some constant $k$. The total time taken to traverse the level (assuming some baseline performance) can be used to evaluate the number of clauses satisfied.

\subsection{Conclusion}
From this section we can see that, after proving a game \textbf{NP}-Hard using the 3-CNFSAT framework, scoring and time can then be taken into consideration to get optimization hardness results with little extra work.

\section{Conclusion}
In this thesis, we surveyed common reduction frameworks for NP and PSPACE Hardness. We have also applied each framework to previously unstudied games. For CircuitSAT, we extended the \textbf{NP}-Complete result of Akari in a square grid to triangular and hexagonal grids. These proofs alter the existing proofs. More substancially, we proved Hexiom to be \textbf{NP}-Complete, which did not yield directly to the conventional CircuitSAT techniques, requiring additional global reasoning to get a complete proof. A simpler proof for HexCells follows directly from Hexiom's.

Reducing 3-CNFSAT, we obtained a Hardness proof for Cut the Rope. Our contribution lies in the fact that Cut the Rope is a physics-based game, a type that has not been studied. One related result is Angry Birds' \textbf{NP}-Completeness~\cite{angrybirds}. This last proof manages to prove Completeness using the conservation of energy in the level - Cut the Rope does not feature such a property, however, so we leave its presence in NP as an open problem.

Reducing TQBF, we proved that Offspring Fling and Back to Bed are \textbf{PSPACE}-Hard. Offspring Fling is a conventional proof using toggle-switches to implement doors. Our contribution in Back to Bed is more significant, since it uses enemy characters patrolling certain areas to construct a door gadget. This can be applied to several other games with enemies within controllable patrolling routes or enemies that chase the player. Games in the stealth genre in particular may be able to use similar door constructions.

\cleardoublepage
%

\pdfbookmark[0]{Bibliography}{bib}

\bibliographystyle{IEEEtran}



\normalem
\bibliography{./IST-Thesis-MSc-Bibliography} 
\ULforem
\cleardoublepage

\appendix
\pdfbookmark[1]{Appendix A}{appendix}
\chapter{Instance Examples}
In this appendix, we present an example reduction for every one of our new results, starting with a formula (or a graph in the case of Super Meat Boy), and creating a level out of it. This will hopefully make it clearer how gadgets connect to one another and how puzzles' solutions are related to formula's.

\section{Cut the Rope}\label{example:cuttherope}
Here we show an example of a Cut the Rope reduction from a 3-CNFSAT instance. The formula represented is $(x \lor y \lor z) \land (\lnot{x} \lor z \lor \lnot{w}).$ 

Hats with the same number are linked. Note that the figures do not represent the full gadgets, only how they connect to one another in a complete reduction. 

\begin{figure}[!ht]
\center
 \subfloat[Start gadget.\label{fig:cuttherope_start}]{%
   \includegraphics[scale=0.43]{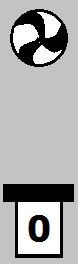}
 }
 \subfloat[Variable X.\label{fig:cuttherope_x}]{%
   \includegraphics[scale=0.17]{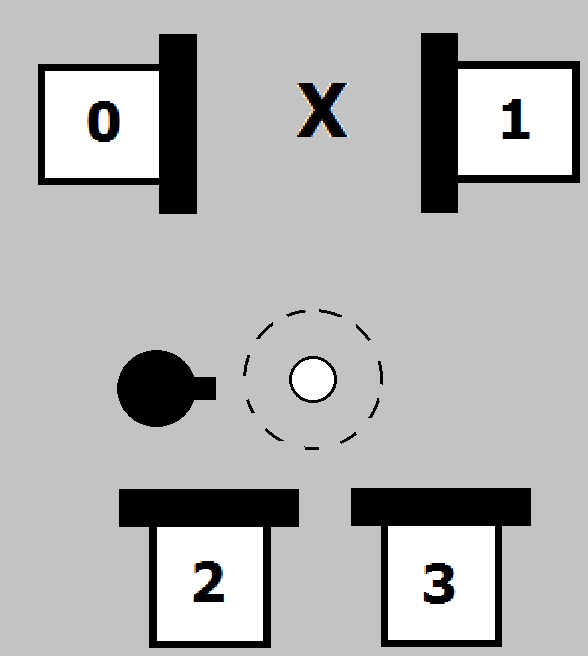}
 }
 \subfloat[Variable Y.\label{fig:cuttherope_y}]{%
   \includegraphics[scale=0.17]{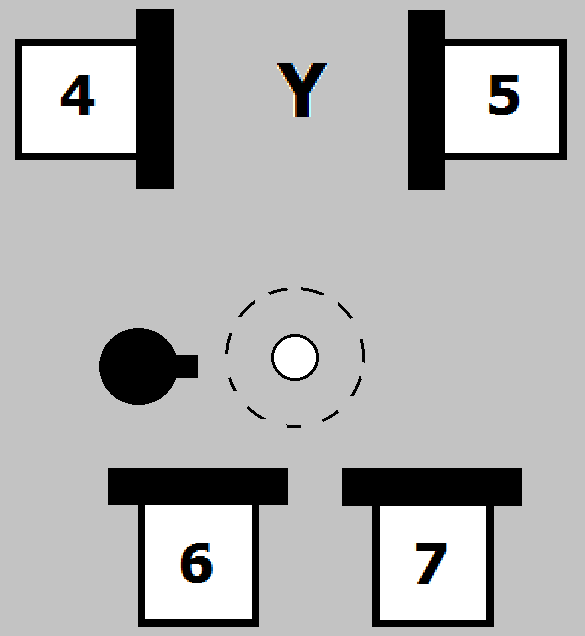}
 }
 \subfloat[Variable Z.\label{fig:cuttherope_z}]{%
   \includegraphics[scale=0.17]{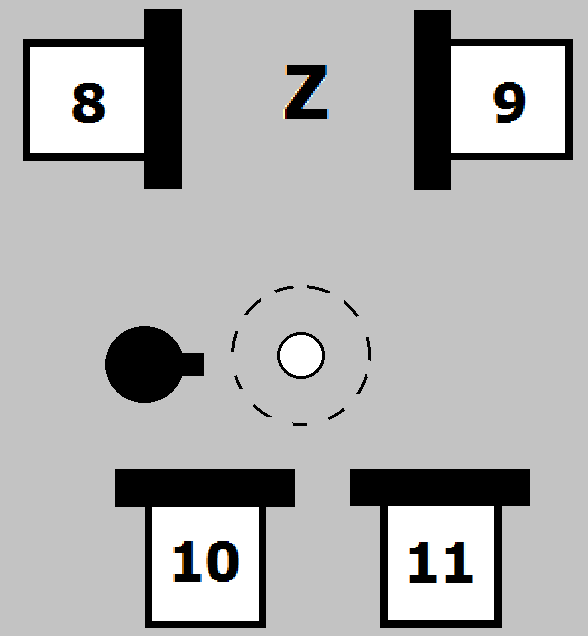}
 }
 \subfloat[Variable W.\label{fig:cuttherope_w}]{%
   \includegraphics[scale=0.16]{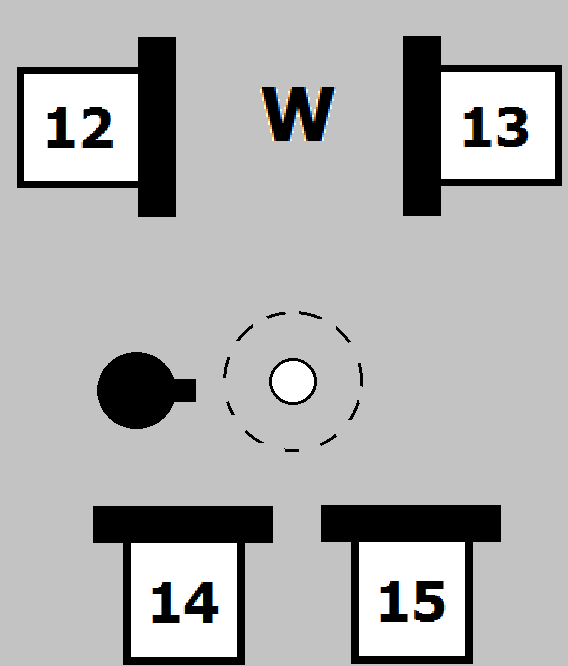}
 }
  \subfloat[Finish gadget.\label{fig:cuttherope_finish}]{%
   \includegraphics[scale=0.37]{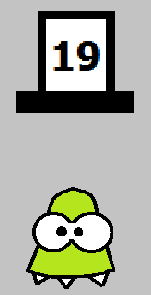}
 }
 
 \subfloat[Clause $x \lor y \lor z$.\label{fig:cuttherope_clause1}]{%
   \includegraphics[scale=0.14]{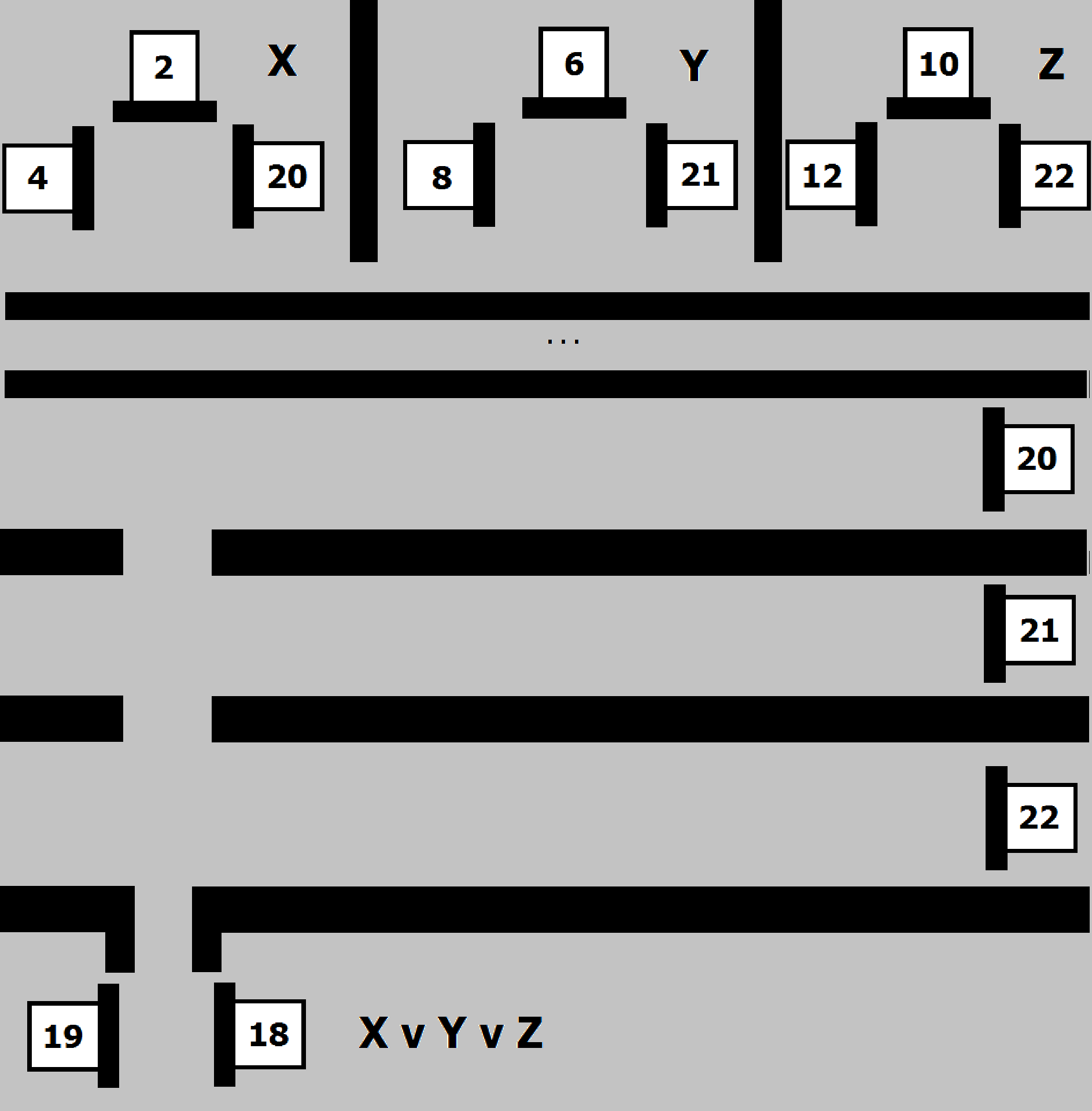}
 }
 \subfloat[Clause $\lnot{x} \lor z \lor \lnot{w}$.\label{fig:cuttherope_clause2}]{%
   \includegraphics[scale=0.14]{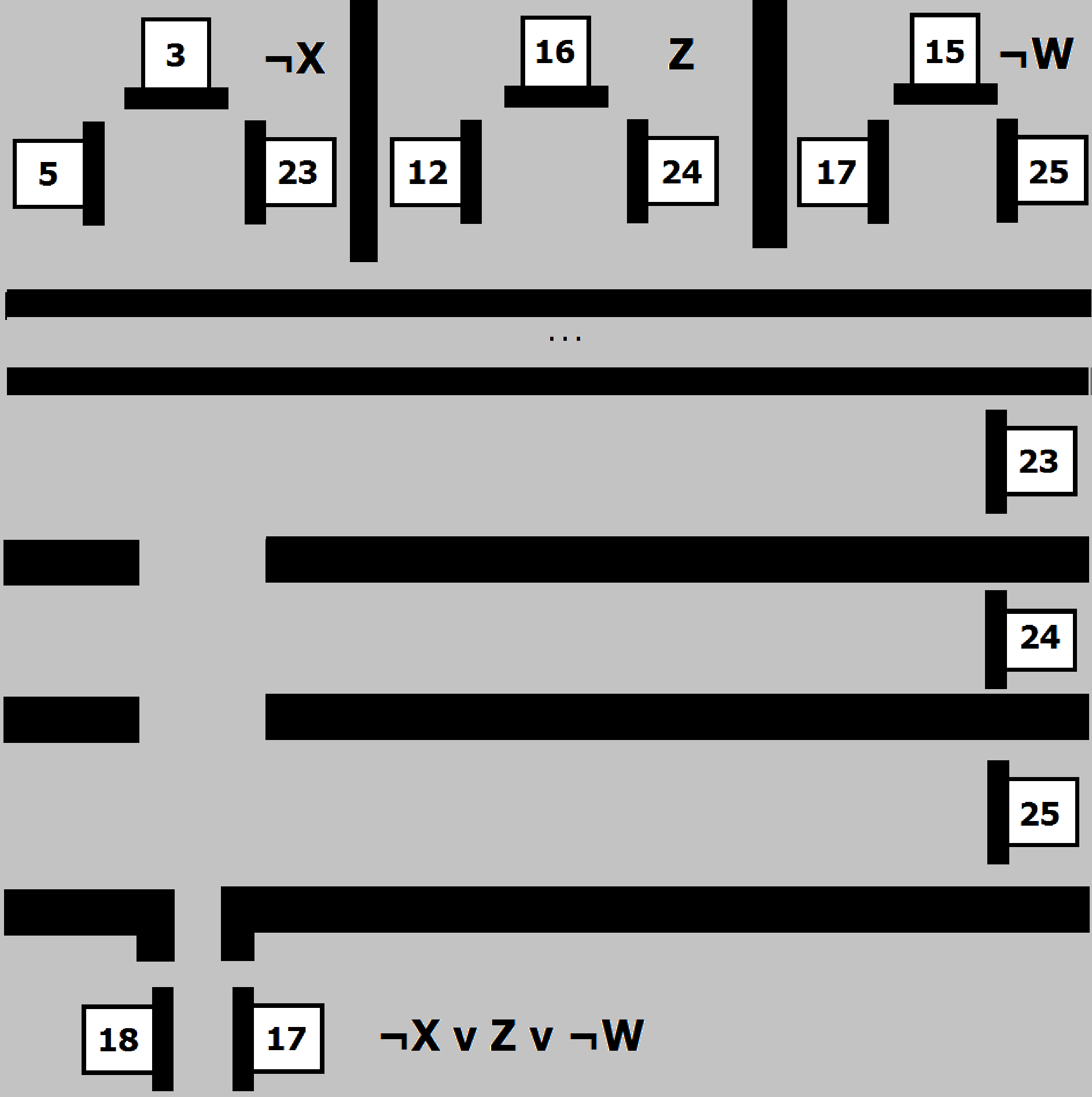}
 }
 \caption{Cut the Rope example scheme.}
 \label{fig:cuttherope_example}
\end{figure}

\section{Hexagonal Akari}\label{example:akarihex}
Here we present an example of a reduction from CircuitSAT to Hexagonal Akari. The formula represented is $(\lnot{x} \land y) \lor x$. The left section represents the AND part of the puzzle. The conversion was done using De Morgan's laws. The top section has the OR as well as the final section that ends the puzzle, requiring the formula to be true. The bottom middle has the variable $x$ to be selected. Note that the signal is propagated from top to bottom in this section, and then turns back up after the FAN-OUT. Variable $y$ is chosen at the lower left corner.

\begin{figure}[!ht]
\center
 \subfloat[Empty puzzle.]{%
   \includegraphics[scale=0.5]{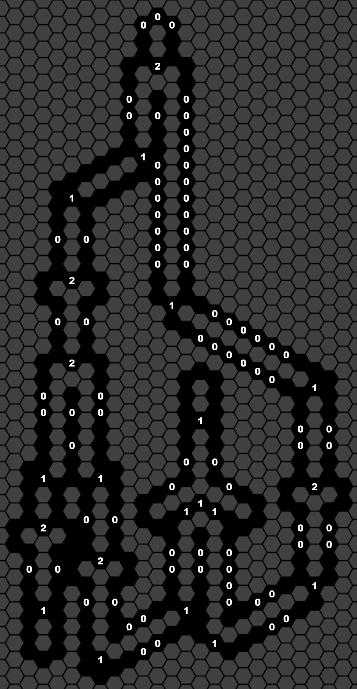}
 }
 \subfloat[Full propagation with x set to false.\label{fig:hexakari_example_notx}]{%
   \includegraphics[scale=0.5]{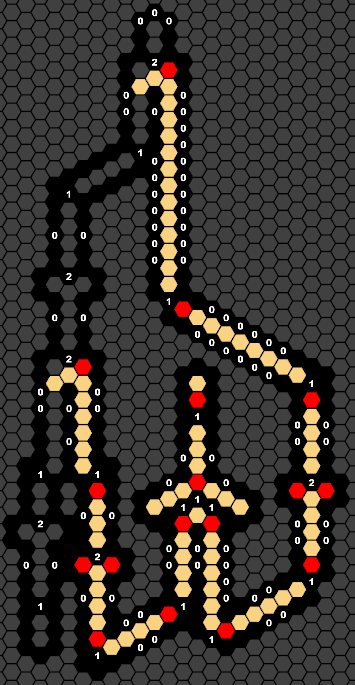}
 }
 \subfloat[Full propagation with x set to true.\label{fig:hexakari_example_x}]{%
   \includegraphics[scale=0.5]{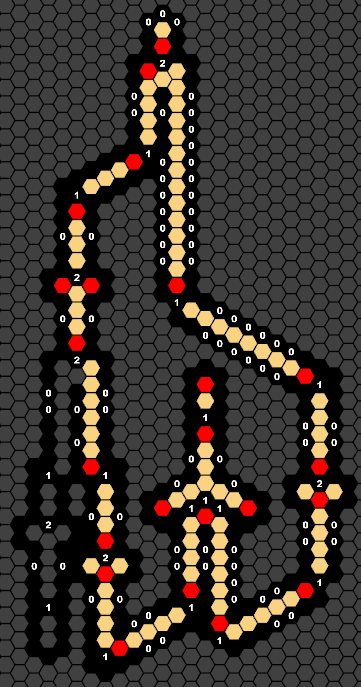}
 }
 \caption{Hexagonal Akari's reduction from $(\lnot{x} \land y) \lor x$.}
 \label{fig:hexakari_example}
\end{figure}

Figure~\ref{fig:hexakari_example_notx} shows all the propagation from assigning $false$ to $x$. From this choice, it is not yet known whether it leads to a valid solution by lighting the top of the gadget. Setting $y$ to $true$ would lead to a valid solution, but not setting $y$ to $false$.

Figure~\ref{fig:hexakari_example_x} shows all the propagation from assigning true to $x$. In this case, we can see from a single variable that the puzzle has a solution, just as setting $x$ to $true$ satisfies the original formula independently of $y$'s value.

\section{Triangular Akari}\label{example:akaritri}
\begin{figure}[!ht]
     \subfloat[Empty TriAkari instance.\label{fig:triakari_instance_empty}]{%
       \includegraphics[scale=0.45]{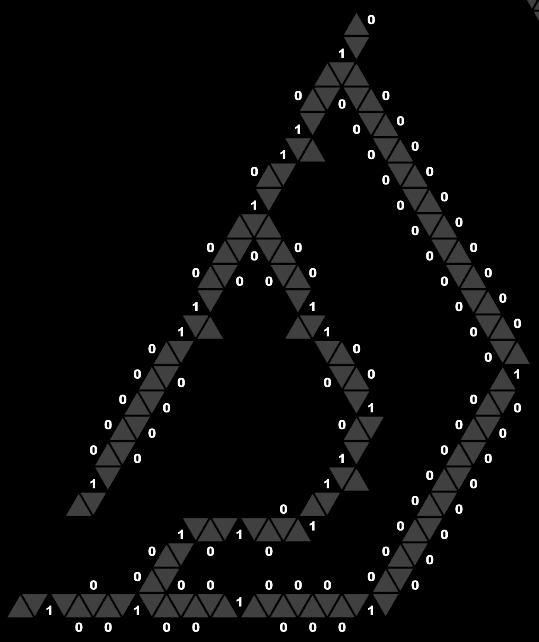}
     }
     \hfill
     \subfloat[Solved TriAkari instance for false $x$ and true $y$.\label{fig:triakari_instance_empty_annotated}]{%
       \includegraphics[scale=0.45]{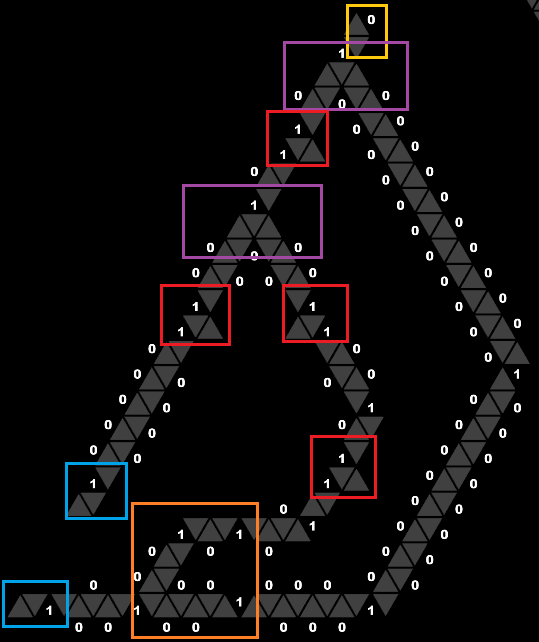}
     }
     \caption{Empty TriAkari instance. Annotated to emphasize different gadgets.}
     \label{fig:triakari_instance}
    \end{figure}

Figure~\ref{fig:triakari_instance_empty} shows an empty instance of Triangular Akari, for the formula $(\lnot x \land y) \lor x$. Figure~\ref{fig:triakari_instance_empty_annotated} shows the same instance with colored rectangles to emphasize each gadget. The color code is as follows:
\begin{enumerate}
    \item Blue represents variable selectors;
    \item Red represents NOTs;
    \item Orange represents FAN-OUTs;
    \item Purple represents ORs;
    \item Yellow represents a cap to force the whole circuit to have a fixed truth value.
\end{enumerate} 
The remaining pieces of the puzzle all form wires and turns that connect the color coded gadgets.

The bottom of the puzzle shows the variable $x$ and its FAN-OUT. The left part of the puzzle shows the variable $y$. Note that the AND gate is the lowest gate on the puzzle, constructed using only an OR gate, with its input and outputs negated.

\begin{figure}[!ht]
     \subfloat[$x=false$ and $y=false$.\label{fig:triakari_instance_notx_noty}]{%
       \includegraphics[scale=0.4]{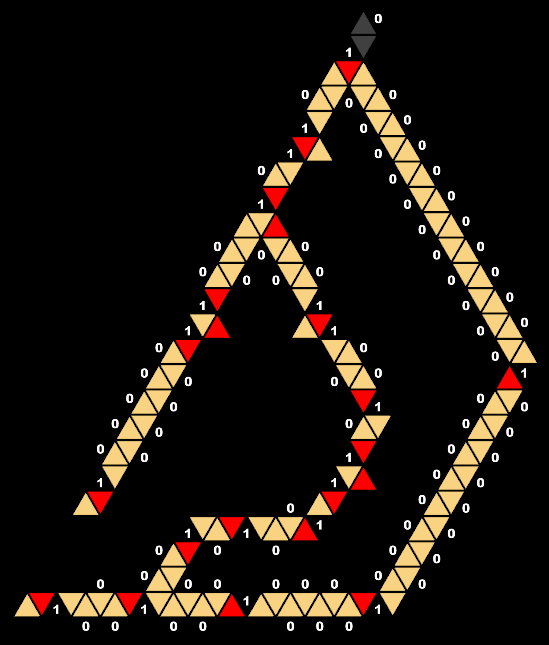}
     }
     \hfill
     \subfloat[$x=false$ and $y=true$.\label{fig:triakari_instance_notx_y}]{%
       \includegraphics[scale=0.4]{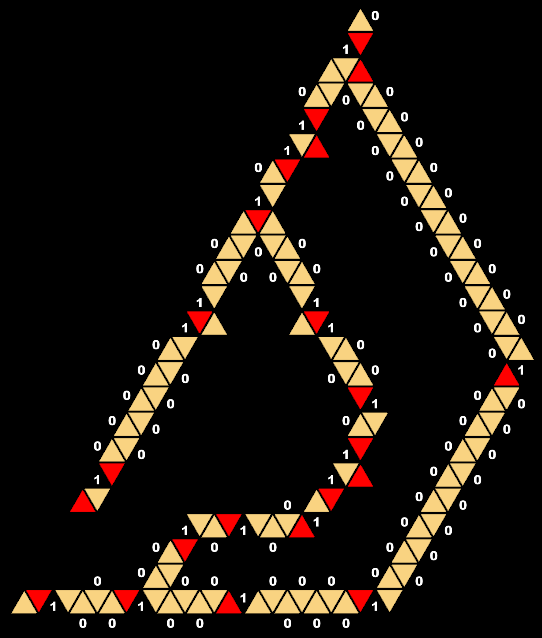}
     }
     \caption{Two different assignments of the variables.}
     \label{fig:triakari_instance_solved}
    \end{figure}

Figure~\ref{fig:triakari_instance_solved} shows one unsatisfiyng and one satisfying assignment of the variables.

\section{Offspring Fling}\label{example:offspringfling}
Here we present an example for the game Offspring Fling (OF). The formula used is $\forall x \exists y \forall z (\lnot x \lor z) \land (x \lor y) \land (y \lor z)$. Although Framework \ref{framework:pspace} assumes a quantified $3$-CNF formula, we use a $2$-CNF formula to keep the reduction size manageable. The generalization of the gadgets to $3$-CNF is straightforward, however. Note also that the various gadgets are connected through simple wires, which are not shown in this example. The crossover gadget is applied to resolve any wire crossing that may appear. ADD A CROSSOVER EXAMPLE -two types.

\begin{figure}[!ht]
    \center
     \subfloat[Universally quantified variable $x$.\label{fig:of_instance_var_x}]{%
       \includegraphics[scale=0.23]{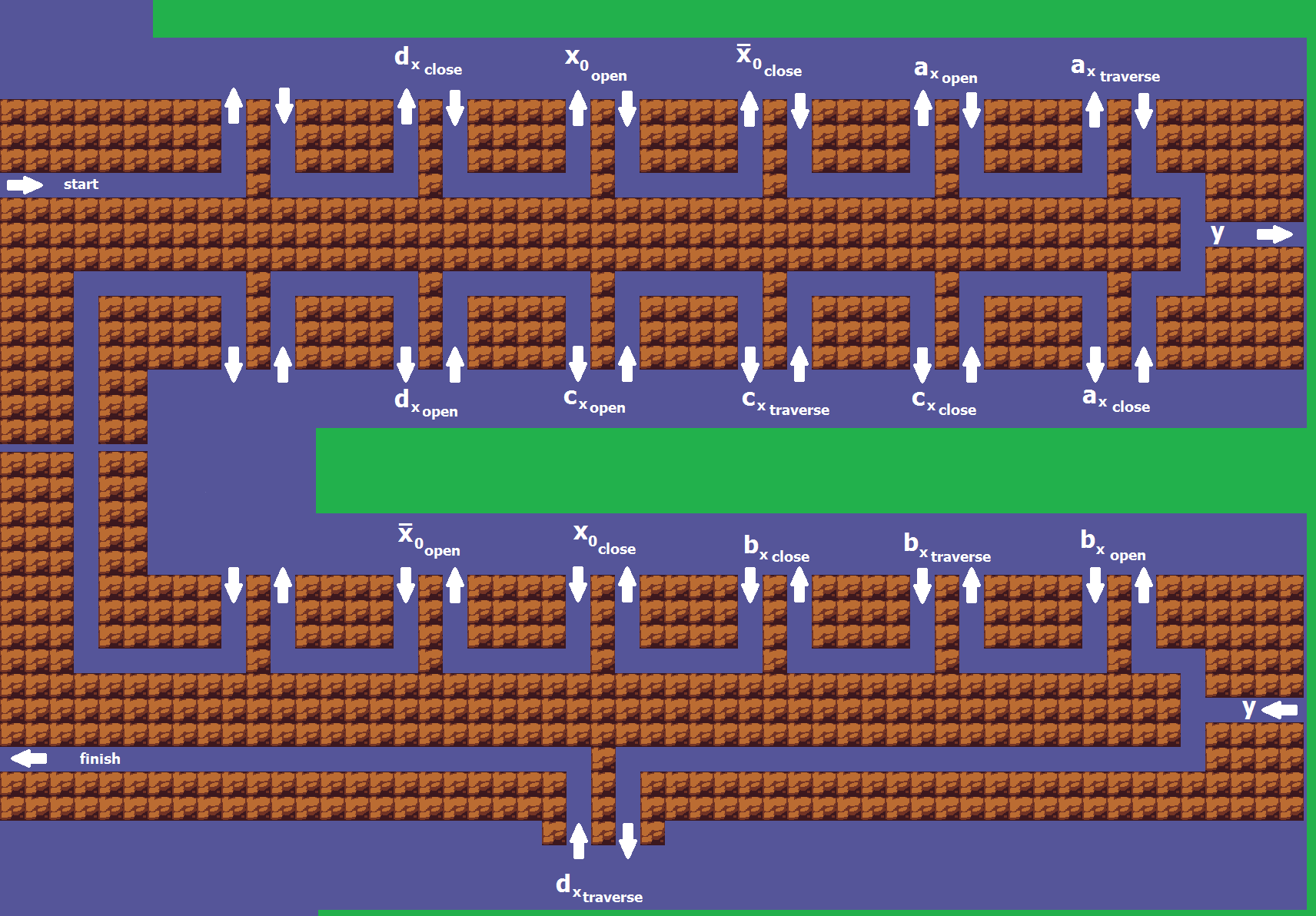}
     }\\
     
     \subfloat[Existentially quantified variable $y$.\label{fig:of_instance_var_y}]{%
       \includegraphics[scale=0.3]{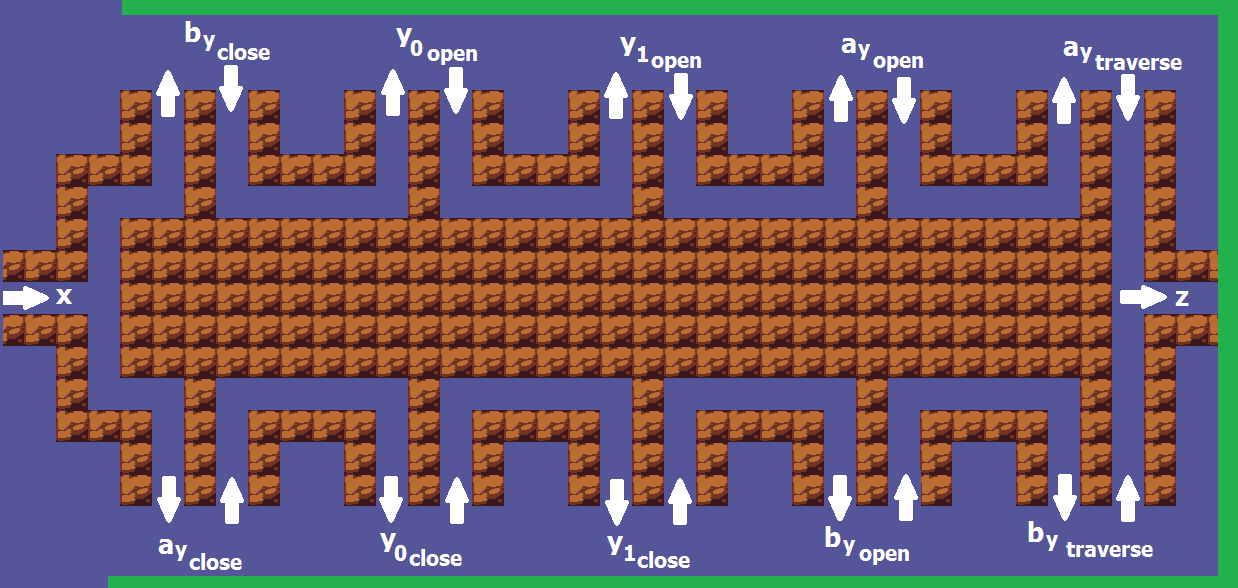}
     }\\
     
     \subfloat[Universally quantified variable $z$.\label{fig:of_instance_var_z}]{%
       \includegraphics[scale=0.23]{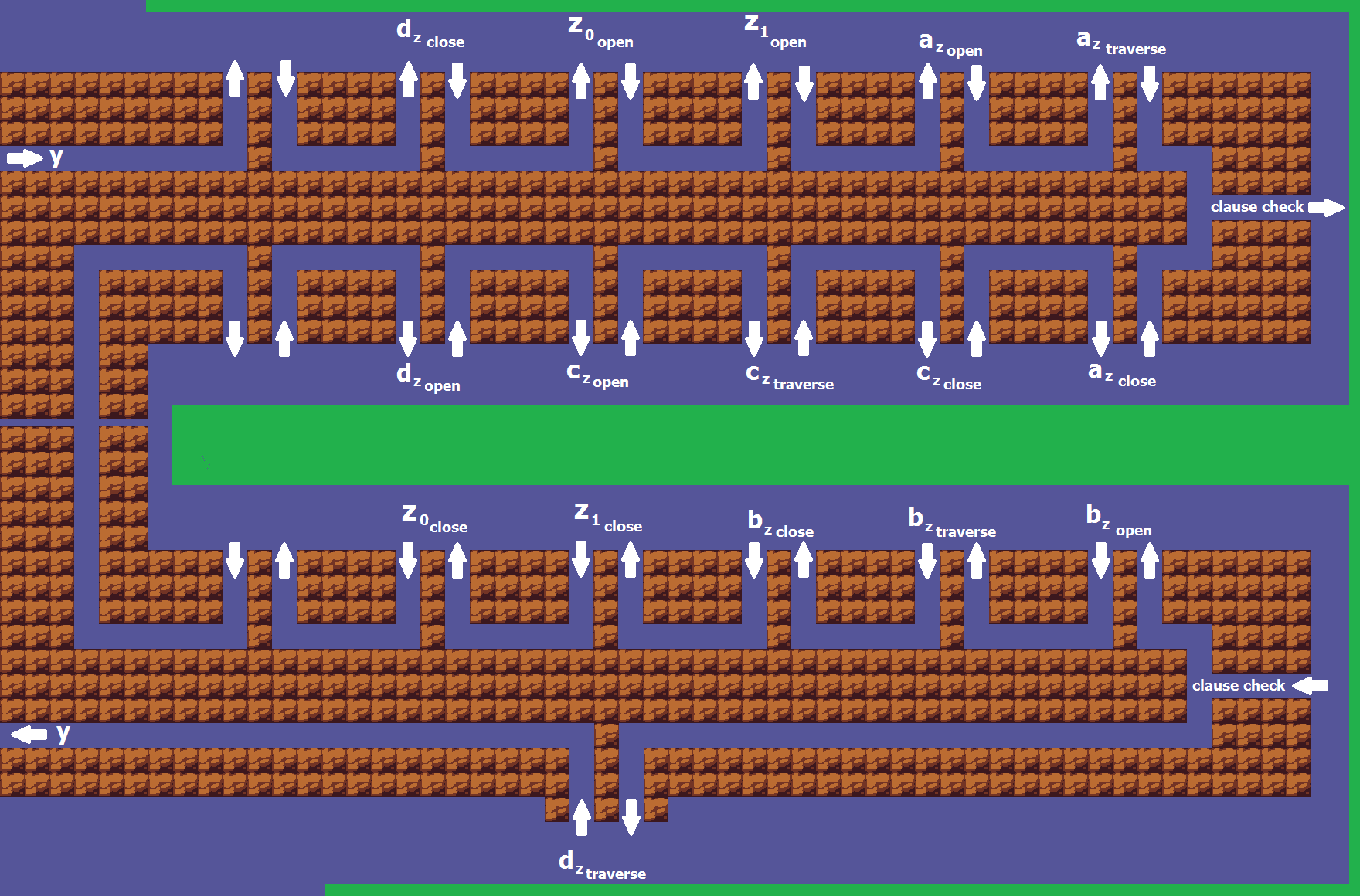}
     }
     \caption{Variable gadgets for OF's instance of formula $\forall x \exists y \forall z (\lnot x \lor z) \land (x \lor y) \land (y \lor z)$.}
     \label{fig:of_instance_var}
    \end{figure}

\begin{figure}[!ht]
    \center
     \subfloat[Open/Close door path for first occurrence of variable x.\label{fig:of_instance_door_x0}]{%
       \includegraphics[scale=0.4]{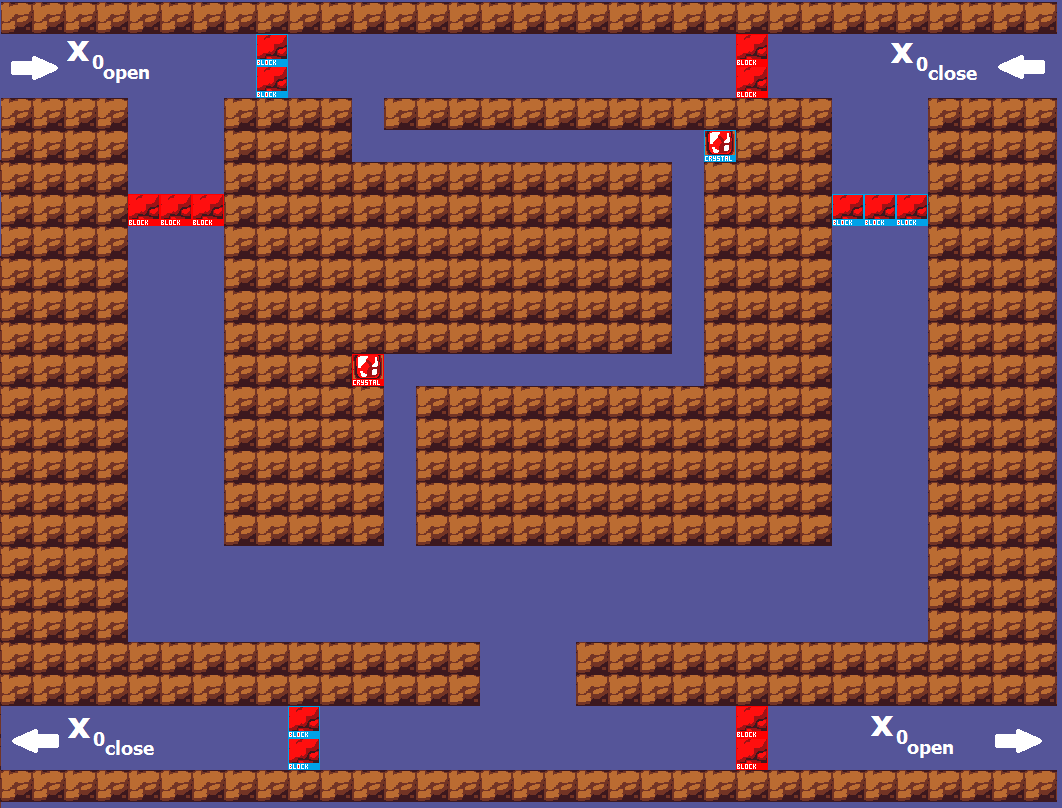}
     }\\
     
     \subfloat[Traverse door path for first occurrence of variable x.\label{fig:of_instance_traverse_x0}]{%
       \includegraphics[scale=0.7]{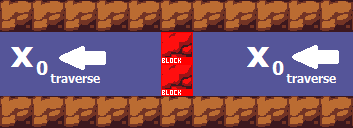}
     }
     \caption{Door example for first occurrence of variable $x$.}
     \label{fig:of_instance_door}
    \end{figure}

Figure~\ref{fig:of_instance_var} shows the universal and existential quantifier gadgets. The gadgets follow the ones presented in Framework~\ref{framework:pspace}, page~\pageref{framework:pspace}. $a, b, c, d, x_i, y_i, z_i$ are all doors. The green areas are used to wire the arrows to the respective door gadget. One door gadget is presented in Figure~\ref{fig:of_instance_door}, for the first appearance of variable $x$, $x0$. Remember that, for the respective variable, the red blocks of the Open/Close and Traverse gadgets share the same state. Every door is the same, with a different annotated name, so the remaining doors are implicit.

\begin{figure}[!ht]
    \center
     \subfloat[Clause $(\lnot x \lor z)$.\label{fig:of_instance_clause0}]{%
       \includegraphics[scale=0.4]{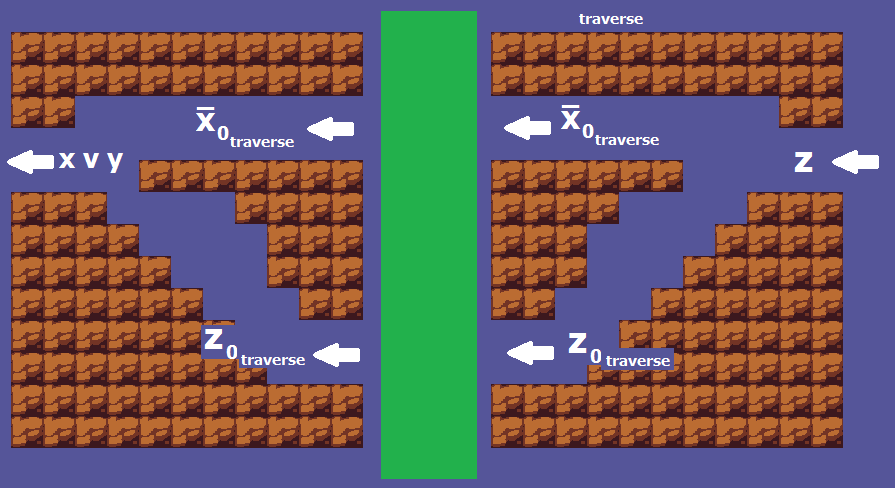}
     }\\
     
     \subfloat[Clause $(x \lor y)$.\label{fig:of_instance_clause1}]{%
       \includegraphics[scale=0.4]{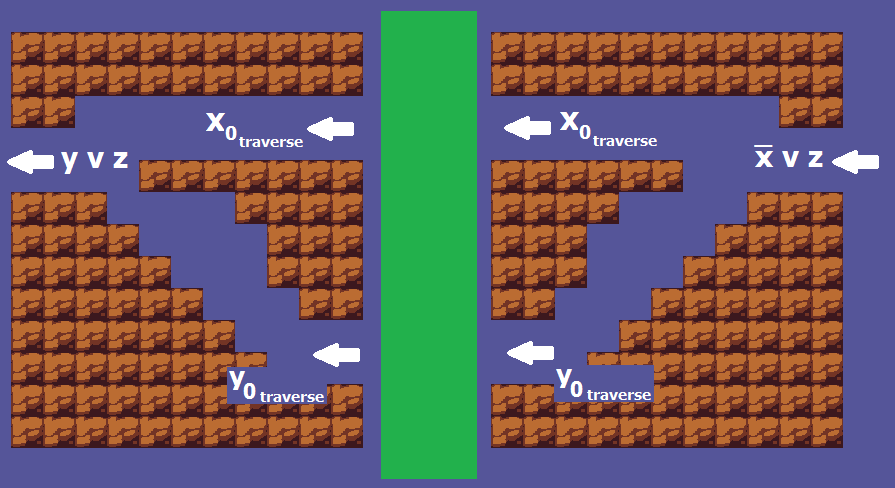}
     }
     
     \subfloat[Clause $( y \lor z)$.\label{fig:of_instance_clause2}]{%
       \includegraphics[scale=0.4]{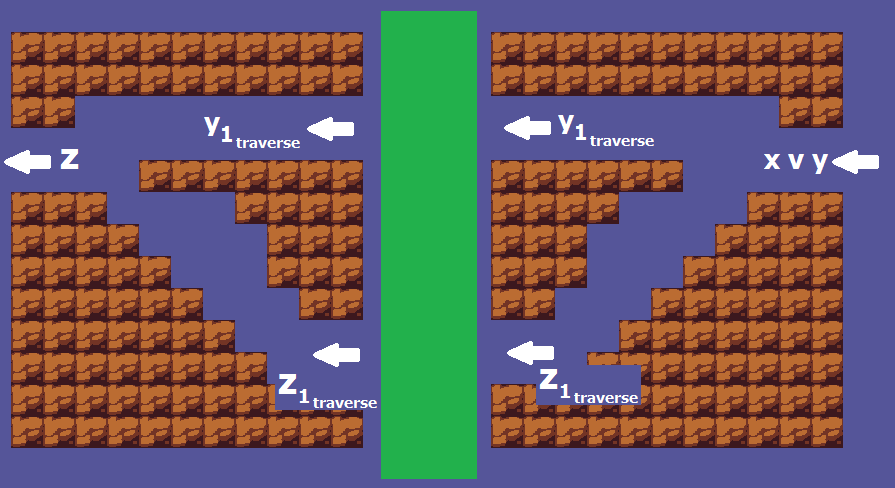}
     }
     \caption{Three clauses of the formula $\forall x \exists y \forall z (\lnot x \lor z) \land (x \lor y) \land (y \lor z)$. }
     \label{fig:of_instance_clauses}
    \end{figure}

Figure~\ref{fig:of_instance_clauses} shows the three clause gadgets. The green areas are used to wire the clause gadgets to the doors' traverse paths.

\section{Super Meat Boy}\label{example:supermeatboy}
\begin{figure}[!htb]
 \center
 \includegraphics[scale=0.4]{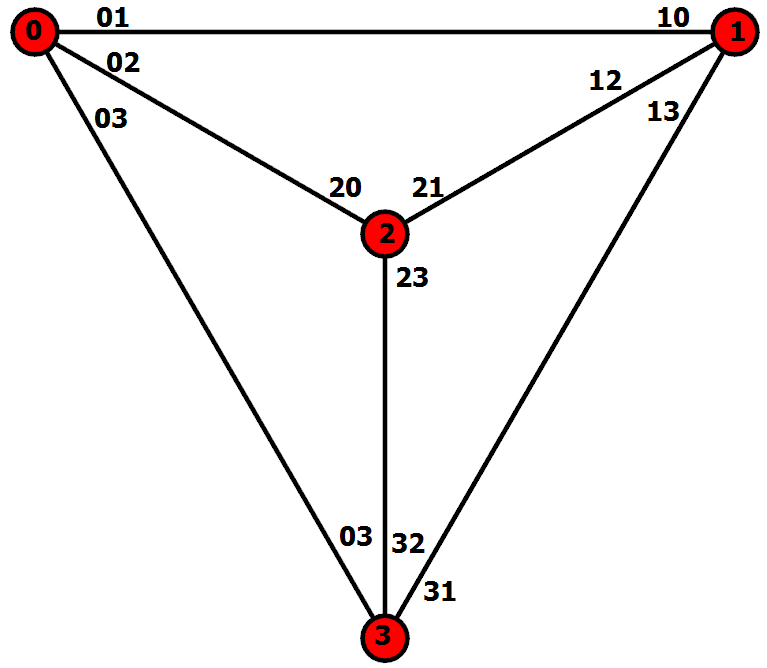}
 \caption{3-regular planar graph.}
 \label{fig:supermeatboy_graph}
\end{figure}

\begin{figure}[!htb]
\center
 \subfloat[Vertex 0.]{%
   \includegraphics[scale=0.3]{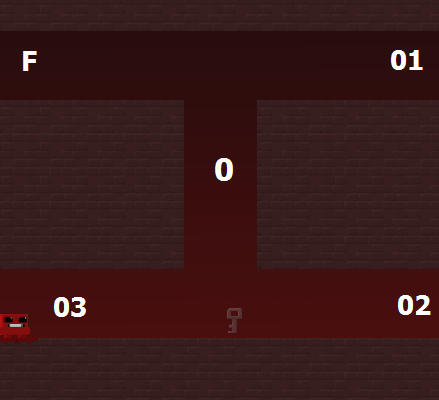}
 }
 \subfloat[Vertex 1.]{%
   \includegraphics[scale=0.3]{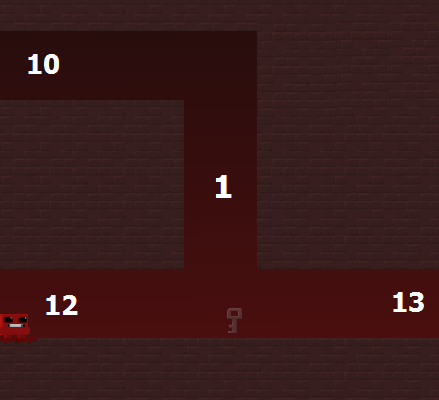}
 }\\
 
 \subfloat[Vertex 2.]{%
   \includegraphics[scale=0.3]{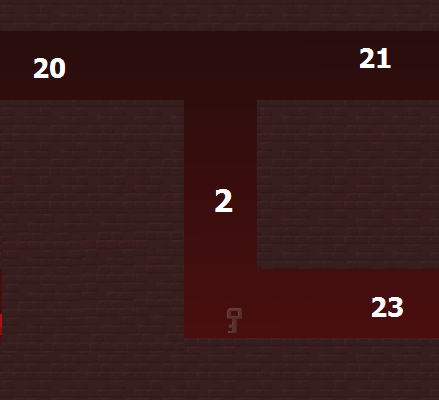}
 }
  \subfloat[Vertex 3.]{%
   \includegraphics[scale=0.3]{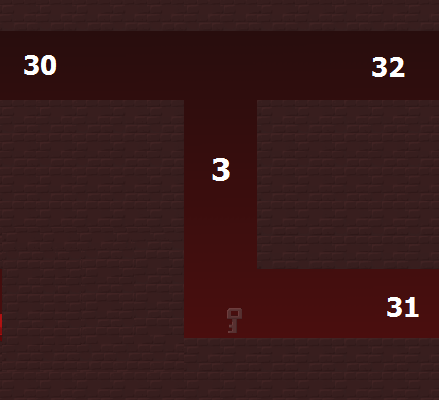}
 }\\
 \subfloat[Edge 01.]{%
   \includegraphics[scale=0.3]{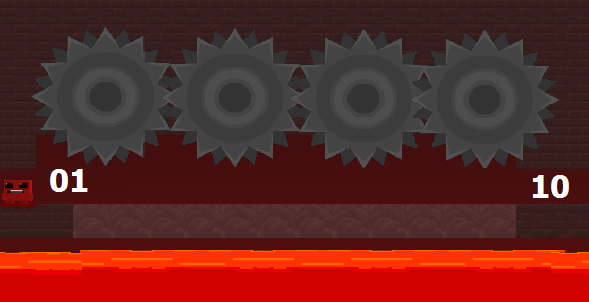}
 }
  \subfloat[Edge 02.]{%
   \includegraphics[scale=0.3]{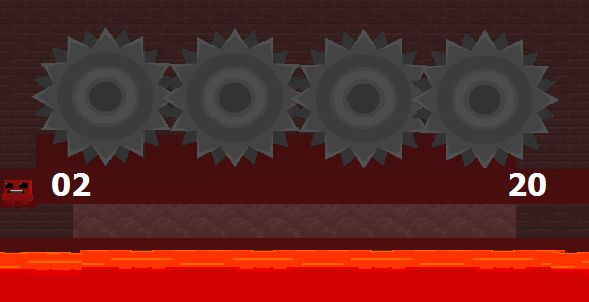}
 }\\
 \subfloat[Edge 03.]{%
   \includegraphics[scale=0.3]{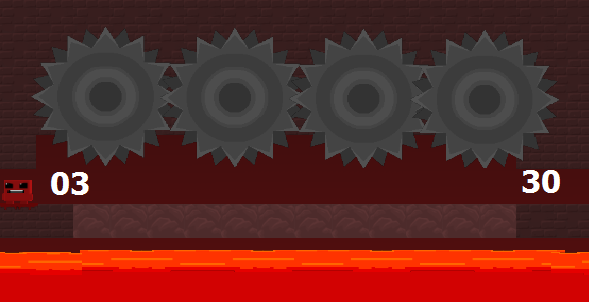}
 }
  \subfloat[Edge 12.]{%
   \includegraphics[scale=0.3]{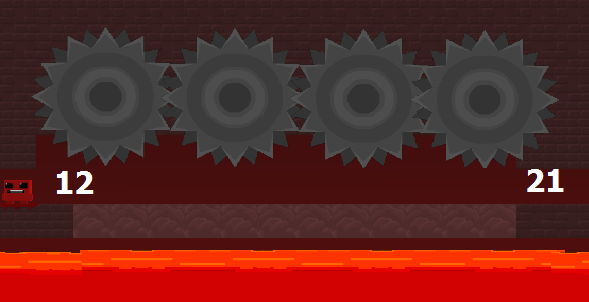}
 }\\
 \subfloat[Edge 13.]{%
   \includegraphics[scale=0.3]{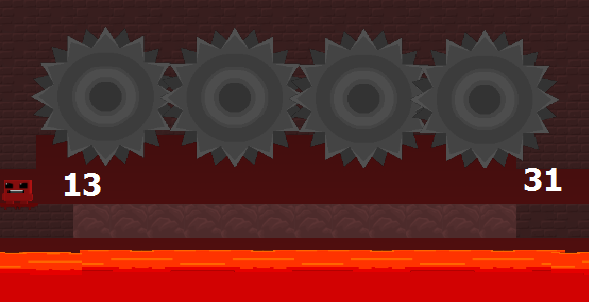}
 }
  \subfloat[Edge 23.]{%
   \includegraphics[scale=0.3]{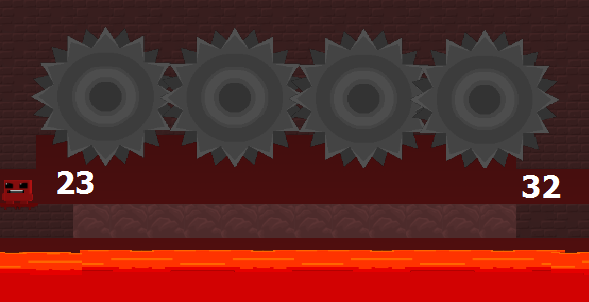}
 }
 \caption{Super Meat Boy instance.}
 \label{fig:supermeatboy_instance}
\end{figure}

Figure~\ref{fig:supermeatboy_graph} shows a 3-regular planar graph with vertices and edges named to make the correspondence with the gadgets easier. The annotated gadgets are shown in Figure~\ref{fig:supermeatboy_instance}. Note that the starting point is fixed in the level. The reduction would consist of $|v|$ levels, one for each vertex. This does not affect the polynomial complexity of the algorithm, as the difference is between O($|x|$) and O($|v|*|x|$), only an order of magnitude higher than the original reduction. 
It would be also possible to add a mechanism of forcing the player to exit from the same vertex from which he entered using keys and one-way paths. It would require crossovers, however; although easily implemented within the game, they do not fit into the framework, so the above solution will be kept.

\section{HexCells}\label{example:hexcells}
\begin{figure}[!ht]
    \center
     \subfloat[Empty Hexiom instance.\label{fig:hexcells_instance_empty}]{%
       \includegraphics[scale=0.65]{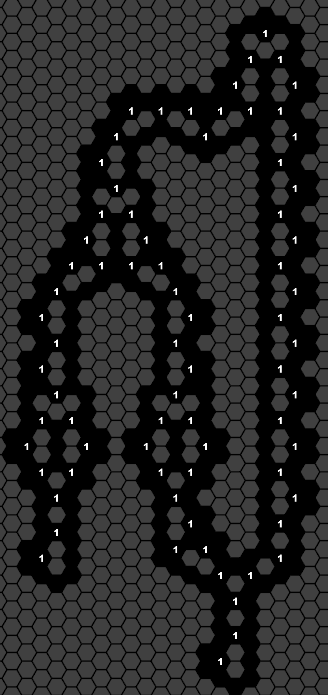}
     }
     \subfloat[Empty Hexiom instance. Annotated to emphasize different gadgets.\label{fig:hexcells_instance_solved}]{%
       \includegraphics[scale=0.65]{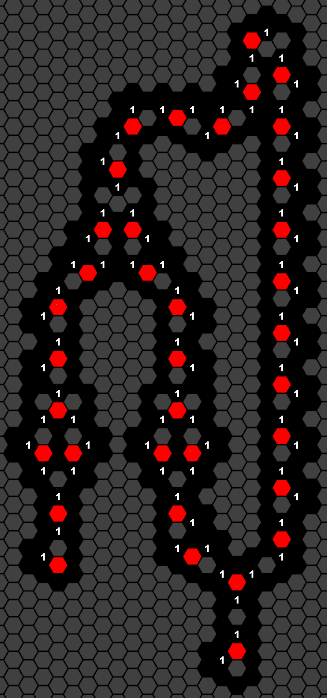}
     }
     \caption{Hexiom instance of the circuit $(\lnot x \land y) \lor x$.}
     \label{fig:hexcells_instance}
    \end{figure}

Figure~\ref{fig:hexcells_instance_empty} shows the HexCells instance from the initial formula $(\lnot x \land y) \lor x$. Note that the construction is exactly the same as Hexiom's, shown in the next section. The difference lies in the pieces that must be placed to solve the problem, and in that Hexiom, due to its additional constraints, requires extra gadgets to accommodate every piece.

Figure~\ref{fig:hexcells_instance_solved} shows one solution to the puzzle, where $x$ is false and $y$ is true. Note that any number of extra bombs can be placed in the empty cells outside the circuit without additional gadgets.

\section{Hexiom}\label{example:hexiom}
\begin{figure}[!ht]
    \center
     \subfloat[Empty Hexiom instance.\label{fig:hexiom_instance_empty}]{%
       \includegraphics[scale=0.65]{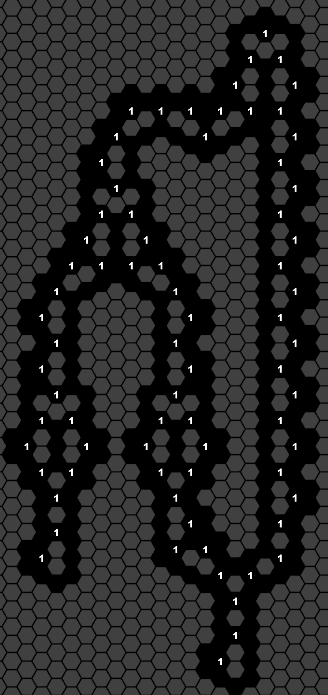}
     }
     \subfloat[Empty Hexiom instance. Annotated to emphasize different gadgets.\label{fig:hexiom_instance_empty_annotated}]{%
       \includegraphics[scale=0.65]{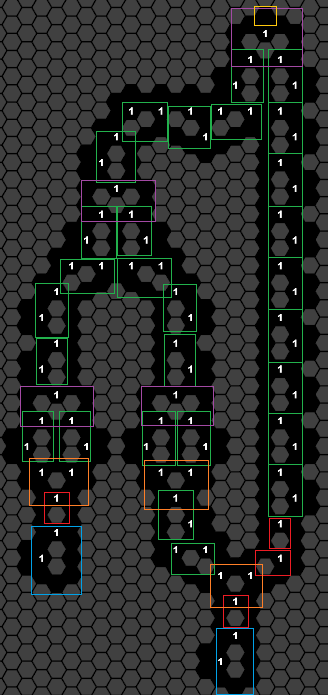}
     }
     \caption{Hexiom instance of the circuit $(\lnot x \land y) \lor x$.}
     \label{fig:hexiom_instance}
    \end{figure}

Figure~\ref{fig:hexiom_instance_empty} shows an example of an Hexiom reduction from the initial formula $(\lnot x \land y) \lor x$.  Figure~\ref{fig:hexiom_instance_empty} has colored rectangles enclosing different gadgets for ease of explanation of the excess problem. The color code is the following:
\begin{enumerate}
    \item Blue represents variable selectors;
    \item Red represents NOTs;
    \item Green represents WIREs and TURNs;
    \item Orange represents FAN-OUTs;
    \item Purple represents NORs;
    \item Yellow represents a cap to force the whole circuit to have a fixed truth value.
\end{enumerate}

\paragraph*{Body of the circuit.}
The left side shows the $(\lnot x \land y)$ portion of the formula, using NOR and Negated FAN-OUT gates exclusively. The right side shows the $x$ portion of the formula as a simple wire propagating to the top. Notice how there are two NOT gates here to shift the length of the wire in order to align everything at the top NOR gate. Finally, the top represents the final OR between the two sub-formulas(e?). Note the yellow rectangle, preventing the signal from propagating further. In this case, the top 1-cell can only be satisfied if the gate's value is false. Because the gate is a NOR, it being false is the same as an OR gate being true. This Hexiom puzzle can only be solved with a solution for the original formula, where the OR is true.
\paragraph*{Excess \& Parity.} After showing how the circuit results in a puzzle, we now need to show to account for the different configurations. First, we count the number of gadgets:
\begin{enumerate}
    \item Variable Selector: 2;
    \item NOT: 4;
    \item WIRE: 28;
    \item FAN-OUT: 3;
    \item NOR: 4;
\end{enumerate}
Then, we need to take into account the number of movable piece that each gadget adds, as presented in Section~\ref{proof:hexiom}, page~\pageref{proof:hexiom}. As a reminder, we list the excess gadgets required:
\begin{enumerate}
    \item Variable Selector: 1-cell or 2-cell, value dependent; Add 2-excess gadget and account for parity;
    \item NOT: 2-cell, value dependent; Add 2-excess gadget.
    \item WIRE: 2-cell always used; No excess gadget required.
    \item FAN-OUT: 3-cell, value dependent; Add 3-or-1 gadget.
    \item NOR: one 2-cell or one 3-cell, value dependent; Add 2-excess and 3-or-1 gadgets.
\end{enumerate}
Referring to the balance table in Section~\ref{proof:hexiom}, page~\pageref{proof:hexiom}, we know that only the Variable Selection, NOT and FAN-OUT gadgets influence the overall parity of the circuit. As a result, the parity of this particular instance is even, and so no parity correcting gadget is required.

\begin{figure}[!ht]
    \center
     \subfloat[Partially solved Hexiom instance for $x=true$.\label{fig:hexiom_instance_xprop}]{%
       \includegraphics[scale=0.65]{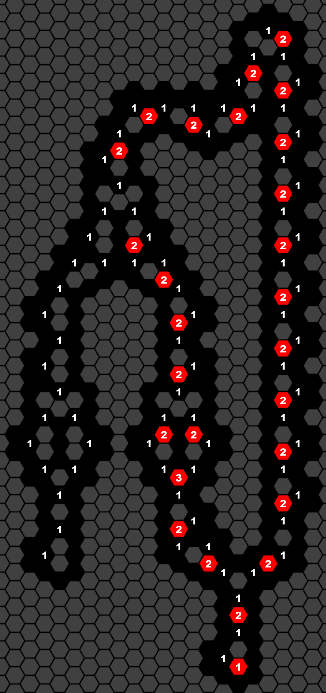}
     }
     \subfloat[Solved Hexiom instance for $x=false$ and $y=true$.\label{fig:hexiom_instance_notx_y}]{%
       \includegraphics[scale=0.65]{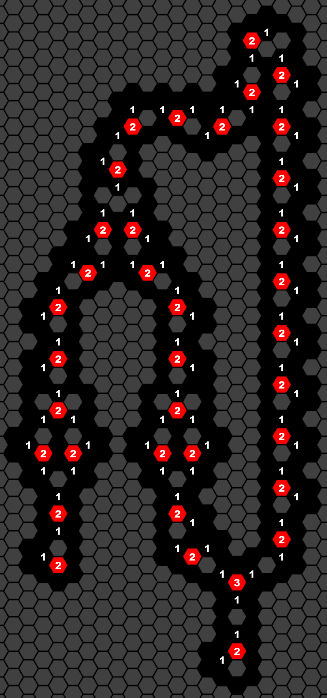}
     }
     \caption{Hexiom instance of the circuit $(\lnot x \land y) \lor x$.}
     \label{fig:hexiom_instance2}
    \end{figure}

To conclude, then, this instance would be composed of the main body shown in Figure~\ref{fig:hexiom_instance} with:
\begin{itemize}
    \item 10x 2-EXCESS gadgets from the main body;
    \item 7x 3-or-\{2,1\} gadgets from the main body;
    \item 7x 2-EXCESS gadget from the extra 3-or-\{2,1\} gadgets;
\end{itemize}
Figure~\ref{fig:hexiom_instance_xprop} shows the main body of the puzzle after choosing true for the value of $x$. Note how it is possible to see that the puzzle as a solution when $x$ is true, regardless of $y$'s value.
Figure~\ref{fig:hexiom_instance_notx_y} shows the other solution, where $x$ is false. Here, $y$ must be true for the puzzle to be solved.

\section{Back to Bed}\label{example:backtobed}
Here we present an instance of our last reduction, from PSPACE to the game Back to Bed, described in Section~\ref{proof:backtobed}, page~\pageref{proof:backtobed}. As in Offspring Fling's proof, the formula used is $\forall x \exists y \forall z (\lnot x \lor z) \land (x \lor y) \land (y \lor z)$. Despite being a quantified 2-CNFSAT formula, the generalization to 3-CNFSAT is straightforward, and in this way we can keep the size of the level more manageable.

The proof is similar to Offspring Fling's, with two significant differences: the first is that doors are implemented with the use of patrolling enemies that can be moved, instead of the use of buttons and open or close paths; the second is that the games has teleports, eliminating the need for a crossover and wiring in general.

We refer back to Figure~\ref{fig:qbf_fw} for the scheme of the level which we implement here. Recall that the player and the NPC that must reach a goal are separate entities and that the player should carry a fish at all times (the fish can bridge gaps of length exactly $2$.

\begin{figure}[!ht]
    \center
     \subfloat[Universally quantified variable x.\label{fig:backtobed_x}]{%
       \includegraphics[scale=0.45]{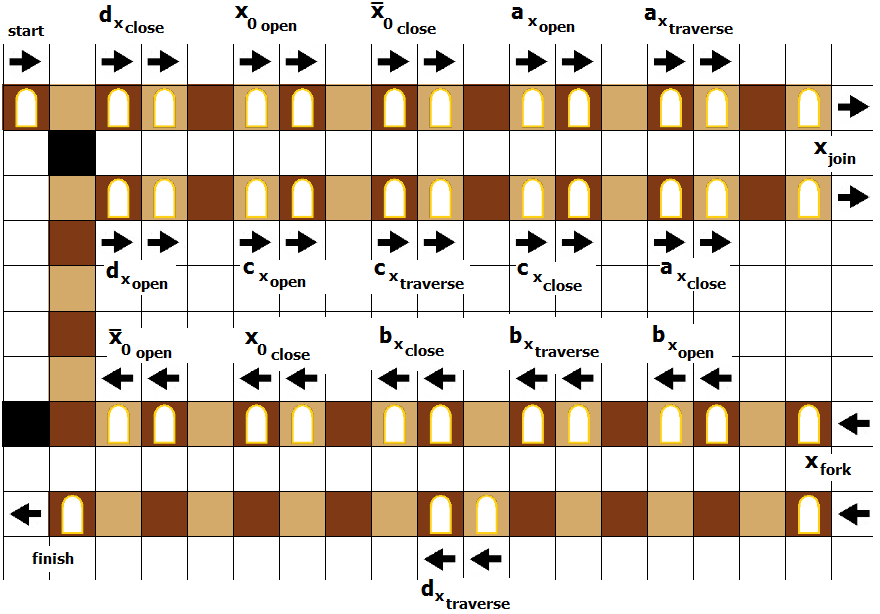}
     }\\
     \subfloat[Existentially quantified variable y.\label{fig:backtobed_y}]{%
       \includegraphics[scale=0.45]{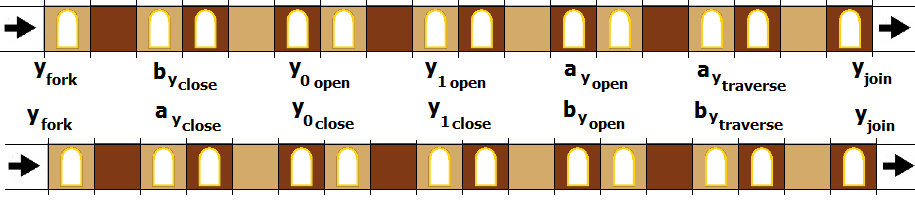}
     }\\
     \subfloat[Universally quantified variable z.\label{fig:backtobed_z}]{%
       \includegraphics[scale=0.45]{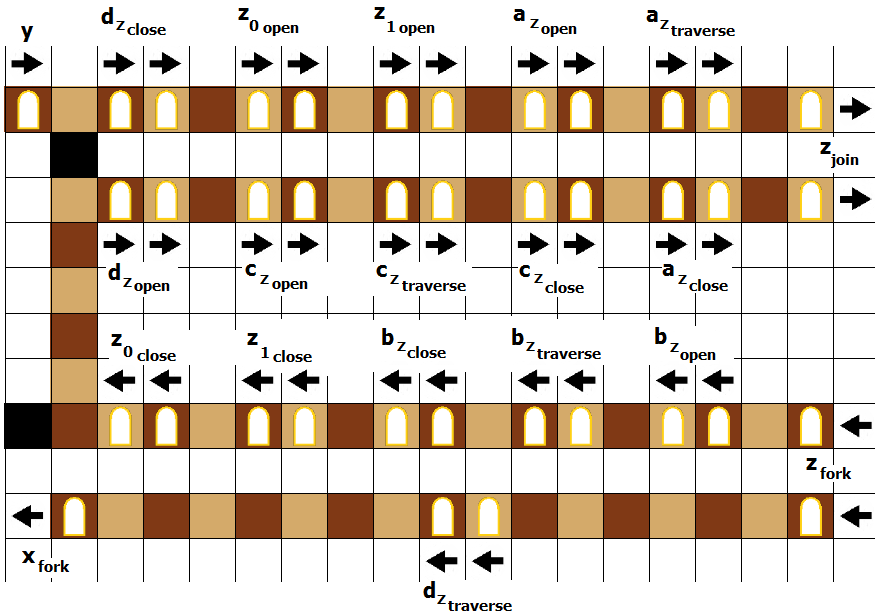}
     }
     \caption{Variables x, y, z.}
     \label{fig:backtobed_variables}
\end{figure}

\begin{figure}[!ht]
    \center
     \subfloat[Fork.\label{fig:backtobed_fork}]{%
       \includegraphics[scale=0.31]{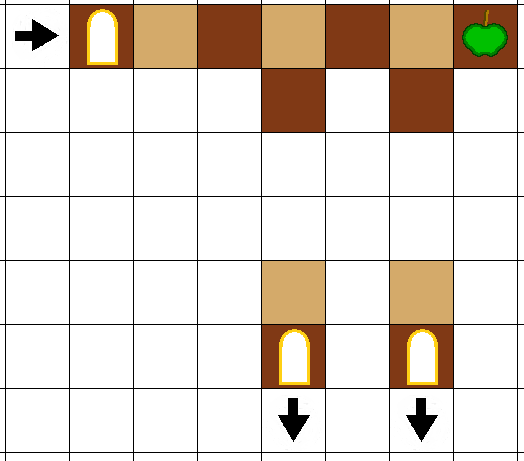}
     }
     \subfloat[Join.\label{fig:backtobed_join}]{%
       \includegraphics[scale=0.5]{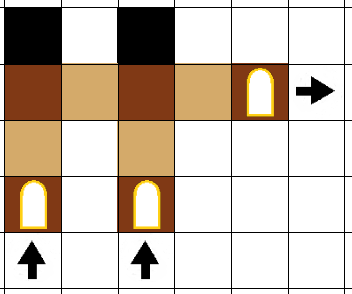}
     }
     \caption{Fork and Join auxiliary gadgets.}
     \label{fig:backtobed_forkjoin}
\end{figure}

Figure~\ref{fig:backtobed_variables} shows the three variables gadgets. The mirror pairs are not necessary, since we could connect one door gadget to another; this representation hopefully helps conforming with the framework presented and makes the door abstraction easier to understand. Note the need to join and fork gadgets when the player is presented with a choice. The two gadgets are shown in Figure~\ref{fig:backtobed_forkjoin}. 

\begin{figure}[!htb]
     \center
     \includegraphics[scale=0.65]{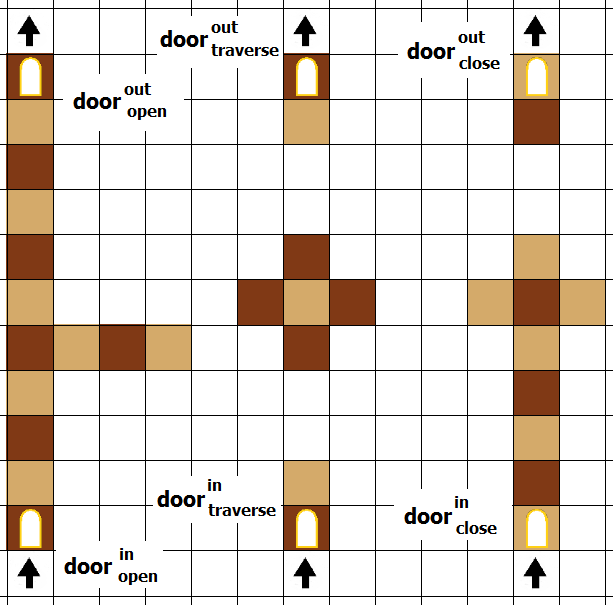}
     \caption{Door.}
     \label{fig:backtobed_door_annotated}
\end{figure}

Figure~\ref{fig:backtobed_door_annotated} shows a generic door. All the gadgets are identical, with the word "door" replaced with the respective variable designation ($v_i$). 

\begin{figure}[!ht]
    \center
     \subfloat[Clause $(\lnot x \lor z)$.\label{fig:backtobed_instance_clause0}]{%
       \includegraphics[scale=0.45]{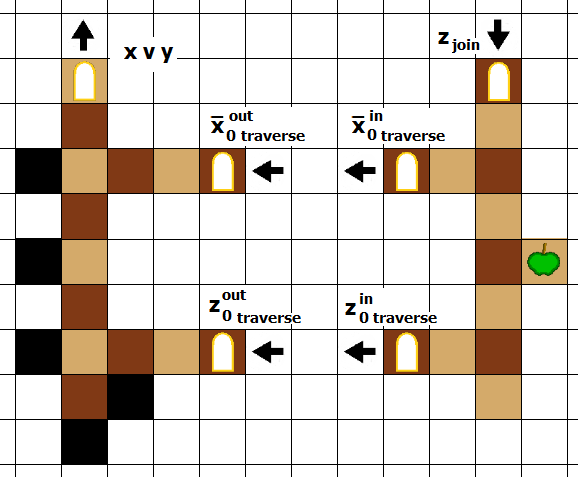}
     }\\
     
     \subfloat[Clause $(x \lor y)$.\label{fig:backtobed_instance_clause1}]{%
       \includegraphics[scale=0.45]{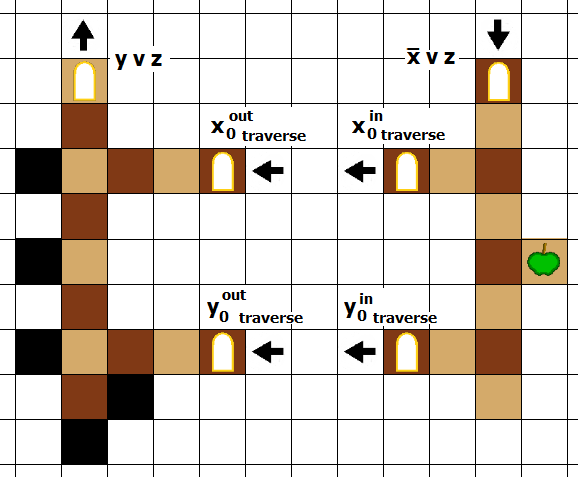}
     }
     
     \subfloat[Clause $( y \lor z)$.\label{fig:backtobed_instance_clause2}]{%
       \includegraphics[scale=0.45]{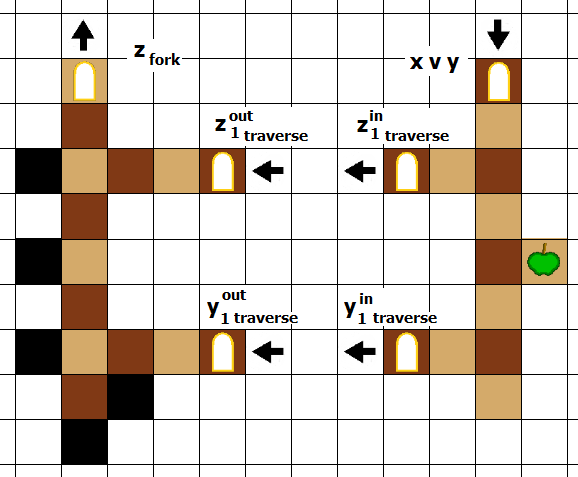}
     }
     \caption{Three clauses of the formula $\forall x \exists y \forall z (\lnot x \lor z) \land (x \lor y) \land (y \lor z)$. }
     \label{fig:backtobed_clauses}
\end{figure}

Figure~\ref{fig:backtobed_clauses} shows the three clauses.

To conclude, recall that between any mirror pair, we add a constant (but large) number of the path segment gadget (see Figure~\ref{fig:backtobed_pathseg}) to prevent the player from moving the apples from one gadget to another.

\cleardoublepage

\end{document}